\documentclass[a4paper,12pt, leqno]{article}
\usepackage{amsmath}
\usepackage{amssymb, latexsym}
\usepackage{amsfonts}
\usepackage{makeidx}
\usepackage{amssymb}

\usepackage{color}     

\newtheorem{lem}{Lemma}[section]
\newtheorem{thm}{Theorem}[section]
\newtheorem{prop}{Proposition}[section]

\numberwithin{equation}{section}
\newcommand{\dbar}{d\!\!\!{\lower-0.6ex\hbox{$-$}}\!}
\newcommand{\dslash}{d\!\!\!{\lower-0.6ex\hbox{$-$}}}
\newcommand{\e}{\varepsilon}
\newcommand{\h}{\hbar}
\newcommand{\ott}{\lower-0.4ex\hbox{${\scriptscriptstyle{\otimes}}$}}
\newcommand{\btt}{\lower-0.2ex\hbox{${\scriptscriptstyle{\bullet}}$}}
\newcommand{\ctt}{\lower-0.2ex\hbox{${\scriptscriptstyle{\circ}}$}}
\newcommand{\dtt}{\lower-0.2ex\hbox{${\scriptscriptstyle{\diamond}}$}}
\newcommand{\odt}{\lower-0.4ex\hbox{${\scriptscriptstyle{\odot}}$}}

\begin{document}

\title{Deformation Expression for Elements of Algebras (IV)\\
--Matrix elements and related integrals--}


\author{
     Hideki Omori\thanks{ Department of Mathematics,
             Faculty of Sciences and Technology,
        Tokyo University of Science, 2641, Noda, Chiba, 278-8510, Japan,
         email: omori@ma.noda.tus.ac.jp}
        \\Tokyo University of Science
\and  Yoshiaki Maeda\thanks{Department of Mathematics,
                Faculty of Science and Technology,
                Keio University, 3-14-1, Hiyoshi, Yokohama,223-8522, Japan,
                email: maeda@math.keio.ac.jp}
          \\Keio University
\and  Naoya Miyazaki\thanks{ Department of Mathematics, Faculty of
Economics, Keio University,  4-1-1, Hiyoshi, Yokohama, 223-8521, Japan,
        email: miyazaki@hc.cc.keio.ac.jp}
        \\Keio University
\and  Akira Yoshioka \thanks{ Department of Mathematics,
          Faculty of Science, Tokyo University of Science,
         1-3, Kagurazaka, Tokyo, 102-8601, Japan,
         email: yoshioka@rs.kagu.tus.ac.jp}
           \\Tokyo University of Science
     }

\maketitle

\tableofcontents

\pagestyle{plain}



\setcounter{equation}{0}

\bigskip 

In this note, we mainly consider the extended Weyl algebra of two 
generators $(u,v)$, that is, the algebra generated by $u,v$ with 
the fundamental commutation relation $u{*}v{-}v{*}u=-i\h$.
Weyl algebra is realized on the space ${\mathbb C}[u,v]$ by 
defining various product $*_{_K}$ depending on a symmetric matrix 
$K$ called the expression parameter. Via such expressions and 
ordinary calculus one can treat various transcendental elements 
such as $*$-exponential functions and elements obtained 
by integrations. 

In particular, setting
$\frac{1}{i\h}u{\ctt}v=\frac{1}{2}(u{*}v{+}v{*}u),$ 
we show that the $*$-exponential function 
$e_*^{z(\frac{1}{i\h}u{\ctt}v{+}\alpha)}$  
has singular points depending on the expression parameters.
The important view point in this chapter is that the expression 
parameter $K$ is moving by the ``individual time parameter'' $\tau$. 

In such an extended Weyl algebra, 
there are three idempotent elements, called  
vacuum, bar-vacuum and pseudo-vacuum, each of them give a matrix 
representation of Weyl algebra. 

Furthermore the $*$-exponential function 
$e_*^{t\frac{1}{i\h}u{\ctt}v}$ is rapidly decreasing on the imaginary axis. 
This defines two different $*$-inverses of $\frac{1}{i\h}u{\ctt}v$.    
By using this, 
the analytic continuation of 
$(\alpha{+}\frac{1}{i\h}u{\ctt}v)^{-1}$ are defined.

Note that the numbers $\alpha$ such 
that $(\alpha{+}\frac{1}{i\h}u{\ctt}v)^{-1}$ does not
exist is called the ``spectre''. It is remarkable that 
the spectra depends on the expression parameters. Several 
interesting properties of elements defined by integrations 
will be given. 
\section{Fundamental facts for star-products}   

For an arbitrary fixed $2{\times}2$-complex symmetric matrix 
$K$, we set $\Lambda=K{+}J$ where $J$ is the standard skew-symmetric
matrix  
$J{=}\tiny{
\begin{bmatrix}
0&-1\\
1&0
\end{bmatrix}}$. 
We define a product ${*}_{_{\Lambda}}$ on the space
of polynomials  ${\mathbb C}[u_1,u_2]$ by the formula 
\begin{equation}
 \label{eq:KK}
 f*_{_{\Lambda}}g=fe^{\frac{i\h}{2}
(\sum\overleftarrow{\partial_{u_i}}
{\Lambda}{}^{ij}\overrightarrow{\partial_{u_j}})}g
=\sum_{k}\frac{(i\h)^k}{k!2^k}
{\Lambda}^{i_1j_1}\!{\cdots}{\Lambda}^{i_kj_k}
\partial_{u_{i_1}}\!{\cdots}\partial_{u_{i_k}}f\,\,
\partial_{u_{j_1}}\!{\cdots}\partial_{u_{j_k}}g.   
\end{equation}
It is known and not hard to prove that 
$({\mathbb C}[u_1,u_2],*_{_{\Lambda}})$ are mutually 
isomorphic associative algebras for every $K$.
The isomorphism class is called the Weyl algebra, denoted by  
$(W_2; *)$. If $K$ is fixed, then every element $A{\in}(W_2; *)$ 
is expressed in the form of ordinary polynomial, which we 
denote by ${:}A{:}_{_K}\in {\mathbb C}[u_1,u_2]$. For instance 
$$
{:}u_1{*}u_1{:}_{_K}=(u_1)^2{+}\frac{i\h}{2}K_{11}, \quad 
{:}u_1{*}u_2{:}_{_K}=u_1u_2{+}\frac{i\h}{2}(K_{12}{-}1), \quad 
{:}u_2{*}u_1{:}_{_K}=u_1u_2{+}\frac{i\h}{2}(K_{12}{+}1). 
$$
Note that $*_K$-product formula gives a way of univalent expression 
for elements of $(W_2; *)$. Via univalent expressions,  
one can consider topological completion of 
the algebra, and transcendental elements.

\medskip
By the formulation of orderings, the intertwiner between
$K$-ordered expression and $K'$-ordered expression 
is explicitly given as follows:
\begin{prop}
\label{intwn}
For every $K, K'\in{\mathfrak S}(2)$, the intertwiner
is defined by  
\begin{equation}
\label{intertwiner}
I_{_K}^{^{K'}}(f)=
\exp\Big(\frac{i\h}{4}\sum_{i,j}(K^{'ij}{-}K^{ij})
\partial_{u_i}\partial_{u_j}\Big)f \,\,
(=I_{0}^{^{K'}}(I_{0}^{^{K}})^{-1}(f)), 
\end{equation}
which gives an isomorphism 
$I_{_K}^{^{K'}}:({\mathbb C}[{\pmb u}]; *_{_{K+J}})\rightarrow 
({\mathbb C}[{\pmb u}]; *_{_{K'+J}})$.
Namely, 
for any $f,g \in {\mathbb C}[{\pmb u}],$ we have  
\begin{equation}\label{intertwiner2}
I_{_K}^{^{K'}}(f*_{_\Lambda}g)=
I_{_K}^{^{K'}}(f)*_{_{\Lambda'}}I_{_K}^{^{K'}}(g),
\end{equation}
where $\Lambda=K{+}J$, $\Lambda'=K'{+}J$.
\end{prop}
Intertwiners do not change the algebraic structure $*$, 
but these change the expression of elements by the ordinary 
commutative structure.  

In what follows, we use the notation $*_{_K}$ instead of 
$*_{_\Lambda}$, since the skew-part $J$ is fixed on the 
standard skew-matrix.
As in the case of one variable, infinitesimal intertwiner 
$$
dI_{_K}(K')=\frac{d}{dt}\Big|_{t=0}I_{_K}^{^{K{+}tK'}}=
\frac{i\h}{4}{K'}_{ij}\partial_{u_i}\partial_{u_j}
$$
is viewed as a flat connection on the trivial bundle 
$\coprod_{K\in{\mathfrak S}(2)}
H{\!o}l({\Bbb C}^{2})$. The equation of parallel translation 
along a curve $K(t)$ is given by 
\begin{equation}\label{parallel}
\frac{d}{dt}f_t=
dI_{_K}(\dot K(t))f_t, \quad 
\dot K(t)=\frac{d}{dt}K(t),
\end{equation}
but this may not have a solution for some initial function.

\bigskip

Associated with $H_*{\in}(W_2; *)$, the $*$-exponential function of $H_*$
is defined by the evolution equation  
\begin{equation}\label{evol111}
\frac{d}{dt}f_t={:}H_*{:}_{_K}{*}f_t,\quad f_0=1.
\end{equation}
If the real analytic solution of (\ref{evol111}) exists then 
the solution is denoted by 
${:}e_*^{tH_*}{:}_{_K}$.  

\medskip
In what follows we set $(u_1,u_2)$ by $(u,v)$.  
Note that a quadratic forms with discriminant $0$ is essentially 
$u^2$ by a linear change of generators, and $*$-exponential
$e_*^{tu^2}$ is treated in \cite{OMMY3}.
Note also that 
$2u{\ctt}v{=}u{*}v{+}v{*}u$ is a representative of 
quadratic forms with discriminant $1$ by a linear change of generators. 
We take our attention to the $*$-exponential function of a quadratic 
form $2u{\ctt}v$ under a general expression parameter  
$K{=}
\tiny
{\begin{bmatrix}
\delta&c\\
c&\delta'
\end{bmatrix}}$.
By noting that 
${:}\frac{1}{i\h}u{\ctt}v{:}_{_K}=\frac{1}{i\h}uv{+}\frac{1}{2}c$, 
the equation \eqref{evol111} for $H_*=\frac{2}{i\h}u{\ctt}v$
is written precisely as 
\begin{equation}\label{bigeqeq}
\begin{aligned}
\frac{d}{dt}f_t(u,v)=(\frac{c}{2}{+}&\frac{1}{i\h}uv)f_t(u,v){+}
((c{+}1)u{+}\delta v)\partial_uf_t(u,v){+}
(\delta'u{+}(c{-}1)v)\partial_vf_t(u,v)\\
{+}&
\frac{i\h}{2}\big(\delta(c{+}1)\partial_u^2f_t(u,v){+}
(\delta\delta'{+}c^2{-}1)\partial_u\partial_vf_t(u,v)
{+}\delta'(c{-}1)\partial_v^2f_t(u,v)\big).
\end{aligned}
\end{equation} 

By setting $\Delta{=}e^t{+}e^{-t}{-}c(e^t{-}e^{-t})$, the solution is given by   
\begin{equation}\label{genericparam00}
{:}e_*^{t\frac{1}{i\h}2u{\ctt}v}{:}_{_{K}}{=}
\frac{2}{\sqrt{\Delta^2{-}(e^t{-}e^{-t})^2\delta\delta'}} 
\,\,e^{\frac{1}{i\h}
\frac{e^t-e^{-t}}{\Delta^2{-}(e^t{-}e^{-t})^2\delta\delta'}
\big((e^t-e^{-t})(\delta' u^2{+}\delta v^2){+}2\Delta uv\big)}.
\end{equation}
${}$\hfill (See \cite{ommy5} to know how to find this formula.)

Setting $\delta\delta'=\rho^2$, we see 
\begin{equation}\label{twolines}
\sqrt{\Delta^2{-}(e^t{-}e^{-t})^2\delta\delta'} =
e^{-t}
\sqrt{((1{-}c{+}\rho)e^{2t}{+}(1{+}c{-}\rho)) 
((1{-}c{-}\rho)e^{2t}{+}(1{+}c{+}\rho))}.
\end{equation}

As ${:}e_*^{z\frac{2}{i\h}u{\ctt}v}{:}_{_K}$ has double
branched singularities, we have to prepare two $\pm$ sheets with slits. 
Hence, we have two origins $0_+$, $0_-$ for ${:}e_*^{z\frac{2}{i\h}u{\ctt}v}{:}_{_K}$. 
Noting that 
${:}e_*^{0\frac{1}{i\h}2u{\ctt}v}{:}_{_{K}}{=}\sqrt{1}$, 
we set 
\begin{equation}\label{seteval}
{:}e_*^{0_+\frac{1}{i\h}u{\ctt}v}{:}_{_K}{=}1,\quad 
{:}e_*^{0_-\frac{1}{i\h}u{\ctt}v}{:}_{_K}{=}-1.
\end{equation}
Note also that 
${:}e_*^{\pm\pi i\frac{1}{i\h}2u{\ctt}v}{:}_{_{K}}{=}\sqrt{1}$, 
but this is {\it not} an absolute scalar.
The $\pm$ sign depends on $K$ and the path form $0$ to $\pi i$ by setting 
${:}e_*^{0\frac{1}{i\h}2u{\ctt}v}{:}_{_{K}}{=}1$.
As this is a scalar-like element belonging to a one parameter subgroup, 
we call it a $q$-{\it scalar}. 

If $t=\pm\frac{\pi i}{2}$, 
then \eqref{genericparam00} is called the {\bf polar element} and denoted by 
${\e}_{00}$ \cite{OMMY4}, \cite{ommy5}: 
\begin{equation}\label{polar}
{:}{\e}_{00}{:}_{_K}=\frac{1}{\sqrt{c^2{-}\delta\delta'}}
e^{\frac{1}{i\h}\frac{-1}{c^2{-}\delta\delta'}
(\delta' u^2{+}\delta v^2 {-}2cuv)}.
\end{equation}

\medskip
For the simplest case  $c=\delta=\delta'=0$ in \eqref{genericparam00} 
is the {\bf Weyl ordered expression}. 
This is not a generic ordered expression having singular 
points on the imaginary axis, and this is $\pi i$-alternating 
periodic. 
 
On the contrary, the {\bf unit ordered expression} is given by 
$K{=}I$, i.e. $\delta{=}\delta'{=}1$, $c{=}0$. 
By \eqref{genericparam00}, we have 
\begin{equation}\label{unitunit}
{:}e_*^{t\frac{1}{i\h}2u{\ctt}v}{:}_{_{I}}=
\frac{2}{\sqrt{4}}e^{\frac{1}{4i\h}
(e^{2t}{+}e^{-2t}{+}2)(u^2{+}v^2){+}2(e^{2t}{-}e^{-2t})uv}.
\end{equation}
This is $\pi i$-periodic and there is no singular point.

The case where $c{=}0$ and $\delta\delta'{\not=}1$,

For the case $\delta=\delta'=0$ but $c\not=0$ 
which involves the {\bf normal ordered expression} for $c=1$, we see that 
\begin{equation}\label{normalantinormal}
{:}e_*^{t\frac{1}{i\h}2u{\ctt}v}{:}_{_{K}}{=}
\frac{2}{\sqrt{\Delta^2}}
\,\,e^{\frac{1}{i\h}
\frac{e^t-e^{-t}}{\Delta^2}\big(2\Delta uv\big)}
=
\frac{2}{\Delta}
\,\,e^{\frac{1}{i\h}
\frac{e^t-e^{-t}}{\Delta}2uv}.
\end{equation} 
This is the case where the singular points are 
not branching ones and they are sitting 
$\pi i$ periodically on a single line parallel to 
the imaginary axis whose real part are given by 
$\log\big|\frac{c+1}{c-1}\big|$. 
We see also that ${:}e_*^{\frac{t}{i\h}2u{\ctt}v}{:}_{_K}$ is 
alternating $\pi i$-periodic along the imaginary axis. 

\medskip
There is another special expression parameter $K$ such that 
$(1{+}c)^2{-}\rho^2{=}0$, which will be called the 
{\bf separating ordered expression}.  
In this case, we have 
\begin{equation}\label{separating}
\sqrt{\Delta^2{-}(e^t{-}e^{-t})^2\delta\delta'} =
\sqrt{4(1{+}c^2){-}4e^{2t}}.
\end{equation}
This is the case where the periodicity of 
${:}e_*^{(s{+}it)\frac{1}{i\h}2u{\ctt}v}{:}_{_K}$ w.r.t. $t$ changes
at $s\lessgtr\log|4(1{+}c^2)|$.

\subsection{Generic properties of $*$-exponential 
functions of quadratic forms}\label{GenRiem}

In the previous note \cite{ommy5} we have studied $*$-exponential functions of
quadratic forms under a fixed expression parameter. In this chapter, 
we fix a quadratic form $\alpha{+}2u{\ctt}v$ for 
$\alpha\in{\mathbb C}$, but move the expression parameter. The
important view point here is that we are thinking that $K$ is moving
by the individual time parameter. 

One parameter subgroup 
$e_*^{z(\alpha{+}2u{\ctt}v)}$ of the $*$-exponential function of a
quadratic form $\alpha{+}2u{\ctt}v$ has remarkable properties that 
the periodicity depends on $\alpha$ and the expression parameters. 
We first summarize generic properties of the $*$-exponential function 
${:}e_*^{z(\alpha{+}\frac{1}{i\h}2u{\ctt}v)}{:}_{_K}$ 
mentioned in \cite{OMMY4}, \cite{ommy5}, where by generic property
we mean properties held in almost all (open dense) expressions. 

Note first that
$\frac{1}{i\h}u{\ctt}v{-}\frac{1}{2}=\frac{1}{i\h}u{*}v$, 
and the fundamental exponential law (cf. ({\bf d}) below) 
$$
e_*^{z(\alpha{+}\frac{1}{i\h}2u{\ctt}v)}=
e^{z\alpha}e_*^{z\frac{1}{i\h}2u{\ctt}v},\quad 
e_*^{z(\alpha{+}\frac{1}{i\h}2u{\ctt}v)}{*}e_*^{z'(\alpha{+}\frac{1}{i\h}2u{\ctt}v)}=
e_*^{(z{+}z')(\alpha{+}\frac{1}{i\h}2u{\ctt}v)}.
$$
Hence the essential part is 
$e_*^{z\frac{1}{i\h}2u{\ctt}v}$, but its periodicity 
depends on $\alpha$. If $\alpha{=}1$, then 
$e_*^{z\frac{1}{i\h}2u{*}v}{=}
e_*^{z({-}1{+}\frac{1}{i\h}2u{\ctt}v)}$. 

\medskip
Let $H\!ol({\mathbb C}^2)$ be the space of all
holomorphic functions of $(u,v)\in {\mathbb C}^2$ with the uniform
convergent topology on each compact subset. 

\noindent
({\bf a})\,\,In generic ordered expressions, one may assume 
there is no singular point on the real axis and the 
pure imaginary axis.

\noindent
({\bf b})\,\,$e_*^{z\frac{2}{i\h}u{\ctt}v}$ is a $H\!ol({\mathbb  C}^2)$-valued 
$2\pi i$-periodic function, i.e. 
${:}e_*^{(z{+}2\pi i)\frac{2}{i\h}u{\ctt}v}{:}_{_K}={:}e_*^{z\frac{2}{i\h}u{\ctt}v}{:}_{_K}$. 
More precisely, 

it is $\pi i$-periodic or 
alternating $\pi i$-periodic. 

\noindent
({\bf c})\,\,$e_*^{z\frac{2}{i\h}u{\ctt}v}$ is rapidly 
decreasing along any line parallel to the real line. 

\noindent
({\bf d})\,\,As a result, 
$e_*^{z\frac{2}{i\h}u{\ctt}v}$ must have periodic singular
points. But the singular points are double 
branched. Hence $e_*^{z\frac{2}{i\h}u{\ctt}v}$ is double-valued with the sign
ambiguity. Singular point set $\Sigma_K$ is 
distributed $\pi i$-periodically along the two lines parallel to the imaginary
axis. In spite of double-valued nature, the exponential law 
$$
e_*^{z\frac{2}{i\h}u{\ctt}v}{*}e_*^{z'\frac{2}{i\h}u{\ctt}v}=
e_*^{(z{+}z')\frac{2}{i\h}u{\ctt}v}
$$
holds under the calculation such that $\sqrt{a}\sqrt{b}=\sqrt{ab}$.

\noindent
({\bf e}) By requesting $1$ at $z=0$,
i.e. ${:}e_*^{0\frac{2}{i\h}u{\ctt}v}{:}_{_K}=1$, the value 
${:}e_*^{[0{\sim}z]\frac{2}{i\h}u{\ctt}v}{:}_{_K}$ is determined
uniquely, where 
$[0{\sim}z]$ is a path from $0$ to $z$ avoiding $\Sigma_K$ 
and evaluating at $z$.

\subsection{Periodicity and the exchanging interval}

In generic ordered expressions 
${:}e_*^{{\pm}\frac{z}{i\h}2u{\ctt}v}{:}_{_K}$, 
is $2\pi i$-periodic along any line parallel to the imaginary axis.
In generic $K$, there are two real number 
$\hat{a}$, $\hat{b}$ $(\hat{a}{<}\hat{b})$ such that the set
$\Sigma_K$ of  singular points 
are lying on the lines $(\hat{a}{+}i{\mathbb R})\cup(\hat{b}{+}i{\mathbb R})$.

Assuming $(1{-}c)^2{-}\rho^2\not=0$ in generic $K$, we set 
$$
a=\log|\frac{1{+}c{-}\rho}{1{-}c{+}\rho}|,\quad b=\log|\frac{1{+}c{+}\rho}{1{-}c{-}\rho}|
$$
in \eqref{twolines}. Then $\hat{a}{=}a{\wedge}b,\,\,\hat{b}{=}a{\vee}b$.
The open interval $I_{\ctt}(K)=(a{\wedge}b,a{\vee}b)$ is called 
the (sheet) {\bf exchanging interval} of $e_*^{z\frac{1}{i\h}2u{\ctt}v}$.
As the exponential law shows that 
$e_*^{z\frac{1}{i\h}2u{*}v}$, $e_*^{z\frac{1}{i\h}2v{*}u}$
have singular points on the same two lines, 
$I_{\ctt}(K)$ is 
called also the exchanging interval of these. 
The separating ordered expression is the case where 
$a\wedge b$ is $-\infty$, i.e. $I_{\ctt}(K){=}(-\infty, a\vee b)$. 

\medskip
As the pattern of periodicity depends on how the 
circle $\{re^{i\theta}; \theta\in \mathbb R\}$ 
round the singular points, it depends delicately on $K$.  
There are three disjoint open subsets 
${\mathfrak K}_{\pm}$ and ${\mathfrak K}_0$ of 
the space of expression parameters such that 
${\mathfrak K}_+\cup{\mathfrak K}_-\cup{\mathfrak K}_0$ is dense.

\noindent
{\bf (1)} If $K{\in}{\mathfrak K}_+$ (resp. ${\mathfrak K}_-$),  
the singular set of 
${:}e_*^{\frac{z}{i\h}2u{\ctt}v}{:}_{_K}$ 
appears $\pi i$-periodically  only in the open right  
(resp. left) half plane, along two lines parallel to the 
imaginary axis, and the $*$-exponential functions 
form a complex semi-group over the left (resp. right) 
half plane without sign ambiguity by 
requesting $1$ at $t{=}0$. 
Moreover, ${:}e_*^{{\pm}z\frac{1}{i\h}2u{\ctt}v}{:}_{_K}$, 
is alternating $\pi i$-periodic on the imaginary axis, and 
${:}e_*^{{\pm}\frac{z}{i\h}2u{\ctt}v}{:}_{_K}$ is 
rapidly decreasing of $e^{-|z|}$ order along any line 
parallel to the real line. 

In particular, we have  
${:}e_*^{[0{\to}\pi]\frac{i}{i\h}2u{\ctt}v}{:}_{_K}=
{:}e_*^{[0{\to}\pi]\frac{-i}{i\h}2u{\ctt}v}{:}_{_K}=-1$, and 
hence ${\e}_{00}^2=-1$ by the $*_{_K}$-product, where  
$[0{\to}a]$ is  
the path starting from the origin $0$ ending at $a$ along the line segment, but 
the $*$-exponential is evaluated at $t{=}a$ by the continuous chase 
from $0$ to $a$ along the path $[0{\to}a]$. 

The special ordered expression $K_s$ used in \cite{OMMY4} is in 
${\mathfrak K}_+$.

\bigskip

\noindent
{\bf(2)} If $K\in{\mathfrak K}_0$, singular set appears 
in both left and right half-planes, 
but not on the imaginary axis. Both of these lines are 
parallel to the imaginary axis. 
In particular, we have  
${:}e_*^{[0{\to}\pi]\frac{i}{i\h}2u{\ctt}v}{:}_{_K}=1$, and 
then ${\e}_{00}^2=1$ by the natural product along 
the imaginary axis. Moreover, 
${:}e_*^{{\pm}\frac{z}{i\h}2u{\ctt}v}{:}_{_K}$, 
is $\pi i$-periodic on the imaginary axis.

The Siegel ordered expression mentioned in \cite{OMMY4}, 
\cite{ommy5}, i.e. 
${\rm{Re}}\,\frac{1}{\h}\langle \xi(iK),\xi\rangle\geq c_{K}|\xi|^2$ for
some $c_{K}{>}0$, is in ${\mathfrak K}_0$.

\bigskip
For every $s\in I_{\ctt}(K)$, by replacing $\frac{2}{i\h}u{\ctt}v$ to 
$\frac{1}{i\h}u{\ctt}v$,   
 ${:}e_*^{(s{+}it)\frac{1}{i\h}u{\ctt}v}{:}_{_K}$ is   
 $2\pi$-periodic w.r.t. $t$, but 
 ${:}e_*^{(s{+}it)\frac{1}{i\h}u{*}v}{:}_{_K}$ is  
 alternating $2\pi$-periodic w.r.t. $t$.
On the contrary, for every 
$s\in{\mathbb R}{\setminus}\overline{I}_{\ctt}(K)$,  
${:}e_*^{(s{+}it)\frac{1}{i\h}u{\ctt}v}{:}_{_K}$ is 
alternating $2\pi$-periodic w.r.t. $t$, and 
${:}e_*^{(s{+}it)\frac{1}{i\h}u{*}v}{:}_{_K}$ is  
$2\pi$-periodic w.r.t. $t$.

\subsubsection{Rules of setting slits and  evaluations } 
As there is no singular point on the real axis, and the pure imaginary axis
in generic ordered expression,  
one can evaluate by \eqref{seteval} 
$${:}e_*^{(s_+)(\frac{1}{i\h}u{\ctt}v{+}\alpha)}{:}_{_K},\quad 
  {:}e_*^{(s_-)(\frac{1}{i\h}u{\ctt}v{+}\alpha)}{:}_{_K}, \quad
{:}e_*^{(is_+)(\frac{1}{i\h}u{\ctt}v{+}\alpha)}{:}_{_K},\quad 
  {:}e_*^{(is_-)(\frac{1}{i\h}u{\ctt}v{+}\alpha)}{:}_{_K}
$$
univalent way for every $s\in{\mathbb R}$,  
We see  
\begin{equation}\label{valpmsheets}
{:}e_*^{(s_-)(\frac{1}{i\h}u{\ctt}v{+}\alpha)}{:}_{_K} 
={-}{:}e_*^{(s_+)(\frac{1}{i\h}u{\ctt}v{+}\alpha)}{:}_{_K},\quad 
{:}e_*^{(is_-)(\frac{1}{i\h}u{\ctt}v{+}\alpha)}{:}_{_K} 
={-}{:}e_*^{(is_+)(\frac{1}{i\h}u{\ctt}v{+}\alpha)}{:}_{_K}
\end{equation} 
 where $(a_+)$, $(a_-)$ are $a$ in the positive, the negative sheet respectively. 

We want to evaluate 
${:}e_*^{(s_{\pm}{+}it)(\frac{1}{i\h}u{\ctt}v{+}\alpha)}{:}_{_K}$ by solving the
differential equation
$$
\frac{d}{dt}f_t={:}i(\frac{1}{i\h}u{\ctt}v{+}\alpha){:}_K{*_K}f_t,\quad 
f_0={:}e_*^{(s_{\pm})(\frac{1}{i\h}u{\ctt}v{+}\alpha)}{:}_{_K}.
$$
If we neglect the sign ambiguity, the solution is given by 
${:}e_*^{(s_{\pm}{+}it)(\frac{1}{i\h}u{\ctt}v{+}\alpha)}{:}_{_K}$. 
As singular points are double branched singular points, 
we have to mind the slits where sheets are changing. 

There are many ways to set slits.  
A typical way is to set slits $2\pi$-periodically between singular points, if $\pm\infty$ 
split in the Riemann surface. In this case, we have
two $-{\infty}_{\pm}$, and two ${\infty}_{\pm}$, where 
$\pm{\infty}_+$ are in positive sheet and 
$\pm{\infty}_-$ are in negative sheet. This will be used for the case
$\alpha{\in}{\mathbb Z}{+}\frac{1}{2}$.    

\medskip 
Another is to set slits between $\pm\infty$ and singular points, where
$\pm{\infty}$ do not split in the Riemann surface.
This will be used for the case $\alpha{\in}{\mathbb Z}$. In any case,
the value ${:}e_*^{z\frac{1}{i\h}u{\ctt}v{+}\alpha)}{:}_{_K}$ changes
sign discontinuously when a point $z$ crossed a slit in the same
sheet. One has to change sheets for the continuous tracing.   

\bigskip
As a standard one, we set slits between singular points 
just as in the vertical segments in the picture below. 
 We have two zero $0_+$, $0_-$ and two $-\infty$;  
$-\infty_+, -\infty_-$ and two $\infty$; 
$\infty_+, \infty_-$.   
The $*$-exponential function 
${:}e_*^{z(\alpha{+}\frac{1}{i\h}2u{\ctt}v)}{:}_{_K}$
is viewed as a single 
$H\!ol({\mathbb C}^2)$-valued 
function on this space. 

\medskip
\begin{center}
\unitlength 0.1in
\begin{picture}( 40.5000, 10.0000)(3.8000,-13.0000)
%
\special{pn 20}%
\special{pa 820 490}%
\special{pa 820 1090}%
\special{fp}%
%
\special{pn 20}%
\special{pa 1620 490}%
\special{pa 1620 1090}%
\special{fp}%
%
\special{pn 20}%
\special{pa 2420 490}%
\special{pa 2420 1090}%
\special{fp}%
%
\special{pn 20}%
\special{pa 3220 490}%
\special{pa 3220 1090}%
\special{fp}%
%
\special{pn 20}%
\special{pa 4020 490}%
\special{pa 4020 1090}%
\special{fp}%
%
\special{pn 20}%
\special{ar 820 500 200 200  1.7359450 4.7123890}%
%
\special{pn 20}%
\special{ar 1620 1100 200 200  1.5707963 4.5472403}%
%
\special{pn 20}%
\special{ar 2420 500 200 200  1.7359450 4.7123890}%
%
\special{pn 20}%
\special{ar 4020 500 200 200  1.7359450 4.7123890}%
%
\special{pn 20}%
\special{ar 3220 1100 200 200  1.5707963 4.5472403}%
%
\special{pn 20}%
\special{ar 2420 1100 200 200  4.8775377 6.2831853}%
\special{ar 2420 1100 200 200  0.0000000 1.5707963}%
%
\special{pn 20}%
\special{ar 4020 1100 200 200  4.8775377 6.2831853}%
\special{ar 4020 1100 200 200  0.0000000 1.5707963}%
%
\special{pn 20}%
\special{ar 810 1080 200 200  4.8775377 6.2831853}%
\special{ar 810 1080 200 200  0.0000000 1.5707963}%
%
\special{pn 20}%
\special{ar 1620 500 200 200  4.7123890 6.2831853}%
\special{ar 1620 500 200 200  0.0000000 1.4056476}%
%
\special{pn 20}%
\special{ar 3220 500 200 200  4.7123890 6.2831853}%
\special{ar 3220 500 200 200  0.0000000 1.4056476}%
%
\special{pn 20}%
\special{pa 820 300}%
\special{pa 1620 310}%
\special{fp}%
\special{pa 2420 300}%
\special{pa 3230 310}%
\special{fp}%
\special{pa 4030 300}%
\special{pa 4430 300}%
\special{fp}%
\special{pa 1620 1300}%
\special{pa 2430 1300}%
\special{fp}%
\special{pa 3220 1300}%
\special{pa 4020 1300}%
\special{fp}%
%
\special{pn 20}%
\special{ar 810 810 130 410  6.2740946 6.2831853}%
\special{ar 810 810 130 410  0.0000000 6.2601125}%
%
\special{pn 20}%
\special{ar 1610 810 130 410  6.2740946 6.2831853}%
\special{ar 1610 810 130 410  0.0000000 6.2601125}%
%
\special{pn 20}%
\special{ar 2410 810 130 410  6.2740946 6.2831853}%
\special{ar 2410 810 130 410  0.0000000 6.2601125}%
%
\special{pn 20}%
\special{ar 3210 810 130 410  6.2740946 6.2831853}%
\special{ar 3210 810 130 410  0.0000000 6.2601125}%
%
\special{pn 20}%
\special{ar 4010 810 130 410  6.2740946 6.2831853}%
\special{ar 4010 810 130 410  0.0000000 6.2601125}%
%
\special{pn 20}%
\special{sh 1}%
\special{ar 690 660 10 10 0  6.28318530717959E+0000}%
\special{sh 1}%
\special{ar 1730 670 10 10 0  6.28318530717959E+0000}%
\special{sh 1}%
\special{ar 2290 660 10 10 0  6.28318530717959E+0000}%
\special{sh 1}%
\special{ar 3330 670 10 10 0  6.28318530717959E+0000}%
\special{sh 1}%
\special{ar 3890 650 10 10 0  6.28318530717959E+0000}%
\special{sh 1}%
\special{ar 4140 940 10 10 0  6.28318530717959E+0000}%
\special{sh 1}%
\special{ar 3090 950 10 10 0  6.28318530717959E+0000}%
\special{sh 1}%
\special{ar 2540 940 10 10 0  6.28318530717959E+0000}%
\special{sh 1}%
\special{ar 1490 950 10 10 0  6.28318530717959E+0000}%
\special{sh 1}%
\special{ar 940 930 10 10 0  6.28318530717959E+0000}%
\special{sh 1}%
\special{ar 940 930 10 10 0  6.28318530717959E+0000}%
%
\special{pn 20}%
\special{pa 860 700}%
\special{pa 1570 700}%
\special{dt 0.054}%
\special{pa 2460 700}%
\special{pa 3170 700}%
\special{dt 0.054}%
\special{pa 3260 900}%
\special{pa 3970 900}%
\special{dt 0.054}%
%
\special{pn 20}%
\special{pa 1660 910}%
\special{pa 2370 910}%
\special{dt 0.054}%
\special{pa 780 890}%
\special{pa 380 890}%
\special{dt 0.054}%
\special{pa 380 890}%
\special{pa 380 890}%
\special{dt 0.054}%
\special{pa 4060 700}%
\special{pa 4430 700}%
\special{dt 0.054}%
\special{pa 4430 1100}%
\special{pa 4430 1100}%
\special{dt 0.054}%
\special{pa 4430 1100}%
\special{pa 4430 1100}%
\special{dt 0.054}%
%
\special{pn 20}%
\special{pa 810 1280}%
\special{pa 380 1280}%
\special{fp}%
%
%
\put(9.9000,-15.0000){\makebox(0,0)[lb]
{\footnotesize{Fig.1\,\,Riemann surface of infinite genus}}}%
\end{picture}%
\end{center}

\bigskip
\noindent
where vertical lines in Fig.1 are near to parallel to the real axis. 

A little care is requested to apply the Cauchy's integral theorem.  
All closed paths drawn in Fig.1  give closed paths 
in the Riemann surface of infinite
genus, where dotted lines are in the negative sheets. On such a curve,
we have to evaluate by continuous tracing along a curve in spite of
the evaluation rule \eqref{valpmsheets}. Note that \eqref{valpmsheets}
is the evaluation along a path from $0_{\pm}$ without crossing slits.

\section{Elements defined by integrals}

Note that in a generic ordered expression 
$e_*^{{\pm}\frac{t}{i\h}2u{\ctt}v}$ is rapidly decreasing with 
the growth order $e^{-|t|}$ along lines parallel to the real axis.
Noting $v{*}u{=}u{\ctt}v{+}\frac{1}{2}i\h$, we see the following:
\begin{prop}\label{existvacuum}
In generic ordered expressions such that there is no 
singular point on the real axis  
$$
\lim_{t{\to}\infty}e_*^{t\frac{1}{i\h}2u{*}v}
{=}0,\quad 
\lim_{t{\to}-\infty}e_*^{t\frac{1}{i\h}2v{*}u}
{=}0, 
$$
but the following limit exists
$$
\lim_{t{\to}-\infty}e_*^{t\frac{1}{i\h}2u{*}v}
{=}{\varpi}_{00},\quad 
\lim_{t{\to}\infty}e_*^{t\frac{1}{i\h}2v{*}u}
{=}\overline{\varpi}_{00}.
$$
\end{prop}

More precisely, in a fixed generic expression parameter 
$K{=}
\tiny{\begin{bmatrix}
\delta&c\\
c&\delta' 
\end{bmatrix}}$, 
${:}e_*^{t\frac{1}{i\h}2u{\ctt}v}{:}_{_{K}}$ is smooth   
rapidly decreasing in $\pm$ directions and \eqref{genericparam00}
gives 
\begin{equation}
  \label{eq:vacuum123123}
  \begin{aligned}
{:}\varpi_{00}{:}_{_K}=
&\lim_{t{\to}-\infty}{:}e_*^{t\frac{1}{i\h}2u{*}v}{:}_{_K}
{=}\frac{2}{\sqrt{(1{+}c)^2{-}\delta\delta'}}
e^{-\frac{1}{i\h}\frac{1}{(1{+}c)^2{-}\delta\delta'}
(\delta u^2{-}(1{+}c)2uv{+}\delta' v^2)},\\
{:}{\overline{\varpi}}_{00}{:}_{_K}
=&\lim_{t{\to}\infty}{:}e_*^{t\frac{1}{i\h}2v{*}u}{:}_{_K}
{=}\frac{2}{\sqrt{(1{-}c)^2{-}\delta\delta'}}
e^{\frac{1}{i\h}\frac{1}{(1{-}c)^2{-}\delta\delta'}
(\delta u^2{+}(1{-}c)2uv{+}\delta' v^2)},\\ 
&
\lim_{t{\to}\infty}{:}e_*^{t\frac{1}{i\h}2u{*}v}{:}_{_K}
{=}0,\quad 
\lim_{t{\to}-\infty}{:}e_*^{t\frac{1}{i\h}2v{*}u}{:}_{_K}
{=}0. 
 \end{aligned}
\end{equation}
without sign ambiguity. 
We call 
$\varpi_{00}$ and $\overline{\varpi}_{00}$ {\bf vacuum} and 
{\bf bar-vacuum} respectively. These are contained in 
the space  ${\mathbb C}e^{Q(u,v)}$ of exponential functions of 
quadratic forms.

\bigskip
As ${\varpi}_{00}$ is defined by the limit, we see 
$u{*}v{*}\varpi_{00}=0=\varpi_{00}{*}u{*}v.$
But the ``bumping identity'' $v{*}f(u{*}v){=}f(v{*}u){*}v$ 
give the following: 
\begin{lem}
  \label{vacvac3} 
$v{*}\varpi_{00}{=}0{=}\varpi_{00}{*}u$, 
$u{*}\overline{\varpi}_{00}{=}0{=}\overline{\varpi}_{00}{*}v$ in generic ordered 
expressions. 
\end{lem}

\noindent
{\bf Proof}\,\, 
Using the continuity of $v{*}$, we see that 
$v{*}\lim_{t{\to}-\infty}e_*^{t\frac{1}{i\h}2u{*}v}{=}
\lim_{t{\to}-\infty}v{*}e_*^{t\frac{1}{i\h}2u{*}v}$. 
Hence, the bumping identity  gives 
$\lim_{t{\to}-\infty}e_*^{t\frac{1}{i\h}2v{*}u}{*}v{=}0$ 
by using Proposition\,\ref{existvacuum}. \hfill $\Box$.

\medskip

The exponential law gives 
$$
\varpi_{00}{*}\varpi_{00}{=}\varpi_{00}, \quad 
\overline{\varpi}_{00}{*}\overline{\varpi}_{00}
{=}\overline{\varpi}_{00}. 
$$ 
However, the product $\varpi_{00}{*}\overline{\varpi}_{00}$ 
diverges whenever this is defined as 
$\lim_{t\to -\infty}e_{*}^{tu{*}v}{*}\overline{\varpi}_{00}$. 
But in fact, it depends how the product is defined, as it 
will be mentioned below it is better so to define that 
$\varpi_{00}{*}\overline{\varpi}_{00}=0$.  

The next identities are easy to see 
$$
\varpi_{00}{*}\frac{1}{i\h}u{\ctt}v=
\frac{1}{2}\varpi_{00},\quad
\frac{1}{i\h}u{\ctt}v{*}\overline{\varpi}_{00}=
{-}\frac{1}{2}\overline{\varpi}_{00}.
$$ 
Note that in order to keep the associativity  
\begin{equation}\label{inorderto}
(\varpi_{00}{*}\frac{1}{i\h}u{\ctt}v)
{*}\overline{\varpi}_{00}=
\varpi_{00}{*}
(\frac{1}{i\h}u{\ctt}v{*}\overline{\varpi}_{00}), 
\end{equation}
we have to define 
$$
\frac{1}{2}\varpi_{00}{*}\overline{\varpi}_{00}=
{-}\frac{1}{2}\varpi_{00}{*}\overline{\varpi}_{00}=0.
$$

Note that there is no sign ambiguity.
But strictly speaking, vacuums should be so defined carefully that 
they do not involve sign ambiguity.

\bigskip
\noindent
{\bf Note}\,\, In the traditional ring theory, vacuums are 
defined as maximal left ideals. But we prefer to 
use the notion of vacuums as physicists uses, to which we 
can not give a mathematical definition, because we do not 
know what is the true nature of the {\it vacuum}.  
Equality in Lemma\,\,\ref{vacvac3} may be understood to give the 
separation of a ``configuration space'' from the 
phase space. It should be noted also that we have 
two sheets and these changes sign in the opposite sheet.

\subsection{Vacuums and pseudo-vacuums}\label{vacuums} 

Note that if $|{\rm{Re}}\,t|$ is sufficiently 
large, then 
$e_*^{(t{+}i\sigma)\frac{1}{i\h}u{*}v}$, 
$e_*^{(t{+}i\sigma)\frac{1}{i\h}v{*}u}$ are both 
$2\pi$-periodic w.r.t. $\sigma$. 
Thus, it is better to define vacuums as the limits of 
period integral: 
\begin{equation}\label{defvacuum}
2\pi\varpi_{00}=\lim_{t\to -\infty} 
\int_{-\pi}^{\pi}e_*^{(t{+}i\sigma)\frac{1}{i\h}u{*}v}d\sigma,
\quad 
2\pi\overline{\varpi}_{00}=\lim_{s\to\infty} 
\int_{-\pi}^{\pi}e_*^{(s{+}i\sigma)\frac{1}{i\h}v{*}u}d\sigma.
\end{equation}
In fact, we have no need to take the limit since these are 
constant if $|t|, |s|$ are sufficiently large. In fact, Cauchy's
integral theorem together with $2\pi$-periodicity gives  
\begin{equation*}
\begin{aligned}
&\frac{1}{2\pi}\int_{-\pi}^{\pi}{:}e_*^{(s{+}i\sigma)\frac{1}{i\h}u{*}v}{:}_{_K}d\sigma
=\left\{
\begin{matrix}
{:}\varpi_{00}{:}_{_K},&  s{<}a\\
0,&    s{>}b
\end{matrix}
\right.\\
&\frac{1}{2\pi}\int_{-\pi}^{\pi}{:}e_*^{(s{+}i\sigma)\frac{1}{i\h}v{*}u}{:}_{_K}d\sigma
=\left\{
\begin{matrix}
0,&  s{<}a\\
{:}\overline{\varpi}_{00}{:}_{_K},& s{>}b.
\end{matrix}
\right.
\end{aligned}
\end{equation*}

\medskip
The product $\varpi_{00}{*}\overline{\varpi}_{00}$ can not be 
defined directly by the definition, but the product 
$$
\int_{-\pi}^{\pi}e_*^{(s{+}i\sigma)\frac{1}{i\h}u{*}v}d\sigma{*}
\int_{-\pi}^{\pi}e_*^{(s'{+}i\sigma')\frac{1}{i\h}v{*}u}d\sigma'
=
\int_{-\pi}^{\pi}\int_{-\pi}^{\pi}
e_*^{(s{+}i\sigma)\frac{1}{i\h}u{*}v{+}(s'{+}i\sigma')\frac{1}{i\h}v{*}u}
d\sigma d\sigma'
$$
can be defined always to give $0$, for by using $\frac{1}{i\h}u{*}v=\frac{1}{i\h}u{\ctt}v{-}\frac{1}{2}$,
and $\frac{1}{i\h}v{*}u=\frac{1}{i\h}u{\ctt}v{+}\frac{1}{2}$, 
the change of variables gives
$$
\int_{-\pi}^{\pi}e^{+i\sigma}d\sigma\int_{-\pi}^{\pi}
e^{\frac{1}{2}(s'{-}s{-}i\tau)}e_*^{(s{+}s'{+}i\tau)\frac{1}{i\h}u{\ctt}v}d\tau=0.
$$
Thus, we have 
\begin{prop}\label{control}
For every polynomial $p(u,v)$, 
$\varpi_{00}{*}p_*(u,v){*}\overline{\varpi}_{00}{=}0{=}
\overline{\varpi}_{00}{*}p(u,v){*}{\varpi}_{00}$ in generic 
ordered expression. 
$($Cf.\,\eqref{inorderto}.$)$
\end{prop}

\bigskip
\unitlength 0.1in
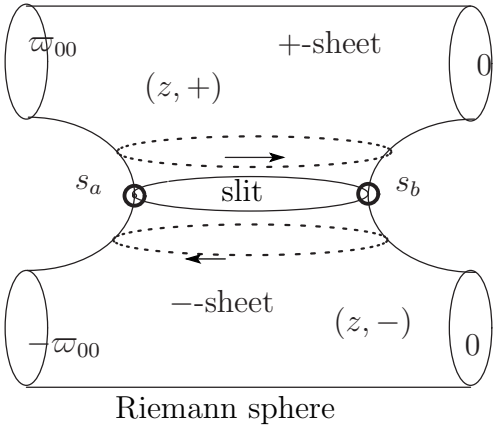
\begin{picture}( 25.2000, 20.2000)(1.2000,-26.3000)
%
\special{pn 8}%
\special{ar 1390 1100 610 90  3.9480720 6.2831853}%
\special{ar 1390 1100 610 90  0.0000000 3.9269908}%
%
\special{pn 8}%
\special{ar 2550 1110 550 400  1.5604017 4.7123890}%
%
\special{pn 8}%
\special{ar 240 1100 550 400  4.7123890 6.2831853}%
\special{ar 240 1100 550 400  0.0000000 1.5811910}%
%
\special{pn 8}%
\special{ar 230 410 110 300  0.0000000 6.2831853}%
%
\special{pn 8}%
\special{ar 2530 420 110 300  0.0000000 6.2831853}%
%
\special{pn 8}%
\special{ar 2530 1800 110 300  0.0000000 6.2831853}%
%
\special{pn 8}%
\special{ar 230 1800 110 300  0.0000000 6.2831853}%
%
\special{pn 8}%
\special{pa 230 2110}%
\special{pa 2530 2110}%
\special{fp}%
\special{pa 2530 120}%
\special{pa 230 120}%
\special{fp}%
%
\special{pn 8}%
\special{ar 790 1110 10 10  0.0000000 6.2831853}%
%
\special{pn 20}%
\special{ar 790 1110 52 52  6.2831853 6.2831853}%
\special{ar 790 1110 52 52  0.0000000 6.0857897}%
%
\special{pn 20}%
\special{ar 2000 1100 52 52  6.2831853 6.2831853}%
\special{ar 2000 1100 52 52  0.0000000 6.0857897}%
\put(2.4000,-3.8000){\makebox(0,0)[lb]{$\varpi_{00}$}}%
\put(25.6000,-4.8000){\makebox(0,0)[lb]{$0$}}%
\put(2.3000,-19.5000){\makebox(0,0)[lb]{$-\varpi_{00}$}}%
\put(25.0000,-19.4000){\makebox(0,0)[lb]{$0$}}%
\put(4.8000,-11.0000){\makebox(0,0)[lb]{$s_a$}}%
\put(21.4000,-11.1000){\makebox(0,0)[lb]{$s_b$}}%
\put(15.3000,-3.8000){\makebox(0,0)[lb]{$+$-sheet}}%
\put(9.7000,-17.3000){\makebox(0,0)[lb]{$-$-sheet}}%
\put(8.4000,-6.3000){\makebox(0,0)[lb]{$(z,+)$}}%
\put(18.2000,-18.6000){\makebox(0,0)[lb]{$(z,-)$}}%
%
\special{pn 13}%
\special{ar 1410 880 710 80  0.1094509 0.1398307}%
\special{ar 1410 880 710 80  0.2309699 0.2613497}%
\special{ar 1410 880 710 80  0.3524889 0.3828687}%
\special{ar 1410 880 710 80  0.4740079 0.5043877}%
\special{ar 1410 880 710 80  0.5955269 0.6259066}%
\special{ar 1410 880 710 80  0.7170459 0.7474256}%
\special{ar 1410 880 710 80  0.8385649 0.8689446}%
\special{ar 1410 880 710 80  0.9600839 0.9904636}%
\special{ar 1410 880 710 80  1.0816028 1.1119826}%
\special{ar 1410 880 710 80  1.2031218 1.2335016}%
\special{ar 1410 880 710 80  1.3246408 1.3550206}%
\special{ar 1410 880 710 80  1.4461598 1.4765396}%
\special{ar 1410 880 710 80  1.5676788 1.5980585}%
\special{ar 1410 880 710 80  1.6891978 1.7195775}%
\special{ar 1410 880 710 80  1.8107168 1.8410965}%
\special{ar 1410 880 710 80  1.9322358 1.9626155}%
\special{ar 1410 880 710 80  2.0537547 2.0841345}%
\special{ar 1410 880 710 80  2.1752737 2.2056535}%
\special{ar 1410 880 710 80  2.2967927 2.3271725}%
\special{ar 1410 880 710 80  2.4183117 2.4486915}%
\special{ar 1410 880 710 80  2.5398307 2.5702104}%
\special{ar 1410 880 710 80  2.6613497 2.6917294}%
\special{ar 1410 880 710 80  2.7828687 2.8132484}%
\special{ar 1410 880 710 80  2.9043877 2.9347674}%
\special{ar 1410 880 710 80  3.0259066 3.0562864}%
\special{ar 1410 880 710 80  3.1474256 3.1778054}%
\special{ar 1410 880 710 80  3.2689446 3.2993244}%
\special{ar 1410 880 710 80  3.3904636 3.4208434}%
\special{ar 1410 880 710 80  3.5119826 3.5423623}%
\special{ar 1410 880 710 80  3.6335016 3.6638813}%
\special{ar 1410 880 710 80  3.7550206 3.7854003}%
\special{ar 1410 880 710 80  3.8765396 3.9069193}%
\special{ar 1410 880 710 80  3.9980585 4.0284383}%
\special{ar 1410 880 710 80  4.1195775 4.1499573}%
\special{ar 1410 880 710 80  4.2410965 4.2714763}%
\special{ar 1410 880 710 80  4.3626155 4.3929953}%
\special{ar 1410 880 710 80  4.4841345 4.5145142}%
\special{ar 1410 880 710 80  4.6056535 4.6360332}%
\special{ar 1410 880 710 80  4.7271725 4.7575522}%
\special{ar 1410 880 710 80  4.8486915 4.8790712}%
\special{ar 1410 880 710 80  4.9702104 5.0005902}%
\special{ar 1410 880 710 80  5.0917294 5.1221092}%
\special{ar 1410 880 710 80  5.2132484 5.2436282}%
\special{ar 1410 880 710 80  5.3347674 5.3651472}%
\special{ar 1410 880 710 80  5.4562864 5.4866661}%
\special{ar 1410 880 710 80  5.5778054 5.6081851}%
\special{ar 1410 880 710 80  5.6993244 5.7297041}%
\special{ar 1410 880 710 80  5.8208434 5.8512231}%
\special{ar 1410 880 710 80  5.9423623 5.9727421}%
\special{ar 1410 880 710 80  6.0638813 6.0942611}%
\special{ar 1410 880 710 80  6.1854003 6.2157801}%
\special{ar 1410 880 710 80  6.3069193 6.3372991}%
%
\special{pn 13}%
\special{ar 1390 1340 710 80  0.1094509 0.1398307}%
\special{ar 1390 1340 710 80  0.2309699 0.2613497}%
\special{ar 1390 1340 710 80  0.3524889 0.3828687}%
\special{ar 1390 1340 710 80  0.4740079 0.5043877}%
\special{ar 1390 1340 710 80  0.5955269 0.6259066}%
\special{ar 1390 1340 710 80  0.7170459 0.7474256}%
\special{ar 1390 1340 710 80  0.8385649 0.8689446}%
\special{ar 1390 1340 710 80  0.9600839 0.9904636}%
\special{ar 1390 1340 710 80  1.0816028 1.1119826}%
\special{ar 1390 1340 710 80  1.2031218 1.2335016}%
\special{ar 1390 1340 710 80  1.3246408 1.3550206}%
\special{ar 1390 1340 710 80  1.4461598 1.4765396}%
\special{ar 1390 1340 710 80  1.5676788 1.5980585}%
\special{ar 1390 1340 710 80  1.6891978 1.7195775}%
\special{ar 1390 1340 710 80  1.8107168 1.8410965}%
\special{ar 1390 1340 710 80  1.9322358 1.9626155}%
\special{ar 1390 1340 710 80  2.0537547 2.0841345}%
\special{ar 1390 1340 710 80  2.1752737 2.2056535}%
\special{ar 1390 1340 710 80  2.2967927 2.3271725}%
\special{ar 1390 1340 710 80  2.4183117 2.4486915}%
\special{ar 1390 1340 710 80  2.5398307 2.5702104}%
\special{ar 1390 1340 710 80  2.6613497 2.6917294}%
\special{ar 1390 1340 710 80  2.7828687 2.8132484}%
\special{ar 1390 1340 710 80  2.9043877 2.9347674}%
\special{ar 1390 1340 710 80  3.0259066 3.0562864}%
\special{ar 1390 1340 710 80  3.1474256 3.1778054}%
\special{ar 1390 1340 710 80  3.2689446 3.2993244}%
\special{ar 1390 1340 710 80  3.3904636 3.4208434}%
\special{ar 1390 1340 710 80  3.5119826 3.5423623}%
\special{ar 1390 1340 710 80  3.6335016 3.6638813}%
\special{ar 1390 1340 710 80  3.7550206 3.7854003}%
\special{ar 1390 1340 710 80  3.8765396 3.9069193}%
\special{ar 1390 1340 710 80  3.9980585 4.0284383}%
\special{ar 1390 1340 710 80  4.1195775 4.1499573}%
\special{ar 1390 1340 710 80  4.2410965 4.2714763}%
\special{ar 1390 1340 710 80  4.3626155 4.3929953}%
\special{ar 1390 1340 710 80  4.4841345 4.5145142}%
\special{ar 1390 1340 710 80  4.6056535 4.6360332}%
\special{ar 1390 1340 710 80  4.7271725 4.7575522}%
\special{ar 1390 1340 710 80  4.8486915 4.8790712}%
\special{ar 1390 1340 710 80  4.9702104 5.0005902}%
\special{ar 1390 1340 710 80  5.0917294 5.1221092}%
\special{ar 1390 1340 710 80  5.2132484 5.2436282}%
\special{ar 1390 1340 710 80  5.3347674 5.3651472}%
\special{ar 1390 1340 710 80  5.4562864 5.4866661}%
\special{ar 1390 1340 710 80  5.5778054 5.6081851}%
\special{ar 1390 1340 710 80  5.6993244 5.7297041}%
\special{ar 1390 1340 710 80  5.8208434 5.8512231}%
\special{ar 1390 1340 710 80  5.9423623 5.9727421}%
\special{ar 1390 1340 710 80  6.0638813 6.0942611}%
\special{ar 1390 1340 710 80  6.1854003 6.2157801}%
\special{ar 1390 1340 710 80  6.3069193 6.3372991}%
%
\special{pn 8}%
\special{pa 1260 910}%
\special{pa 1560 910}%
\special{fp}%
\special{sh 1}%
\special{pa 1560 910}%
\special{pa 1494 890}%
\special{pa 1508 910}%
\special{pa 1494 930}%
\special{pa 1560 910}%
\special{fp}%
%
\special{pn 8}%
\special{pa 1260 1440}%
\special{pa 1060 1440}%
\special{fp}%
\special{sh 1}%
\special{pa 1060 1440}%
\special{pa 1128 1460}%
\special{pa 1114 1440}%
\special{pa 1128 1420}%
\special{pa 1060 1440}%
\special{fp}%
\put(12.4000,-11.4000){\makebox(0,0)[lb]{slit}}%
\put(6.9000,-23.0000){\makebox(0,0)[lb]{Riemann sphere}}%
\end{picture}%
\hfill
\parbox[b]{.45\linewidth}
{Keeping the periodicity in mind, the Riemann surface of $e_*^{z\frac{1}{i\h}u{*}v}$ 
may be viewed as the Riemann sphere, where $s_a$, $s_b$ in 
the figure are singular points. Vertical circuit between these 
correspond to the lines parallel to the imaginary axis. 
The complex number 
$z{\in}{\mathbb C}$ is expressed in the figure by 
$(z,+)$ and $(z,-)$ in $\pm$-sheet respectively.
Note also that the integral along a such vertical circuit 
vanishes by the alternating periodicity.

In such a compact Riemann surface, homological cycles play important 
role. But some of them are not related to closed one parameter
subgroups of $e_*^{t(\alpha{+}\frac{1}{i\h}u{\ctt}v)}$.
 }

Noting
${:}e_*^{(s{+}i\sigma)\frac{1}{i\h}u{*}v}{:}_{_K}$ is $4\pi$-periodic
for $|s|\gg 0$, and
$\int_{-2\pi}^{2\pi}{:}e_*^{(s{+}i\sigma)\frac{1}{i\h}u{*}v}{:}_{_K}d\sigma=0$
for $s\in I_{\ctt}(K)$ by the alternating $2\pi$-periodicity, we see 
\begin{equation}\label{1steqeq}
\begin{aligned}
&\frac{1}{4\pi}\int_{-2\pi}^{2\pi}{:}e_*^{(s{+}i\sigma)\frac{1}{i\h}u{*}v}{:}_{_K}d\sigma
=\left\{
\begin{matrix}
{:}\varpi_{00}{:}_{_K},&  s<a\\
0,& a<s
\end{matrix}
\right.\\
&\frac{1}{4\pi}\int_{-2\pi}^{2\pi}{:}e_*^{(s{+}i\sigma)\frac{1}{i\h}v{*}u}{:}_{_K}d\sigma
=\left\{
\begin{matrix}
0,&  s<b\\
{:}\overline{\varpi}_{00}{:}_{_K},& b<s,
\end{matrix}
\right.
\end{aligned}
\end{equation}
Hence we see 
$$
\frac{1}{i\h}u{*}v{*}
\frac{1}{4\pi}\int_{-2\pi}^{2\pi}{:}e_*^{(s{+}i\sigma)\frac{1}{i\h}u{*}v}{:}_{_K}d\sigma
{=}{-}\delta(s{-}a)\varpi_{00}.
$$
It is easy to see that first equality of \eqref{1steqeq}
  is the solution of 
$$
\frac{d}{ds}f(s)={-}\delta(s{-}a)\varpi_{00}{*}f(s),\quad f(-\infty)=\varpi_{00}.
$$

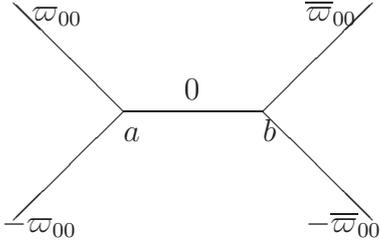
\setlength{\unitlength}{0.8mm}
\begin{picture}(30,20)(0,-9)
\put(25,20){\line(1,0){23}}
\put(25,20){\line(-1,1){18}}
\put(25,20){\line(-1,-1){18}}
\put(48,20){\line(1,1){18}}
\put(48,20){\line(1,-1){18}}
\put(35,22){$0$}
\put(10,35){$\varpi_{00}$}
\put(5,0){$-\varpi_{00}$}
\put(55,35){$\overline{\varpi}_{00}$}
\put(55,0){$-\overline{\varpi}_{00}$}
\put(25,15){$a$}
\put(48,15){$b$}
\end{picture}
\hfill
\parbox[b]{.55\linewidth}
{Minding the negative sheet, the integral 
$$
\frac{1}{4\pi}\int_{-2\pi}^{2\pi}
{:}\Big(e_*^{(s{+}i\sigma)\frac{1}{i\h}u{*}v}{+}e_*^{(s{+}i\sigma)\frac{1}{i\h}v{*}u}\Big){:}_{_K}d\sigma
$$
may be expressed symbolically by the l.h.s.\!\! picture. 
Note that if $a<s<b$, then ${:}e_*^{(s{+}i\sigma)\frac{1}{i\h}u{*}v}{:}_{_K}$,
$\sigma{\in}[-\pi,\pi]$ is not a closed path.}

\medskip
If $a<s<b$, the integral on half-period depends on $s$ 
\begin{equation}
\begin{aligned}
\frac{d}{ds}\int_{0}^{2\pi}\!\!e_*^{(s{+}i\sigma)\frac{1}{i\h}u{*}v}d\sigma
=&\int_{0}^{2\pi}\!\!(-i)\frac{d}{d\sigma}e_*^{(s{+}i\sigma)\frac{1}{i\h}u{*}v}d\sigma
=\frac{1}{i}(e_*^{(s{+}i2\pi)\frac{1}{i\h}u{*}v}{-}e_*^{s\frac{1}{i\h}u{*}v})={2i}e_*^{s\frac{1}{i\h}u{*}v}.
\end{aligned}
\end{equation}

\subsubsection{The pseudo-vacuum} 

Let $I_{\ctt}(K)=[a,b]$ be the exchanging interval of 
${:}e_*^{z\frac{1}{i\h}u{\ctt}v}{:}_{_K}$.  The periodicity of 
$e_*^{(s{+}it)\frac{1}{i\h}u{\ctt}v}$ with respect to $t$ depends on 
$s{\in}{\mathbb R}$. $e_*^{(s{+}it)\frac{1}{i\h}u{\ctt}v}$ is  
$2\pi i$-periodic in $t$ if $a<s<b$, 
and alternating $2\pi i$-periodic, if $s<a$ or
$b<s$. For $a{<}s{<}b$, we set  
\begin{equation}\label{pseudoquasivac}
{:}\varpi_*(s){:}_{_K}{=}\frac{1}{2\pi}\int_0^{2\pi}{:}e_*^{(s{+}it)(\frac{1}{i\h}u{\ctt}v)}{:}_{_K}dt.
\end{equation}
This is independent of $s$ whenever $a<s<b$ and  
$(\frac{1}{i\h}u{\ctt}v){*}\int_0^{2\pi}e_*^{(s{+}it)\frac{1}{i\h}u{\ctt}v}=0$.

Suppose $K\in {\mathfrak K}_0$. Then 
$e_*^{it\frac{1}{i\h}u{\ctt}v}$ is $2\pi$-periodic  and 
$\frac{1}{2\pi}\int_0^{2\pi}{:}e_*^{it(\frac{1}{i\h}u{\ctt}v)}{:}_{_K}dt$
has the idempotent property. 
It is easy to see that
$\frac{1}{2\pi}\int_0^{2\pi}{:}e_*^{it(\frac{1}{i\h}u{\ctt}v)}{:}_{_K}dt=
\frac{1}{4\pi}\int_0^{4\pi}{:}e_*^{it(\frac{1}{i\h}u{\ctt}v)}{:}_{_K}dt.$

\medskip
\noindent
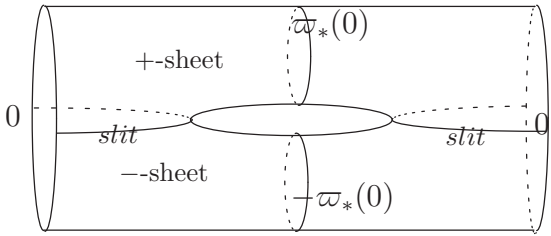
\unitlength 0.1in
\begin{picture}( 28.1100, 12.7100)(  8.1000,-21.0100)
%
\special{pn 8}%
\special{ar 3562 1498 60 588  4.7123890 6.2831853}%
\special{ar 3562 1498 60 588  0.0000000 1.5707963}%
%
\special{pn 8}%
\special{ar 1016 1502 60 588  4.7123890 6.2831853}%
\special{ar 1016 1502 60 588  0.0000000 1.5707963}%
%
\special{pn 8}%
\special{ar 1010 1498 60 588  1.5707963 4.7123890}%
%
\special{pn 8}%
\special{pa 1016 916}%
\special{pa 3554 916}%
\special{fp}%
\special{pa 3548 2084}%
\special{pa 1010 2084}%
\special{fp}%
%
\special{pn 8}%
\special{ar 3554 1508 46 594  1.4442042 1.4817629}%
\special{ar 3554 1508 46 594  1.5944389 1.6319976}%
\special{ar 3554 1508 46 594  1.7446737 1.7822324}%
\special{ar 3554 1508 46 594  1.8949084 1.9324671}%
\special{ar 3554 1508 46 594  2.0451432 2.0827019}%
\special{ar 3554 1508 46 594  2.1953779 2.2329366}%
\special{ar 3554 1508 46 594  2.3456127 2.3831713}%
\special{ar 3554 1508 46 594  2.4958474 2.5334061}%
\special{ar 3554 1508 46 594  2.6460821 2.6836408}%
\special{ar 3554 1508 46 594  2.7963169 2.8338756}%
\special{ar 3554 1508 46 594  2.9465516 2.9841103}%
\special{ar 3554 1508 46 594  3.0967864 3.1343450}%
\special{ar 3554 1508 46 594  3.2470211 3.2845798}%
\special{ar 3554 1508 46 594  3.3972558 3.4348145}%
\special{ar 3554 1508 46 594  3.5474906 3.5850493}%
\special{ar 3554 1508 46 594  3.6977253 3.7352840}%
\special{ar 3554 1508 46 594  3.8479601 3.8855188}%
\special{ar 3554 1508 46 594  3.9981948 4.0357535}%
\special{ar 3554 1508 46 594  4.1484296 4.1859882}%
\special{ar 3554 1508 46 594  4.2986643 4.3362230}%
\special{ar 3554 1508 46 594  4.4488990 4.4864577}%
\special{ar 3554 1508 46 594  4.5991338 4.6366925}%
%
\special{pn 8}%
\special{ar 2292 1508 522 82  3.9327711 6.2831853}%
\special{ar 2292 1508 522 82  0.0000000 3.9269908}%
%
\special{pn 8}%
\special{ar 2332 1176 60 256  4.7123890 6.2831853}%
\special{ar 2332 1176 60 256  0.0000000 1.5707963}%
%
\special{pn 8}%
\special{ar 2318 1834 60 256  4.7123890 6.2831853}%
\special{ar 2318 1834 60 256  0.0000000 1.5707963}%
%
\special{pn 8}%
\special{ar 2332 1176 60 256  1.5707963 1.6469868}%
\special{ar 2332 1176 60 256  1.8755582 1.9517487}%
\special{ar 2332 1176 60 256  2.1803201 2.2565106}%
\special{ar 2332 1176 60 256  2.4850820 2.5612725}%
\special{ar 2332 1176 60 256  2.7898439 2.8660344}%
\special{ar 2332 1176 60 256  3.0946059 3.1707963}%
\special{ar 2332 1176 60 256  3.3993678 3.4755582}%
\special{ar 2332 1176 60 256  3.7041297 3.7803201}%
\special{ar 2332 1176 60 256  4.0088916 4.0850820}%
\special{ar 2332 1176 60 256  4.3136535 4.3898439}%
\special{ar 2332 1176 60 256  4.6184154 4.6946059}%
%
\special{pn 8}%
\special{ar 2318 1834 60 256  1.5707963 1.6467457}%
\special{ar 2318 1834 60 256  1.8745938 1.9505432}%
\special{ar 2318 1834 60 256  2.1783913 2.2543406}%
\special{ar 2318 1834 60 256  2.4821887 2.5581381}%
\special{ar 2318 1834 60 256  2.7859862 2.8619356}%
\special{ar 2318 1834 60 256  3.0897837 3.1657330}%
\special{ar 2318 1834 60 256  3.3935811 3.4695305}%
\special{ar 2318 1834 60 256  3.6973786 3.7733280}%
\special{ar 2318 1834 60 256  4.0011761 4.0771254}%
\special{ar 2318 1834 60 256  4.3049735 4.3809229}%
\special{ar 2318 1834 60 256  4.6087710 4.6847204}%
%
\special{pn 8}%
\special{ar 1010 1508 768 70  6.2295782 6.2831853}%
\special{ar 1010 1508 768 70  0.0000000 1.4833099}%
%
\special{pn 8}%
\special{ar 3580 1508 766 72  3.0976827 3.1263566}%
\special{ar 3580 1508 766 72  3.2123781 3.2410519}%
\special{ar 3580 1508 766 72  3.3270734 3.3557473}%
\special{ar 3580 1508 766 72  3.4417688 3.4704426}%
\special{ar 3580 1508 766 72  3.5564641 3.5851379}%
\special{ar 3580 1508 766 72  3.6711594 3.6998333}%
\special{ar 3580 1508 766 72  3.7858548 3.8145286}%
\special{ar 3580 1508 766 72  3.9005501 3.9292240}%
\special{ar 3580 1508 766 72  4.0152455 4.0439193}%
\special{ar 3580 1508 766 72  4.1299408 4.1586146}%
\special{ar 3580 1508 766 72  4.2446361 4.2733100}%
\special{ar 3580 1508 766 72  4.3593315 4.3880053}%
\special{ar 3580 1508 766 72  4.4740268 4.5027007}%
\special{ar 3580 1508 766 72  4.5887222 4.6173960}%
%
\special{pn 8}%
\special{ar 3562 1508 748 60  1.5071264 3.0603763}%
%
\special{pn 8}%
\special{ar 1010 1508 764 62  4.6482256 4.6773519}%
\special{ar 1010 1508 764 62  4.7647305 4.7938567}%
\special{ar 1010 1508 764 62  4.8812353 4.9103616}%
\special{ar 1010 1508 764 62  4.9977402 5.0268664}%
\special{ar 1010 1508 764 62  5.1142451 5.1433713}%
\special{ar 1010 1508 764 62  5.2307499 5.2598761}%
\special{ar 1010 1508 764 62  5.3472548 5.3763810}%
\special{ar 1010 1508 764 62  5.4637596 5.4928858}%
\special{ar 1010 1508 764 62  5.5802645 5.6093907}%
\special{ar 1010 1508 764 62  5.6967693 5.7258955}%
\special{ar 1010 1508 764 62  5.8132742 5.8424004}%
\special{ar 1010 1508 764 62  5.9297790 5.9589053}%
\special{ar 1010 1508 764 62  6.0462839 6.0754101}%
\special{ar 1010 1508 764 62  6.1627887 6.1885323}%
\put(27.0000,-9.3000){\makebox(0,0)[rt]{$\varpi_{*}(0)$}}%
\put(22.9000,-20.0000){\makebox(0,0)[lb]{$-\varpi_{*}(0)$}}%
\put(8.1000,-15.4000){\makebox(0,0)[lb]{$0$}}%
\put(35.5000,-15.8000){\makebox(0,0)[lb]{$0$}}%
\put(12.9000,-16.3000){\makebox(0,0)[lb]{\footnotesize{$slit$}}}%
\put(30.9000,-16.4000){\makebox(0,0)[lb]{\footnotesize{$slit$}}}%
\put(14.8000,-12.5000){\makebox(0,0)[lb]{\footnotesize{$+$-sheet}}}%
\put(14.0000,-18.4000){\makebox(0,0)[lb]{\footnotesize{$-$-sheet}}}%
\end{picture}%
\hfill
\parbox[b]{.55\linewidth}
{This is called the {\bf pseudo-vacuum}. We denote this by ${:}\varpi_*(0){:}_{_K}$, 
but note that the pseudo-vacuum is expressed only by expression
parameter $K\in{\mathfrak K}_0$.
Minding the existence of opposite sheet and the 
$2\pi i$-alternating periodicity for $|s|\gg 0$, we set the 
slits between singular points and $\pm\infty$.  
Consider now the integral 
$\frac{1}{4\pi}\int_0^{4\pi}{:}e_*^{(s{+}it)(\frac{1}{i\h}u{\ctt}v)}{:}_{_K}dt$. 
Putting the periodicity in mind, Cauchy's integral theorem gives}  
\begin{equation}\label{pseudovac}
\frac{1}{4\pi}\int_0^{4\pi}{:}e_*^{(s{+}it)(\frac{1}{i\h}u{\ctt}v)}{:}_{_K}dt=
\left\{
\begin{matrix}
0,& s<a&{ }\\
{:}\varpi_*(0){:}_{_K},&a<s<b,&\quad K\in{\mathfrak K}_0\\
0,& b<s&{ }
\end{matrix}
\right.
\end{equation}
\unitlength 0.1in
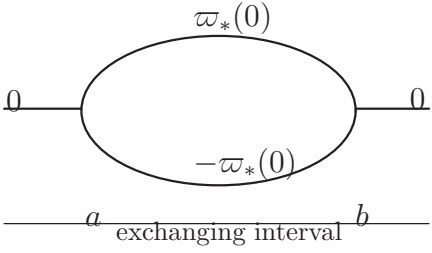
\begin{picture}(22.0000, 11.6000)(1.2000,-15.6000)
%
\special{pn 13}%
\special{ar 1230 670 710 390  0.0253467 6.2831853}%
\special{ar 1230 670 710 390  0.0000000 0.0178199}%
%
\special{pn 13}%
\special{pa 520 660}%
\special{pa 120 660}%
\special{fp}%
\special{pa 1940 660}%
\special{pa 2320 660}%
\special{fp}%
%
\special{pn 8}%
\special{pa 120 1260}%
\special{pa 2320 1260}%
\special{fp}%
\put(19.4000,-12.7000){\makebox(0,0)[lb]{$b$}}%
\put(5.4000,-12.7000){\makebox(0,0)[lb]{$a$}}%
\put(7.0000,-13.7000){\makebox(0,0)[lb]{\footnotesize{exchanging interval}}}%
\put(1.3000,-6.7000){\makebox(0,0)[lb]{$0$}}%
\put(22.2000,-6.6000){\makebox(0,0)[lb]{$0$}}%
\put(11.0,-2.7000){\makebox(0,0)[lb]{$\varpi_*(0)$}}%
\put(11.0,-10.3000){\makebox(0,0)[lb]{$-{\varpi_*(0)}$}}%
\end{picture}%
\hfill
\parbox[b]{.55\linewidth}
{This formula may be expressed symbolically by the l.h.s. picture.  
Since $\varpi_{00}$, $\overline{\varpi}_{00}$ are defined respectively by integrals 
$\frac{1}{2\pi}\int_{-\pi}^{\pi}
{:}e_*^{(s{+}i\sigma)(\frac{1}{i\h}u{\ctt}v\mp\frac{1}{2})}{:}_{_K}d\sigma$, 
for $s<\!<0$ and $s>\!>0$,  
the direct computation via changing variables gives also 
$$
\begin{aligned}
&{:}\varpi_{00}{*}\varpi_*(0){:}_{_K}=0={:}\varpi_*(0){*}\varpi_{00}{:}_{_K},
\quad K{\in}{\mathfrak K}_0,\\ 
&{:}\overline{\varpi}_{00}{*}\varpi_*(0){:}_{_K}=0={:}\varpi_*(0){*}\overline{\varpi}_{00}{:}_{_K},
\quad K{\in}{\mathfrak K}_0.
\end{aligned}
$$
}

Note that ${\varpi}_{00}, \overline{\varpi}_{00}$ are defined also by 
$\frac{1}{4\pi}\int_{-2\pi}^{2\pi}
{:}e_*^{(s{+}i\sigma)(\frac{1}{i\h}u{\ctt}v\mp\frac{1}{2})}{:}_{_K}d\sigma$, 
for $s<\!<0$ and $s>\!>0$. 
Using these we have also 
\begin{prop}\label{control00}
If $K{\in}{\mathfrak K}_0$, then 
for every polynomial $p(u,v)$, 
$$
\varpi_{*}(0){*}p_*(u,v){*}{\varpi}_{00}{=}0{=}
{\varpi}_{00}{*}p_*(u,v){*}{\varpi}_{*}(0),\quad   
\varpi_{*}(0){*}p_*(u,v){*}\overline{\varpi}_{00}{=}0{=}
\overline{\varpi}_{00}{*}p_*(u,v){*}{\varpi}_{*}(0).
$$
\end{prop}

\noindent
{\bf Proof}\,\,Note first that
$(u{\ctt}v){*}{\varpi}_{*}(0){=}0{=}{\varpi}_{*}(0){*}(u{\ctt}v)$, and  
$\varpi_{00}{*}u{=}0{=}v{*}\varpi_{00}$. Hence we have only to show 
$$
{\varpi}_{*}(0){*}u{*}{\varpi}_{00}=0={\varpi}_{00}{*}v{*}{\varpi}_{*}(0).
$$
 The bumping identity gives  ${\varpi}_{*}(0){*}u=
u{*}\frac{1}{4\pi}\int_{-2\pi}^{2\pi}e_*^{t(\frac{1}{i\h}u{\ctt}v{+}1)}dt$. Thus, 
for $s\ll 0$,
$$
16\pi^2{\varpi}_{*}(0){*}u{*}{\varpi}_{00}=
u{*}\int_{-2\pi}^{2\pi}\int_{-2\pi}^{2\pi}
e_*^{it(\frac{1}{i\h}u{\ctt}v{+}1)}{*}e_*^{(s{+}i\tau)\frac{1}{i\h}u{\ctt}v}dtd\tau
$$
Exponential law and changing variables make this integral vanish.
${}$\hfill$\Box$

\subsubsection{No other idempotent element}
In general, ${:}e_*^{it\frac{1}{i\h}u{\ctt}v}{:}_{_K}$ is $4\pi$-periodic one
parameter subgroup in $t{\in}{\mathbb R}$ under a generic ordered expression. 
Hence the exponential law gives for every rational number $\frac{q}{p}$, 
${:}e_*^{it(\frac{1}{i\h}u{\ctt}v{\pm}\frac{q}{p})}{:}_{_K}$ is
$4p\pi$-periodic one parameter subgroup and the period integral
$$
\varpi_{*}(\pm\frac{q}{p})=
\frac{1}{4p\pi}\int_0^{4p\pi}e_*^{it(\frac{1}{i\h}u{\ctt}v{\pm}\frac{q}{p})}dt
$$
has the idempotent property 
$\varpi_{*}(\pm\frac{q}{p}){*}\varpi_{*}(\pm\frac{q}{p})=\varpi_{*}(\pm\frac{q}{p})$.
However, we show in what follows that this is nontrivial only if $\frac{q}{p}{=}0$.

Cauchy's integral theorem shows that 
if $0<a$, then ${:}\varpi_{*}(\frac{q}{p}){:}_{_K}=0$ for 
$\frac{q}{p}>{-}\frac{1}{2}$,  and if $b<0$, then  
${:}\varpi_{*}(\frac{q}{p}){:}_{_K}=0$  for 
$\frac{q}{p}<\frac{1}{2}$. 
Recalling \eqref{shiftvacuum}, we see that 
pseudo-vacuums appears only for the case $a<0<b$, and 
the case
$|\frac{q}{p}|\leq \frac{1}{2}$ is essential by a suitable shift of
integers via bumping identity.  

If $a<0<b$, i.e. $K\in{\mathfrak K}_0$, then 
${:}e_*^{i\sigma(\frac{1}{i\h}u{\ctt}v{+}\frac{q}{p})}{:}_{_K}$ is 
$2p\pi$-periodic, but 
if $p$ is an even integer
(hence $q$ is an odd integer), then 
${:}e_*^{i\sigma(\frac{1}{i\h}u{\ctt}v{+}\frac{q}{p})}{:}_{_K}$ is 
alternating $p\pi$ periodic, and 
\begin{equation}\label{indep-p}
\frac{1}{2p\pi}\int_{0}^{2p\pi}
{:}e_*^{(s{+}i\sigma)(\frac{1}{i\h}u{\ctt}v{+}\frac{q}{p})}{:}_{_K}d\sigma=0.
\end{equation}
In fact, we see a stronger result as follows:
\begin{thm}\label{thepseudovacuum}
If ${-}\frac{1}{2}{<}\frac{q}{p}{<}\frac{1}{2}$, then under the
$K$-expression,   
$\varpi_{*}(\pm\frac{q}{p}){=}
\frac{1}{2p\pi}\int_0^{2p\pi}e_*^{it(\frac{1}{i\h}u{\ctt}v{\pm}\frac{q}{p})}dt{=}0$
except the case $\frac{q}{p}{=}0$, where 
$\varpi_{*}(0){=}
\frac{1}{4\pi}\int_0^{4\pi}e_*^{it(\frac{1}{i\h}u{\ctt}v)}dt{=}
\frac{1}{2\pi}\int_0^{2\pi}e_*^{it(\frac{1}{i\h}u{\ctt}v)}dt$.
\end{thm}

\noindent
{\bf Proof}\,\,
$e_*^{it(\frac{q}{p}{+}\frac{1}{i\h}u{\ctt}v)}$ is $2p\pi$-periodic,
As $e_*^{it(\frac{q}{p}{+}\frac{1}{i\h}u{\ctt}v)}$ and $e^{it\frac{1}{p}}$ are $2p\pi$-periodic, 
the Fourier expansion of this is given by 
$$
e_*^{i\tau(\frac{q}{p}{+}\frac{1}{i\h}u{\ctt}v)}=
\frac{1}{2p\pi}\sum_{k\in\mathbb Z}\int_0^{2p\pi}
e_*^{it(\frac{q}{p}{+}\frac{1}{i\h}u{\ctt}v{-}\frac{k}{p})}dt\,\,e^{i\frac{k}{p}\tau}.
$$
This converges in the $C^{\infty}$-topology on $S^1$.
Set the r.h.s. by 
$f(\tau)=\sum a_ke^{i\frac{k}{p}\tau}$.
Note the $2p\pi$-periodicity, and also that $f(\tau)$ has the property 
$f(\tau{+}2\pi)=e^{2\pi i\frac{q}{p}}f(\tau)$. 
This gives 
$$
f(\tau{+}2\pi)=\sum a_ke^{i\frac{k}{p}2\pi}e^{i\frac{k}{p}\tau}=
e^{2\pi i\frac{q}{p}}\sum a_ke^{i\frac{k}{p}\tau}.
$$
The uniqueness of Fourier coefficients gives 
$a_ke^{i\frac{k}{p}2\pi}=e^{2\pi i\frac{q}{p}}a_k$ and hence
$a_k$ vanishes if $k\not=q{+}p\ell$, $\ell\in{\mathbb Z}$.
Hence the Fourier series of $f(\tau)$ is 
\begin{equation}\label{Fourierexp}
\begin{aligned}
e_*^{i\tau(\frac{q}{p}{+}\frac{1}{i\h}u{\ctt}v)}=
\sum_{\ell}\frac{1}{2p\pi}\int_0^{2p\pi}
e_*^{it(\frac{1}{i\h}u{\ctt}v{-}\ell)}dt\,\,
e^{i(\frac{q}{p}{+}\ell)\tau}
=
\sum_{\ell}\frac{1}{2\pi}\int_0^{2\pi}
e_*^{it(\frac{1}{i\h}u{\ctt}v{-}\ell)}dt\,\,
e^{i(\frac{q}{p}{+}\ell)\tau},
\end{aligned}
\end{equation}
for the integrand is $2\pi$-periodic.  Componentwise integration
gives  that if $\frac{q}{p}{+}\ell\not=0$, then 
$$
\varpi_{*}(\frac{q}{p})=
\frac{1}{2p\pi}\int_0^{2p\pi}e_*^{i\tau(\frac{q}{p}{+}\frac{1}{i\h}u{\ctt}v)}d\tau=
\sum_{\ell}\frac{1}{2pi}\int_0^{2\pi}
e_*^{it(\frac{1}{i\h}u{*}v{-}\ell)}dt
\int_0^{2p\pi}e^{i(\frac{q}{p}{+}\ell)\tau}d\tau=0.
$$ 
This is nontrivial if and only if $\frac{q}{p}=0$ and $\ell{=}0$.
${ }$\hfill $\Box$


\subsection{Two inverses in generic ordered expressions}

Recall the formula \eqref{genericparam00} shows that   
${:}e_*^{\zeta\frac{1}{i\h}u{\ctt}v}{:}_{_K}$ is 
$4\pi i$-periodic along any line parallel to the pure imaginary axis, and 
rapidly decreasing along any line parallel to the real line.  

Similar to Jacobi's theta function in \cite{OMMY3}, we have a
convergence of the infinite summation  
\begin{equation}\label{Theta}
\Theta_{L}(\zeta;*)=
\sum_{k=-\infty}^{\infty}e_*^{(\zeta{+}kL)\frac{1}{i\h}u{\ctt}v},\quad 
\Theta_{L}(\zeta;K)=
\sum_{k=-\infty}^{\infty}{:}e_*^{(\zeta{+}kL)\frac{1}{i\h}u{\ctt}v}{:}_{_K}
\end{equation}
whenever there is no singular point on the line 
$\zeta{+}{\mathbb R}$. Hence we have two different inverses of 
$1{-}e_*^{\zeta\frac{1}{i\h}u{\ctt}v}$:
$$
\sum_{k=0}^{\infty}e_*^{(\zeta{+}kL)\frac{1}{i\h}u{\ctt}v},\quad 
\sum_{k=-\infty}^{-1}e_*^{(\zeta{+}kL)\frac{1}{i\h}u{\ctt}v}.
$$
At the first glance, $\Theta_{L}(\zeta;*)$ looks to be double
periodic such that 
$$
\Theta_{L}(\zeta{+}L;*)=\Theta_{L}(\zeta;*), \quad 
\Theta_{L}(\zeta{+}4\pi i;*)=\Theta_{L}(\zeta;*). 
$$  
However, there is no rule to evaluate 
$\Theta_{L}(\zeta;*)$ at $\zeta=x{+}iy$ univalent way if 
$\zeta=x{+}iy{+}kL$ is on the slit. Hence, $\Theta_{L}(\zeta;*)$ is  
well-defined only in the case that the exchanging interval $I_{\ctt}(K)$ 
is empty (eg. the case $\rho=0$ in \eqref{normalantinormal}). 
If this is the case, $\Theta_{L}(\zeta;*)$ is a double periodic
element having essential singular points. Thus, we have to develop
some other theory for such an ``elliptic'' element, which will be
discussed in forthcoming paper.

\medskip
Similarly, supposing there is no singular point on 
$iy{+}{\mathbb R}$, \eqref{genericparam00} gives for
$|{\rm{Re}}\,z|<\frac{1}{2}$  
that 
$$
D(iy, z{+}\frac{1}{i\h}u{\ctt}v, K)=
 \int_{-\infty}^{\infty}{:}e_*^{(x{+}iy)(z{+}\frac{1}{i\h}u{\ctt}v)}{:}_{_K}dx
$$ 
is an element of $H{\!o}l({\mathbb C}^2)$ in generic $K$. 
By Cauchy's integral theorem $D(iy, z{+}\frac{1}{i\h}u{\ctt}v, K)$
does not depend on $iy$ if $|y|$ is sufficiently small. But, the value
jumps discontinuously when $iy{+}{\mathbb R}$ hits the singularity.  

Note that integrals  
$\int_{-\infty}^{0}e_*^{t(z{+}\frac{1}{i\h}u{\ctt}v)}dt,
\,\,\,
-\int_{0}^{\infty}e_*^{t(z{+}\frac{1}{i\h}u{\ctt}v)}dt$
give inverses of $z{+}\frac{1}{i\h}u{\ctt}v$ in generic ordered 
expressions, which are denoted by 
$(z{+}\frac{1}{i\h}u{\ctt}v)_{*+}^{-1}$, 
$(z{+}\frac{1}{i\h}u{\ctt}v)_{*-}^{-1}$ respectively.  
The following may be viewed as a Sato's hyperfunction:     
\begin{prop}
  \label{property}
If $-\frac{1}{2}<{\rm {Re}}\,z<\frac{1}{2}$, then the difference 
of the two inverses is given by 
\begin{equation}
  \label{eq:zuv}
(z{+}\frac{1}{i\h}u{\ctt}v)_{*+}^{-1}
{-}(z{+}\frac{1}{i\h}u{\ctt}v)_{*-}^{-1}{=}
\int_{-\infty}^{\infty}e_*^{t(z{+}\frac{1}{i\h}u{\ctt}v)}dt,  
\end{equation}
which is holomorphic on this strip in generic 
ordered expressions. 
This will be called $*$-delta function and denoted by 
$\delta_*({-}iz{+}\frac{1}{\h}u{\ctt}v)$. 
\end{prop}

Note also that 
$$
{:}\delta_*({-}iz{+}\frac{1}{\h}u{\ctt}v){:}_{_K}=
D(0, z{+}\frac{1}{i\h}u{\ctt}v, K).
$$
Recall that 
$(z{+}\frac{1}{i\h}u{\ctt}v){*}\delta_*({-}iz{+}\frac{1}{\h}u{\ctt}v)=0$.
Thus, we have to set 
$$
e_*^{is(z{+}\frac{1}{i\h}u{\ctt}v)}{*}\delta_*({-}iz{+}\frac{1}{\h}u{\ctt}v){=}
\delta_*({-}iz{+}\frac{1}{\h}u{\ctt}v)
$$
whenever the $*$-product $e_*^{isH}{*}f$ is defined as the real
analytic solution of the evolution equation $\frac{d}{ds}f_s=iH{*}f_s$,
$f_0=f$. In spite of this, the integral element  
$D(is, z{+}\frac{1}{i\h}u{\ctt}v, K)$ is not continuous in $s$ in general. 

\bigskip
For $a,b$ such that $a\not=b$, four elements 
\begin{equation}\label{resolventcal}
\frac{1}{b-a}
\{(a+\frac{1}{i\h}u{\ctt}v)_{*\pm}^{-1}-
  (b+\frac{1}{i\h}u{\ctt}v)_{*\pm}^{-1}\}\quad 
(\text{independent}\,\,\pm),
\end{equation}
give respectively inverses of 
$$
(a+\frac{1}{i\h}u{\ctt}v){*}(b+\frac{1}{ i\h}u{\ctt}v).
$$
Thus, we define the $*$-product 
$(a+\frac{1}{i\h}u{\ctt}v)_{*\pm}^{-1}{*}
(b+\frac{1}{i\h}u{\ctt}v)_{*\pm}^{-1}$ by this 
formula. Note that $*$-multiplications are 
calculated by summations. Similarly, we define 
$$
(a+\frac{1}{i\h}u{\ctt}v)_{*\pm}^{-1}{*}
(a+\frac{1}{i\h}u{\ctt}v)_{*\pm}^{-1}{=}
-\frac{d}{da}(a+\frac{1}{i\h}u{\ctt}v)_{*\pm}^{-1},\quad (\pm\,\text{ respectively})
$$
but $(a+\frac{1}{i\h}u{\ctt}v)_{*\pm}^{-1}{*}
(a+\frac{1}{i\h}u{\ctt}v)_{*\mp}^{-1}$ diverges. 
We refer such calculations 
to {\it resolvent calculus}. (Cf.\cite{OMMY3}.)

\bigskip
On the other hand, note that a change of variables $t\to -t$ gives 
$$
((-z){+}\frac{1}{i\h}u{\ctt}v)_{*-}^{-1}{=}
{-}\int_0^{\infty}e_*^{{-}t(z{-}\frac{1}{i\h}u{\ctt}v)}dt 
{=}{-}\int_{-\infty}^{0}e_*^{(z{-}\frac{1}{i\h}u{\ctt}v)}dt.
$$ 
Thus, in generic ordered expressions, we see that 
\begin{equation}
  \label{eq:notation}
(z{-}\frac{1}{i\h}u{\ctt}v)_{*-}^{-1}{=} 
{-}((-z){+}\frac{1}{i\h}u{\ctt}v)_{*-}^{-1}.   
\end{equation}

\begin{prop}\label{twoinv}
In generic ordered expressions, integrals 
$\int_{-\infty}^0e_*^{t(z{+}\frac{1}{i\h}u{\ctt}v)}
\quad{\text{and}}\,\, 
\int_{-\infty}^0e_*^{t(z{-}\frac{1}{i\h}u{\ctt}v)}$
converge on the domain ${\rm {Re}}\,z{>}-\frac{1}{2}$.
\end{prop}

Using this, note first that 
$(z{\pm}\frac{1}{i\h}u{\ctt}v)_{*\pm}^{-1}$ are holomorphic 
on the domain ${\rm{Re}}\,z >{-}\frac{1}{2}$ 
in generic ordered expressions. It is natural 
to expect that 
$$
(z{\pm}\frac{1}{i\h}u{\ctt}v)_{*\pm}^{-1}
{=}C(C(z{\pm}\frac{1}{i\h}u{\ctt}v))_{*\pm}^{-1}
$$ 
for any non-zero constant $C$. But this holds 
only for $C=re^{i\theta}$ such that $|\theta|$ is small 
enough, because of the singular point of 
$e_*^{t(z{+}\frac{1}{i\h}u{\ctt}v)}$.
To confirm this, we set
$C{=}re^{i\theta}$ and consider the integral 
$$ 
re^{i\theta}\int_{-\infty}^{0}
e_*^{re^{i\theta}t(z{\pm}\frac{1}{i\h}u{\ctt}v)}dt.
$$
It is easy to see that its $r$-derivative vanishes for $r>0$.
For the $\theta$-derivative, note that in generic $K$-ordered
expressions, the phase part of the integrand is bounded in 
$t$ and the amplitude is  
$$
\frac{2e^{i\theta}tz}{(1{-}\kappa)
e^{e^{i\theta}t/2}+(1{+}\kappa)e^{-e^{i\theta}t/2}},
\quad \kappa{\not=}1.
$$
The integral converges whenever 
${\rm{Re}}\,e^{i\theta}(z{\pm}\frac{1}{2})>0$, 
and $-te^{i\theta}$ does not hit the singular points.
Then, the integration by parts gives that its 
$\theta$-derivative vanishes.  

It follows that in generic ordered expressions, 
$(z{\pm}\frac{1}{i\h}u{\ctt}v)_{*\pm}^{-1}$ are holomorphic 
on an overhanged sectoral domain  
$$
{\mathbb C}{\setminus}\{-D{-}\frac{1}{2}\},\quad 
D=\{z=re^{i\theta}, |\theta|<\frac{\pi}{2}-\e\}.
$$ 

\noindent
{\bf Remark}\,\,If  $a{>}0$, then $a\big((a(z{\pm}\frac{1}{i\h}u{\ctt}v)_{*\pm})^{-1}\big){=}
(z{\pm}\frac{1}{i\h}u{\ctt}v)_{*\pm}^{-1}$, 
but $a\big((a(z{\pm}\frac{1}{i\h}u{\ctt}v)_{*\pm})^{-1}\big){=}
(z{\pm}\frac{1}{i\h}u{\ctt}v)_{*\mp}^{-1}$ for $a{<}0$.

\bigskip
Next, it is natural to expect that the bumping identity 
$(u{\ctt}v){*}v{=}v{*}(u{\ctt}v{-}i\h)$ 
gives the following ``sliding identities"
$$
v_{*+}^{-1}{*}(z{+}\frac{1}{i\h}u{\ctt}v)_{*+}^{-1}{*}v{=}
(z{-}1{+}\frac{1}{i\h}u{\ctt}v)_{+*}^{-1}, \quad 
v_{*+}^{-1}{*}(z{-}\frac{1}{i\h}u{\ctt}v)_{*-}^{-1}{*}v{=}
(z{+}1{-}\frac{1}{i\h}u{\ctt}v)_{*-}^{-1}
$$ 
whenever one can use the inverse of $v$ in a suitable ordered 
expression. 
In \S\,\ref{analconi}, analytic continuation will be produced via these 
sliding identities.  

\medskip
However, the existence of $v_{*+}^{-1}$ is not a generic property.
In this note, we use the sliding identity by using,   
instead of $v_{*+}^{-1}$, the left inverse $v^{\ctt}$ of 
$v$ given by \eqref{half-inv}. 

\noindent
{\bf Remark}\, There is a $K$-ordered expression such that 
${:}\int_{-\infty}^0e_*^{tv}dt{:}_{_K}$ 
converges to give an inverse of ${:}v_{*+}^{-1}{:}_{_K}$ of 
$v$ (cf. \cite{OMMY3}).  
But we easily see ${:}v_{*+}^{-1}{*}\varpi_{00}{:}_{_K}$ must  
diverge, for $v{*}\varpi_{00}=0$ in generic ordered expression.

\subsubsection{Remarks on ordinary calculus}

Next, we note that \eqref{genericparam00} gives in generic 
ordered expressions that the integrals 
$$
(u{*}v)_{*-}^{-1}{=}
-\frac{1}{i\h}\int_0^{\infty}e_*^{s\frac{1}{i\h}u{*}v}ds,
\quad
(v{*}u)_{*+}^{-1}{=}
\frac{1}{i\h}\int_{-\infty}^0e_*^{s\frac{1}{i\h}v{*}u}ds
$$
exist to give inverses of $u{*}v$, $v{*}u$ respectively.
Hence the next ones give left/right inverses of $u,v$ 
\begin{equation}\label{half-inv}
v^{\ctt}{=}u{*}(v{*}u)_{*+}^{-1},\quad  
u^{\btt}{=}v{*}(u{*}v)_{*-}^{-1},
\end{equation}
for it is easy to see that
$$
v{*}v^{\ctt}{=}1,\quad v^{\ctt}{*}v{=}1{-}\varpi_{00}, \quad
u{*}u^{\btt}{=}1,\quad u^{\btt}{*}u{=}
1{-}\overline{\varpi}_{00}.
$$
The bumping identity $u{*}f_*(v{*}u)=f_*(u{*}v){*}u$ gives 
$$
v{*}(z{+}\frac{1}{i\h}u{\ctt}v){*}v^{\ctt}
{=}z{+}1{+}\frac{1}{i\h}u{\ctt}v, \quad 
v^{\ctt}{*}(z{+}\frac{1}{i\h}u{\ctt}v){*}v{=}
(1{-}\varpi_{00}){*}(z{-}1{+}\frac{1}{i\h}u{\ctt}v).
$$
$$
u{*}(z{+}\frac{1}{i\h}u{\ctt}v){*}u^{\btt}
{=}z{-}1{+}\frac{1}{i\h}u{\ctt}v, \quad 
u^{\btt}{*}(z{+}\frac{1}{i\h}u{\ctt}v){*}u{=}
(1{-}\overline{\varpi}_{00}){*}
(z{+}1{+}\frac{1}{i\h}u{\ctt}v).
$$

The successive use of the bumping identity gives  
the following useful formula:
\begin{equation}
  \label{eq:powerinv2}
\begin{aligned}
&(v^{\ctt})_*^n{*}\varpi_{00}
{=}(u{*}(v{*}u)_{*+}^{-1})^n{*}\varpi_{00}
{=}\frac{1}{n!}(\frac{1}{i\h}u)_*^n{*}\varpi_{00},\\
&(u^{\btt})_*^n{*}\overline{\varpi}_{00}
{=}(v{*}(u{*}v)_{*-}^{-1})^n{*}\overline{\varpi}_{00}
{=}\frac{1}{n!}(\frac{1}{i\h}v)_*^n{*}\overline{\varpi}_{00}.
\end{aligned}
\end{equation}

Next Proposition shows that the vacuum representation involves
ordinary elementary calculus. 
\begin{prop}\label{elemcalcal0}
In generic ordered expressions, we have  
$$
\frac{1}{i\h}v{*}f(u){*}\varpi_{00}=f'(u){*}\varpi_{00}, \quad 
v^{\ctt}{*}f(u){*}\varpi_{00}=\frac{1}{i\h}\int_{-\infty}^uf(x)dx{*}\varpi_{00}.
$$
That is, $\frac{1}{i\h}v{*}$ represents the differentiation, and its 
right inverse $v^{\ctt}{*}$ represents the integration. 
\end{prop}

\noindent
{\bf Proof}\,\,Since $v{*}f(u)=f(u){*}v{+}i\h f'(u)$, first one is
easy to see. For the second,
$[\frac{1}{i\h}v{*}u,u]=u$ and $v{*}\varpi_{00}=0$ gives that 
$$
e_*^{t\frac{1}{i\h}v{*}u}{*}f(u){*}e_*^{-t\frac{1}{i\h}v{*}u}=f(e^tu),\quad 
e_*^{t\frac{1}{i\h}v{*}u}{*}\varpi_{00}=e^t{*}\varpi_{00}. 
$$
It follows 
$v^{\ctt}{*}f(u){*}\varpi_{00}=\frac{1}{i\h}\int_{-\infty}^0e^{t}u{*}f(e^tu)dt{*}\varpi_{00}.$
Set $x=e^tu$ to obtain the result. \hfill $\Box$

\bigskip
\fbox{\parbox[c]{.9\linewidth}{
By Proposition\,\ref{elemcalcal0}, one can begin the whole story with
the algebra of ordinary elementary calculus of differentiation and
integration. This implies that all 
strange phenomena mentioned in \cite{OMMY3}, \cite{OMMY4} and
\cite{ommy5} are already involved in the  ordinary elementary calculus.}}

\subsection{Integrals along closed paths}

All closed paths drawn in the
Fig.1 give closed paths in the Riemann surface with infinite
genus, where dotted lines are in the negative sheets. 
Let $C$ be a one of such closed path. Then, the integral 
$\int_C{:}e_*^{z(\alpha{+}\frac{1}{i\h}u{\ctt}v)}dz{:}_{_K}$ must 
satisfy 
\begin{equation}\label{circlezero}
{:}(\alpha{+}\frac{1}{i\h}u{\ctt}v)*\!
\int_Ce_*^{z(\alpha{+}\frac{1}{i\h}u{\ctt}v)}dz{:}_{_K}=
{:}\int_C\frac{d}{dz}e_*^{z(\alpha{+}\frac{1}{i\h}u{\ctt}v)}dz{:}_{_K}=0.
\end{equation}

\noindent
{\bf Residue-like integrals}\,\,\,
Now, let $C$ be a closed path avoiding the singular points 
and intersecting the slit even (possibly zero) times.  
At a first glance, the integral  
 $\int_Ce_*^{z(\frac{1}{i\h}u{\ctt}v{+}\alpha)}dz$
looks to relate to the residues of singular points sitting in the inside 
of $C$. In fact, this integral is nothing to do with residues. 
We call this a residue-like integral. The residue-like integral 
of $e_*^{z(\frac{1}{i\h}u{\ctt}v{+}\alpha)}$ 
relates to the global topological nature. 
On the contrary, the residues of 
$e_*^{z(\alpha{+}\frac{1}{i\h}u{\ctt}v)}$ does not  
have a global nature. 

\bigskip
The bumping identity $u{*}f_*(v{*}u)=f_*(u{*}v){*}u$ gives 
\begin{equation}\label{slide01}
u^n{*}e_*^{z(\frac{1}{i\h}u{\ctt}v{+}\alpha)}{=}
e_*^{z(\frac{1}{i\h}u{\ctt}v{+}\alpha{-}n)}{*}u^n,
 \quad 
v^n{*}e_*^{z(\frac{1}{i\h}u{\ctt}v{+}\alpha)}{=}
e_*^{z(\frac{1}{i\h}u{\ctt}v{+}\alpha{+}n)}{*}v^n.
\end{equation}
Hence  
$$
\begin{aligned}
\int_{C}
{:}u{*}e_*^{z(\frac{1}{i\h}u{\ctt}v{+}\alpha)}{*}v{:}_{_K}dz
=
\int_{C}{:}(u{\ctt}v{-}\frac{i\h}{2}){*}e_*^{z(\frac{1}{i\h}u{\ctt}v{+}\alpha{-}1)}{:}_{_K}dz
=
(i\h)(\frac{1}{2}{-}\alpha)
\int_{C}
{:}e_*^{z(\frac{1}{i\h}u{\ctt}v{+}\alpha{-}1)}{:}_{_K}dz.
\end{aligned}
$$
Repeat this to obtain 
\begin{equation*}
\int_{C}{:}u^n{*}e_*^{z(\frac{1}{i\h}u{\ctt}v{+}\alpha)}{*}v^ndz{:}_{_K}
=
(i\h)^n(\frac{1}{2}{-}\alpha)_n
{:}\int_{C}e_*^{z(\frac{1}{i\h}u{\ctt}v{+}\alpha{-}n)}dz{:}_{_K},
\end{equation*}
where $(a)_n=a(a{+}1)\cdots(a{+}n{-1})$, $(a)_0=1$. 
Similarly we have 
\begin{equation*}
\begin{aligned}
{:}v^n{*}\!\int_{C}\!\!e_*^{z(\frac{1}{i\h}u{\ctt}v{+}\alpha)}{:}_{_K}dz{*}u^n{:}_{_K}
{=}&
(i\h)^n(-\frac{3}{2}{-}\alpha)(-\frac{5}{2}{-}\alpha)\cdots(-\frac{2n{+}1}{2}{-}\alpha)
{:}\int_{C}\!\!e_*^{z(\frac{1}{i\h}u{\ctt}v{+}\alpha{+}n)}dz{:}_{_K}\\
{=}&
(\frac{1}{2}{-}\alpha)_{-n}{:}\int_{C}\!e_*^{z(\frac{1}{i\h}u{\ctt}v{+}\alpha{+}n)}dz{:}_{_K},
\end{aligned}
\end{equation*}
by extending the convention $(a)_{-n}=(a{-}1)(a{-}2)\cdots(a{-}n)$. 
If we use the convention    
\begin{equation}\label{pmconvention}
\zeta^k=
\left\{
\begin{matrix}
u^k,& k\geq 0\\
v^{|k|},& k<0,
\end{matrix}
\right.\qquad 
{\hat\zeta}^{\ell}=
\left\{
\begin{matrix}
v^\ell,& \ell\geq 0\\
u^{|\ell|},& \ell<0,
\end{matrix}
\right.
\end{equation}
then we have 
\begin{equation}\label{shiftvacuum}
{:}\zeta^n{*}\int_{C}e_*^{z(\frac{1}{i\h}u{\ctt}v{+}\alpha)}dz{*}\hat\zeta^n{:}_{_K}
=(\frac{1}{2}{-}\alpha)_n{:}\int_{C}e_*^{z(\frac{1}{i\h}u{\ctt}v{+}\alpha{-}n)}dz{:}_{_K}, \quad
n{\in}{\mathbb Z} 
\end{equation}
Hence the essential part of the integral \eqref{shiftvacuum} 
is reduced to the case $|{\rm{Re}}(\alpha)|\leq\frac{1}{2}$.

\medskip
Moreover, recall the associativity  
$$
\begin{aligned}
(e_*^{s(\alpha{+}\frac{1}{i\h}u{\ctt}v)}{*}\frac{1}{i\h}u{\ctt}v){*}
e_*^{t(\beta{+}\frac{1}{i\h}u{\ctt}v)}{=}
e_*^{s(\alpha{+}\frac{1}{i\h}u{\ctt}v)}{*}(\frac{1}{i\h}u{\ctt}v {*}
e_*^{t(\beta{+}\frac{1}{i\h}u{\ctt}v)})
{=}
\frac{1}{i\h}u{\ctt}v{*}(e_*^{s(\alpha{+}\frac{1}{i\h}u{\ctt}v)}{*}
e_*^{t(\beta{+}\frac{1}{i\h}u{\ctt}v)})
\end{aligned}
$$
proved by the formal associativity theorem
(cf.\cite{ommy5}).  By the fundamental product formula
(cf.\cite{ommy5}), we see the product remains 
in the space ${\mathbb C}[u,v]e^{Q(u,v)}$. 
Since this is continuous w.r.t $(s,t)$, the integration by
$dsdt$ on compact domain $C{\times}C$ in \eqref{circlezero} gives that 
$$
\iint_{C{\times}C}(e_*^{z(\alpha{+}\frac{1}{i\h}u{\ctt}v)}{*}\frac{1}{i\h}u{\ctt}v)*
e_*^{\zeta(\beta{+}\frac{1}{i\h}u{\ctt}v)}dzd\zeta=
\iint_{C{\times}C}e_*^{z(\alpha{+}\frac{1}{i\h}u{\ctt}v)}{*}(\frac{1}{i\h}u{\ctt}v{*}
e_*^{\zeta(\beta{+}\frac{1}{i\h}u{\ctt}v)})dzd\zeta.
$$
Since $C{\times}C$ is compact, the uniform continuity gives 
$$
\int_{C}(e_*^{z(\alpha{+}\frac{1}{i\h}u{\ctt}v)}{*}\frac{1}{i\h}u{\ctt}v)dz{*}
\int_{C}e_*^{\zeta(\beta{+}\frac{1}{i\h}u{\ctt}v)}d\zeta=
\int_{C}e_*^{z(\alpha{+}\frac{1}{i\h}u{\ctt}v)}dz{*}
\int_{C}(\frac{1}{i\h}u{\ctt}v{*}
e_*^{\zeta(\beta{+}\frac{1}{i\h}u{\ctt}v)})d\zeta.
$$  
Hence we see 
$$
\begin{aligned}
{:}{-}\alpha\int_Ce_*^{z(\alpha{+}\frac{1}{i\h}u{\ctt}v)}dz*\!
\int_Ce_*^{\zeta(\beta{+}\frac{1}{i\h}u{\ctt}v)}d\zeta{:}_{_K}&=
{:}\int_Ce_*^{z(\alpha{+}\frac{1}{i\h}u{\ctt}v)}dz{*}\frac{1}{i\h}u{\ctt}v*\!
\int_Ce_*^{\zeta(\beta{+}\frac{1}{i\h}u{\ctt}v)}d\zeta{:}_{_K}\\
&={:}{-}\beta\int_Ce_*^{z(\alpha{+}\frac{1}{i\h}u{\ctt}v)}dz*\!
\int_Ce_*^{\zeta(\beta{+}\frac{1}{i\h}u{\ctt}v)}d\zeta{:}_{_K}.
\end{aligned}
$$
Thus, if $\alpha\not=\beta$, then
\begin{equation}\label{mmm}
{:}\int_Ce_*^{z(\alpha{+}\frac{1}{i\h}u{\ctt}v)}dz*\!
\int_Ce_*^{\zeta(\beta{+}\frac{1}{i\h}u{\ctt}v)}d\zeta{:}_{_K}=0 
\quad (\alpha\not=\beta).
\end{equation}

In the case $\alpha=0$ in generic $K$, we see easily

\noindent
\unitlength 0.1in
\begin{picture}( 33.0700, 10.3300)(  2.0000,-13.5900)
%
\special{pn 8}%
\special{pa 356 392}%
\special{pa 3482 392}%
\special{fp}%
\special{pa 3482 874}%
\special{pa 364 874}%
\special{fp}%
\special{pa 364 1354}%
\special{pa 3482 1354}%
\special{fp}%
%
\special{pn 8}%
\special{pa 356 706}%
\special{pa 1564 706}%
\special{fp}%
\special{pa 1564 1186}%
\special{pa 356 1186}%
\special{fp}%
%
\special{pn 8}%
\special{pa 2402 584}%
\special{pa 3482 584}%
\special{fp}%
\special{pa 3482 1066}%
\special{pa 2402 1066}%
\special{fp}%
\special{pa 2402 1066}%
\special{pa 2402 1066}%
\special{fp}%
%
\special{pn 8}%
\special{pa 1954 1354}%
\special{pa 1954 392}%
\special{fp}%
\special{pa 1996 392}%
\special{pa 1996 1354}%
\special{fp}%
%
\special{pn 20}%
\special{sh 1}%
\special{ar 2410 584 10 10 0  6.28318530717959E+0000}%
\special{sh 1}%
\special{ar 2410 1066 10 10 0  6.28318530717959E+0000}%
\special{sh 1}%
\special{ar 1564 1186 10 10 0  6.28318530717959E+0000}%
\special{sh 1}%
\special{ar 1564 706 10 10 0  6.28318530717959E+0000}%
\special{sh 1}%
\special{ar 1564 706 10 10 0  6.28318530717959E+0000}%
%
\special{pn 8}%
\special{pa 1504 392}%
\special{pa 1116 392}%
\special{fp}%
\special{sh 1}%
\special{pa 1116 392}%
\special{pa 1182 412}%
\special{pa 1168 392}%
\special{pa 1182 372}%
\special{pa 1116 392}%
\special{fp}%
\special{pa 1116 1354}%
\special{pa 1512 1354}%
\special{fp}%
\special{sh 1}%
\special{pa 1512 1354}%
\special{pa 1446 1334}%
\special{pa 1460 1354}%
\special{pa 1446 1374}%
\special{pa 1512 1354}%
\special{fp}%
\special{pa 1954 1034}%
\special{pa 1954 714}%
\special{fp}%
\special{sh 1}%
\special{pa 1954 714}%
\special{pa 1934 780}%
\special{pa 1954 766}%
\special{pa 1974 780}%
\special{pa 1954 714}%
\special{fp}%
\special{pa 1996 714}%
\special{pa 1996 1034}%
\special{fp}%
\special{sh 1}%
\special{pa 1996 1034}%
\special{pa 2016 966}%
\special{pa 1996 980}%
\special{pa 1976 966}%
\special{pa 1996 1034}%
\special{fp}%
\special{pa 1996 1034}%
\special{pa 1996 1034}%
\special{fp}%
%
\special{pn 8}%
\special{pa 2688 1354}%
\special{pa 3032 1354}%
\special{fp}%
\special{sh 1}%
\special{pa 3032 1354}%
\special{pa 2966 1334}%
\special{pa 2980 1354}%
\special{pa 2966 1374}%
\special{pa 3032 1354}%
\special{fp}%
%
\special{pn 8}%
\special{pa 3016 392}%
\special{pa 2670 392}%
\special{fp}%
\special{sh 1}%
\special{pa 2670 392}%
\special{pa 2736 412}%
\special{pa 2722 392}%
\special{pa 2736 372}%
\special{pa 2670 392}%
\special{fp}%
%
\special{pn 8}%
\special{pa 528 552}%
\special{pa 356 552}%
\special{fp}%
\special{sh 1}%
\special{pa 356 552}%
\special{pa 422 572}%
\special{pa 408 552}%
\special{pa 422 532}%
\special{pa 356 552}%
\special{fp}%
%
\special{pn 8}%
\special{pa 528 1034}%
\special{pa 356 1034}%
\special{fp}%
\special{sh 1}%
\special{pa 356 1034}%
\special{pa 422 1054}%
\special{pa 408 1034}%
\special{pa 422 1014}%
\special{pa 356 1034}%
\special{fp}%
%
\special{pn 8}%
\special{pa 3308 1194}%
\special{pa 3482 1194}%
\special{fp}%
\special{sh 1}%
\special{pa 3482 1194}%
\special{pa 3414 1174}%
\special{pa 3428 1194}%
\special{pa 3414 1214}%
\special{pa 3482 1194}%
\special{fp}%
%
\special{pn 8}%
\special{pa 3308 738}%
\special{pa 3482 738}%
\special{fp}%
\special{sh 1}%
\special{pa 3482 738}%
\special{pa 3414 718}%
\special{pa 3428 738}%
\special{pa 3414 758}%
\special{pa 3482 738}%
\special{fp}%
\put(19.7000,-14.1800){\makebox(0,0)[lb]{$s$}}%
\put(15.6400,-14.9000){\makebox(0,0)[lb]{$a$}}%
\put(24.0000,-14.700){\makebox(0,0)[lb]{$b$}}%
\put(35.0700,-12.4200){\makebox(0,0)[lb]{$0$}}%
\put(34.9000,-8.0900){\makebox(0,0)[lb]{$0$}}%
\put(2.0000,-5.9200){\makebox(0,0)[lb]{$0$}}%
\put(2.2600,-10.8900){\makebox(0,0)[lb]{$0$}}%
\put(17.4600,-6.9600){\makebox(0,0)[lb]{$C_+$}}%
\put(18.4600,-4.0000){\makebox(0,0)[lb]{$s{+}4\pi i$}}%
\put(20.8200,-10.2500){\makebox(0,0)[lb]{$C_-$}}%
%
\special{pn 8}%
\special{pa 3300 1354}%
\special{pa 3300 392}%
\special{dt 0.045}%
%
\special{pn 8}%
\special{pa 538 392}%
\special{pa 538 1354}%
\special{dt 0.045}%
%
\special{pn 8}%
\special{pa 540 390}%
\special{pa 540 550}%
\special{fp}%
\special{pa 540 1190}%
\special{pa 540 1350}%
\special{fp}%
\special{pa 3300 1350}%
\special{pa 3300 1070}%
\special{fp}%
\special{pa 3300 580}%
\special{pa 3300 390}%
\special{fp}%
\end{picture}%
\hfill
\parbox[b]{.45\linewidth}{  
$$
{:}\int_{C_+}e_*^{z\frac{1}{i\h}u{\ctt}v}dz{:}_{_K}=4\pi{:}\varpi_{*}(s){:}_{_K}
$$
$$
{:}\int_{C_-}e_*^{z\frac{1}{i\h}u{\ctt}v}dz{:}_{_K}=-4\pi{:}\varpi_{*}(s){:}_{_K}
$$
for $\lim_{\tau\to\pm\infty}e_*^{(\tau{+}it)\frac{1}{i\h}u{\ctt}v}=0$.}

\bigskip
On the other hand, as 
$\lim_{\tau\to
  -\infty}e_*^{(\tau{+}it)\frac{1}{i\h}u{*}v}=\varpi_{00}$,
$\lim_{\tau\to
  \infty}e_*^{(\tau{+}it)\frac{1}{i\h}u{*}v}=0$, we have

\bigskip
\noindent
\unitlength 0.1in
\begin{picture}( 28.0000,  8.1000)(2.3000,-15.7000)
%
\special{pn 8}%
\special{pa 430 370}%
\special{pa 2830 370}%
\special{fp}%
\special{pa 2830 370}%
\special{pa 2830 1170}%
\special{fp}%
\special{pa 2830 1170}%
\special{pa 430 1170}%
\special{fp}%
\special{pa 430 1170}%
\special{pa 430 370}%
\special{fp}%
%
\special{pn 8}%
\special{pa 1200 880}%
\special{pa 2000 680}%
\special{fp}%
%
\special{pn 20}%
\special{sh 1}%
\special{ar 1980 680 10 10 0  6.28318530717959E+0000}%
\special{sh 1}%
\special{ar 1220 880 10 10 0  6.28318530717959E+0000}%
\special{sh 1}%
\special{ar 1220 880 10 10 0  6.28318530717959E+0000}%
%
\special{pn 8}%
\special{pa 2830 940}%
\special{pa 2830 740}%
\special{fp}%
\special{sh 1}%
\special{pa 2830 740}%
\special{pa 2810 808}%
\special{pa 2830 794}%
\special{pa 2850 808}%
\special{pa 2830 740}%
\special{fp}%
\special{pa 1860 370}%
\special{pa 1460 370}%
\special{fp}%
\special{sh 1}%
\special{pa 1460 370}%
\special{pa 1528 390}%
\special{pa 1514 370}%
\special{pa 1528 350}%
\special{pa 1460 370}%
\special{fp}%
\special{pa 430 570}%
\special{pa 430 770}%
\special{fp}%
\special{sh 1}%
\special{pa 430 770}%
\special{pa 450 704}%
\special{pa 430 718}%
\special{pa 410 704}%
\special{pa 430 770}%
\special{fp}%
\special{pa 1460 1170}%
\special{pa 1860 1170}%
\special{fp}%
\special{sh 1}%
\special{pa 1860 1170}%
\special{pa 1794 1150}%
\special{pa 1808 1170}%
\special{pa 1794 1190}%
\special{pa 1860 1170}%
\special{fp}%
\put(25.5000,-5.3000){\makebox(0,0)[lb]{$C$}}%
\put(12.1000,-12.6000){\makebox(0,0)[lb]{$a$}}%
\put(19.8000,-12.8000){\makebox(0,0)[lb]{$b$}}%
%
\special{pn 8}%
\special{pa 590 550}%
\special{pa 230 550}%
\special{fp}%
\special{sh 1}%
\special{pa 230 550}%
\special{pa 298 570}%
\special{pa 284 550}%
\special{pa 298 530}%
\special{pa 230 550}%
\special{fp}%
\special{pa 2630 950}%
\special{pa 3030 950}%
\special{fp}%
\special{sh 1}%
\special{pa 3030 950}%
\special{pa 2964 930}%
\special{pa 2978 950}%
\special{pa 2964 970}%
\special{pa 3030 950}%
\special{fp}%
\put(4.7000,-6.8000){\makebox(0,0)[lb]{$\varpi_{00}$ or $0$}}%
\put(13.7000,-8.0000){\makebox(0,0)[lb]{\footnotesize{slit}}}
\put(15.7000,-3.5000){\makebox(0,0)[lb]{$s{+}2\pi i$}}
\put(26.9000,-11.4000){\makebox(0,0)[lb]{$0$ or $\overline{\varpi}_{00}$}}%
\end{picture}%
\hfill
\parbox[b]{.5\linewidth}{  
$$
{:}\int_{C}e_*^{z\frac{1}{i\h}u{*}v}dz{:}_{_K}={-}2\pi{:}\varpi_{00}{:}_{_K}.
$$  
Similarly, we have 
$$
{:}\int_{C}e_*^{z\frac{1}{i\h}v{*}u}dz{:}_{_K}=2\pi{:}\overline{\varpi}_{00}{:}_{_K}.
$$  
}

The next one is given by $2\pi i$-alternating periodicity and by Cauchy's integral theorem:

\bigskip
\noindent
\unitlength 0.1in
\begin{picture}( 25.9000,  6.0600)(  2.0000, -7.8600)
%
\special{pn 8}%
\special{pa 200 188}%
\special{pa 2790 188}%
\special{dt 0.045}%
%
\special{pn 8}%
\special{pa 2790 786}%
\special{pa 200 786}%
\special{fp}%
%
\special{pn 8}%
\special{pa 200 610}%
\special{pa 1196 610}%
\special{fp}%
\special{pa 1794 410}%
\special{pa 2790 410}%
\special{fp}%
%
\special{pn 8}%
\special{pa 788 786}%
\special{pa 788 610}%
\special{fp}%
\special{pa 2084 786}%
\special{pa 2084 410}%
\special{fp}%
\special{pa 2084 410}%
\special{pa 2084 410}%
\special{fp}%
%
\special{pn 8}%
\special{pa 790 190}%
\special{pa 790 610}%
\special{dt 0.045}%
\special{pa 2080 410}%
\special{pa 2080 190}%
\special{dt 0.045}%
%
\special{pn 20}%
\special{sh 1}%
\special{ar 1790 410 10 10 0  6.28318530717959E+0000}%
\special{sh 1}%
\special{ar 1180 610 10 10 0  6.28318530717959E+0000}%
\special{sh 1}%
\special{ar 1180 610 10 10 0  6.28318530717959E+0000}%
\put(11.8000,-8.5000){\makebox(0,0)[lb]{$a$}}%
\put(17.0000,-8.9000){\makebox(0,0)[lb]{$b$}}%
%
\special{pn 8}%
\special{pa 2030 370}%
\special{pa 2032 338}%
\special{pa 2032 306}%
\special{pa 2018 276}%
\special{pa 1994 256}%
\special{pa 1964 250}%
\special{pa 1930 254}%
\special{pa 1900 260}%
\special{sp}%
%
\special{pn 8}%
\special{pa 1920 250}%
\special{pa 1720 250}%
\special{fp}%
\special{sh 1}%
\special{pa 1720 250}%
\special{pa 1788 270}%
\special{pa 1774 250}%
\special{pa 1788 230}%
\special{pa 1720 250}%
\special{fp}%
%
\special{pn 8}%
\special{pa 2020 650}%
\special{pa 2022 682}%
\special{pa 2022 714}%
\special{pa 2008 744}%
\special{pa 1984 764}%
\special{pa 1954 770}%
\special{pa 1920 768}%
\special{pa 1890 760}%
\special{sp}%
%
\special{pn 8}%
\special{pa 830 340}%
\special{pa 828 308}%
\special{pa 830 276}%
\special{pa 842 246}%
\special{pa 868 226}%
\special{pa 898 220}%
\special{pa 930 224}%
\special{pa 960 230}%
\special{sp}%
%
\special{pn 8}%
\special{pa 840 630}%
\special{pa 838 662}%
\special{pa 840 694}%
\special{pa 852 724}%
\special{pa 878 744}%
\special{pa 908 750}%
\special{pa 940 748}%
\special{pa 970 740}%
\special{sp}%
%
\special{pn 8}%
\special{pa 830 330}%
\special{pa 830 480}%
\special{fp}%
\special{sh 1}%
\special{pa 830 480}%
\special{pa 850 414}%
\special{pa 830 428}%
\special{pa 810 414}%
\special{pa 830 480}%
\special{fp}%
%
\special{pn 8}%
\special{pa 950 750}%
\special{pa 1160 750}%
\special{fp}%
\special{sh 1}%
\special{pa 1160 750}%
\special{pa 1094 730}%
\special{pa 1108 750}%
\special{pa 1094 770}%
\special{pa 1160 750}%
\special{fp}%
\special{pa 1160 750}%
\special{pa 1160 750}%
\special{fp}%
%
\special{pn 8}%
\special{pa 2020 690}%
\special{pa 2020 490}%
\special{fp}%
\special{sh 1}%
\special{pa 2020 490}%
\special{pa 2000 558}%
\special{pa 2020 544}%
\special{pa 2040 558}%
\special{pa 2020 490}%
\special{fp}%
\put(9.9000,-3.5000){\makebox(0,0)[lb]{$C$}}%
\put(21.9000,-3.5000){\makebox(0,0)[lb]{\footnotesize{slit}}}%
\end{picture}%
\hfill
\parbox[b]{.7\linewidth}{
$$
{:}\int_{C}e_*^{z\frac{1}{i\h}u{\ctt}v}dz{:}_{_K}
=2\int_{-\infty}^{\infty}e_*^{s\frac{1}{i\h}u{\ctt}v}ds{:}_{_K}
{=}2{:}\delta_*(\frac{1}{\h}u{\ctt}v){:}_{_K}.
$$}

\subsubsection{Residues and secondary residues}\label{Defressevval}
Let $\hat{a}{+}i\sigma_{\hat a}$, $\hat{b}{+}i\sigma_{\hat b}$ be the
singular points such that 
$0<\sigma_{\hat a}{<}2\pi$, $0<\sigma_{\hat b}{<}2\pi$. 

Now, let $\sigma$ be the representative of the isolated  
singular points $s_a{+}2\pi i\ell$ or $s_b{+}2\pi i\ell$.  
We  make a double covering space 
(the Riemann surface) of a neighborhood $D$ of $\sigma$ to
make a single-valued holomorphic function, and then we take 
its Laurent expansion. 

Recall generic $K$-ordered expression of $*$-exponential 
function $e_*^{z(\frac{1}{i\h}u{\ctt}v{+}\alpha)}$ is written 
in the form 
$$
{:}e_*^{z(\frac{1}{i\h}u{\ctt}v{+}\alpha)}{:}_{_K}
=\frac{e^{\alpha z}}{\sqrt{g(z)}}e^{\frac{1}{i\h}H(z,u,v)}.
$$  

By the first comment in \S\,\ref{GenRiem}, 
 $\sigma$ $g(z)$, $H(z,u,v)$ are given on a neighborhood of an isolated singular point 
 by using a holomorphic functions $h(z),\, a(z,u,v),\, b(z,u,v)$ as 
$$
\begin{aligned}
&g(z)=(z{-}\sigma)h(z),\quad h(\sigma)\not=0,\\ 
&H(z,u,v)=\frac{a(z,u,v)}{z{-}\sigma}{+}b(z,u,v), 
\quad a(\sigma,u,v)\not=0,\quad H(z,0,0)=0.
\end{aligned}
$$

Hence, setting $z=\sigma{+}s^2$, the Laurent series of
$\frac{1}{\sqrt{g(z)}}$ at $s=0$ is given by 
$$
\frac{1}{\sqrt{g(z)}}=
\frac{1}{s}(h_0{+}h_1s^2{+}h_2s^4{+}\cdots),\,\, h_0\not=0,
$$
without terms of even degree. If $h_k=0$ for all $k\geq 1$, then 
there is only one singular point $s=0$. Hence, if $g(z)$ has many
zeros, then  $\frac{1}{\sqrt{g(z)}}$ has terms of positive degree. 
The Laurent series of 
${:}e_*^{z\frac{1}{i\h}u{\ctt}v}{:}_{_K}$ at the singular point 
$\sigma$ is written as  
\begin{equation}\label{onlyeven}
\begin{aligned}
&{:}e_*^{(\sigma{+}s^2)(\frac{1}{i\h}u{\ctt}v{+}\alpha)}{:}_{_K}=\\ 
&\cdots{+}\frac{a_{-(2k+1)}(\sigma,K)}{s^{2k+1}}{+}\cdots{+}
\frac{a_{-1}(\sigma,K)}{s}{+}a_1(\sigma,K)s
{+}\cdots {+}a_{2k+1}(\sigma,K)s^{2k+1}{+}\cdots
\end{aligned}
\end{equation}
without terms of even degree, where 
\begin{equation}\label{residuform}
\begin{aligned}
&\frac{1}{2\pi i}\int_{\tilde{C}}\!s^{{-2k}}
{:}e_*^{(\sigma{+}s^2)(\frac{1}{i\h}u{\ctt}v{+}\alpha)}{:}_{_K}ds{=}
{\rm{Res}}_{s=0}(
{:}s^{{-2k}}e_*^{(\sigma{+}s^2)(\frac{1}{i\h}u{\ctt}v{+}\alpha)}{:}_{_K})
{=}a_{2k{-}1}(\sigma,K)\\
&\frac{1}{2\pi i}\int_{\tilde{C}}\!s^{{-2k-1}}
{:}e_*^{(\sigma{+}s^2)(\frac{1}{i\h}u{\ctt}v{+}\alpha)}{:}_{_K}ds{=}
{\rm{Res}}_{s=0}(
{:}s^{{-2k{-}1}}e_*^{(\sigma{+}s^2)(\frac{1}{i\h}u{\ctt}v{+}\alpha)}{:}_{_K}){=}0.
\end{aligned}
\end{equation}
by setting $\tilde{C}$ a small circle with the center at $0$. As it is
mentioned above, there is nontrivial positive terms of even degree, as 
${:}e_*^{z\frac{1}{i\h}u{\ctt}v}{:}_{_K}$ has infinitely many singular points.

\bigskip 
The next one the fundamental property of residues 
given by the integration by parts: 

\begin{prop}\label{propresidue}
In a generic ordered expression $K$ which is fixed, the Laurent coefficients
$a_{(2k{-}1)}(\sigma,K)$ of  
${:}e_*^{(\sigma{+}s^2)(\alpha{+}\frac{1}{i\h}u{\ctt}v)}{:}_{_K}$
satisfy  
\begin{equation}\label{niceformula111}
\begin{aligned}
2\pi i{:}&(\alpha{+}\frac{1}{i\h}u{\ctt}v){:}_{_K}
{*_{_K}}a_{2k{-}1}(\sigma,K){=}
\int_{\tilde{C}}\!\frac{1}{2}s^{-2k-1}
{:}\frac{d}{ds}
e_*^{(\sigma{+}s^2)(\alpha{+}\frac{1}{i\h}u{\ctt}v)}{:}_{_K}ds\\
&{=}
-\frac{2k{+}1}{2}
\int_{\tilde{C}}\!s^{-2(k{+}1)}
{:}e_*^{(\sigma{+}s^2)\frac{1}{i\h}u{\ctt}v}{:}_{_K}ds
 =-2\pi i\frac{2k{+}1}{2}a_{2k+1}(\sigma,K).
\end{aligned}
\end{equation}
\end{prop}

\bigskip
\noindent
{\bf Calculation on $\pm$-sheets}

Although we use only the case $p=2$, it is useful 
to fix a method to compute the residue at a $p$-branching 
($\sqrt[p]{z}$-) singular point by using $p$-sheets  
together with $\omega=e^{\frac{2\pi i}{p}}$. 

To treat $\frac{1}{\sqrt[p]{z}}$ as a single valued 
function, we take $p$-covering $z=s^p$ and regard   
$\frac{1}{\sqrt[p]{z}}$ as 
$$
\frac{1}{\sqrt[p]{z}}=\{\frac{1}{s},\,\, \frac{1}{\omega s},
\,\,\cdots,\,\, 
\frac{1}{\omega^{p-1}s}\}
$$ 
preparing $p$-sheets, and 
$$
\frac{1}{\sqrt[p]{z}}d\sqrt[p]{z}=
\{\frac{1}{s}ds,\,\, \frac{1}{\omega s}d(\omega s),
\,\,\cdots,\,\, 
\frac{1}{\omega^{p-1}s}d(\omega^{p-1}s)\}.
$$
Hence computing the contour $\int_{\tilde{C}}\frac{1}{s}ds$ 
is nothing but the calculation 
$\int_C\frac{1}{\sqrt[p]{z}}d\sqrt[p]{z}$ on each  

\bigskip
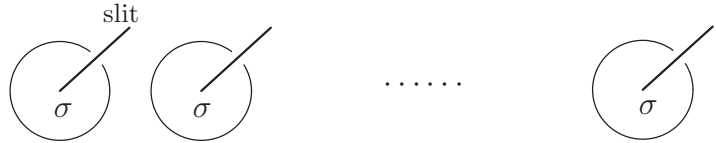
\begin{center}
\unitlength 0.1in
\begin{picture}(36.3700, 13.6800)(  5.2400,-15.9000)
%
\special{pn 8}%
\special{ar 782 552 258 258  5.7291696 6.2831853}%
\special{ar 782 552 258 258  0.0000000 5.3610095}%
%
\special{pn 13}%
\special{pa 782 552}%
\special{pa 1142 222}%
\special{fp}%
\put(7.5000,-6.9000){\makebox(0,0)[lb]{$\sigma$}}%
%
\special{pn 8}%
\special{ar 1514 552 260 258  5.7283931 6.2831853}%
\special{ar 1514 552 260 258  0.0000000 5.3603103}%
%
\special{pn 13}%
\special{pa 1514 552}%
\special{pa 1874 222}%
\special{fp}%
\put(14.7000,-6.9000){\makebox(0,0)[lb]{$\sigma$}}%
%
\special{pn 8}%
\special{ar 3800 546 260 258  5.7263604 6.2831853}%
\special{ar 3800 546 260 258  0.0000000 5.3582638}%
%
\special{pn 13}%
\special{pa 3800 546}%
\special{pa 4162 216}%
\special{fp}%
\put(37.8000,-6.7000){\makebox(0,0)[lb]{$\sigma$}}%
\put(10.0600,-1.9200){\makebox(0,0)[lb]{\footnotesize{slit}}}%
%
\put(12.7000,-15.9000){\makebox(0,0)[lb]{}}%
\put(24.5400,-5.5800){\makebox(0,0)[lb]{$\cdots\cdots$}}%
\end{picture}%
\end{center}

\vspace{-1.5cm}

\noindent
sheet where the existence of slit keeps the integrand 
single value. But as a result, the 
integral equals formally $p\int_C\frac{1}{pz}dz=\int_C\frac{1}{z}dz$. 

In the case of $e_*^{z\frac{1}{i\h}u{\ctt}v}$, 
we have only to use  two sheets,
and we can apply this method to treat several 
singular points at the same time. 
The next one gives the periodical properties of residues of 
${*}$-exponential functions: Let 
${\rm{Res}}(\sigma{+}2\pi ik,\alpha,K)$ be the residue of 
$e_*^{z(\frac{1}{i\h}u{\ctt}v{+}\alpha)}$ at $z=\sigma{+}2\pi ik$.
\begin{lem}\label{residuelem}
If $\alpha$ is an integer or a half-integer, then 
in generic ordered expression, the residue has the 
alternating $2\pi i$-periodicity  
$$
{\rm{Res}}(\sigma{+}2\pi ik,\alpha,K)
{=}(-1)^k{\rm{Res}}(\sigma,\alpha,K).
$$ 
\end{lem}
\noindent
{\bf Proof}\,\,is given by showing 
${\rm{Res}}(\sigma,\alpha,K){+}
{\rm{Res}}(\sigma{+}2\pi i,\alpha,K)
=\frac{1}{2}
({\rm{Res}}(\sigma,\alpha,K){+}
{\rm{Res}}(\sigma{+}2\pi i,\alpha,K))$.

\medskip
\unitlength 0.1in
\begin{picture}( 12.1900, 25.6000)(4.4100,-30.1000)
%
\special{pn 8}%
\special{ar 1758 1160 948 902  3.1415927 4.6098863}%
%
\special{pn 8}%
\special{pa 808 150}%
\special{pa 808 2710}%
\special{dt 0.045}%
%
\special{pn 8}%
\special{ar 1758 2298 948 902  3.1415927 4.6098863}%
%
\special{pn 8}%
\special{ar 808 2312 322 306  5.0093728 6.2831853}%
\special{ar 808 2312 322 306  0.0000000 4.7769543}%
%
\special{pn 8}%
\special{ar 808 2320 366 348  5.0175665 6.2831853}%
\special{ar 808 2320 366 348  0.0000000 4.7731660}%
%
\special{pn 8}%
\special{ar 822 1110 322 306  5.0093728 6.2831853}%
\special{ar 822 1110 322 306  0.0000000 4.7769543}%
%
\special{pn 8}%
\special{ar 822 1116 366 348  5.0175665 6.2831853}%
\special{ar 822 1116 366 348  0.0000000 4.7731660}%
\put(11.0000,-4.2000){\makebox(0,0)[lb]{\footnotesize{slit}}}%
\put(11.3000,-9.2000){\makebox(0,0)[lb]{$2C$}}%
\put(4.7000,-16.9000){\makebox(0,0)[lb]{l.h.s.}}%
\put(8.9000,-11.7000){\makebox(0,0)[lb]
{$\sigma{+}2\pi i$}}%
\put(8.5000,-23.7000){\makebox(0,0)[lb]{$\sigma$}}%
\put(11.1000,-25.7000){\makebox(0,0)[lb]{$2{\hat{C}}$}}%
\end{picture}%
\hfill
\parbox[b]{.7\linewidth}{The l.h.s.(resp.r.h.s.) 
of the vertical dotted line is the region that 
$e_*^{z\frac{1}{i\h}u{\ctt}v}$ is  $2\pi i$-periodic 
(resp. alternating $2\pi i$-periodic). Double circles 
are the paths of integrals 
$$
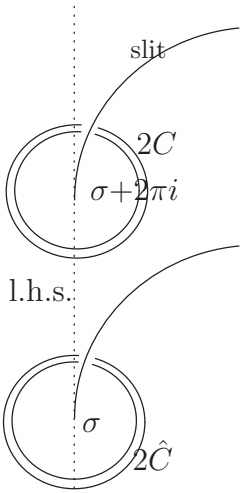
\int_{C^2}e_*^{z(\alpha{+}\frac{1}{i\h}u{\ctt}v)}
d{\sqrt{z{-}\sigma}},\quad 
\int_{\hat{C}^2}
e_*^{z(\alpha{+}\frac{1}{i\h}u{\ctt}v)}
d{\sqrt{z{-}\sigma{-}2\pi i}}
$$
in each sheet. But in fact these are 
\begin{equation}\label{resintegral}
2\int_{C}e_*^{z(\alpha{+}\frac{1}{i\h}u{\ctt}v)}
d{\sqrt{z{-}\sigma}},\quad 
2\int_{\hat{C}}
e_*^{z(\alpha{+}\frac{1}{i\h}u{\ctt}v)}
d{\sqrt{z{-}\sigma{-}2\pi i}}.
\end{equation}
Sum up all integrals to obtain 
$$
2\pi i{\rm{Res}}(\sigma,\alpha,K){+}
2\pi i{\rm{Res}}(\sigma{+}2\pi i,\alpha,K),
$$
but the alternating $2\pi i$-periodicity of 
of the integrand on the r.h.s. or on the 
l.h.s. erases the half of the quantity.  
\hfill$\Box$}

\medskip 
\noindent
Remark here by minding about slits that the 
integrals \eqref{resintegral} may be replaced by 
$$
\int_{C}e_*^{z(\alpha{+}\frac{1}{i\h}u{\ctt}v)}
\frac{dz}{\sqrt{z{-}\sigma}},
\quad 
\int_{\hat{C}}
e_*^{z(\alpha{+}\frac{1}{i\h}u{\ctt}v)}
\frac{dz}{\sqrt{z{-}\sigma{-}2\pi i}}.
$$

\bigskip
In this way of expressing of integrals, we can consider a wider 
region beyond neighborhoods of singular points. 
Fix a singular point $\sigma$ and consider a contour integral 
\begin{equation}\label{resintres}
\int_{C^2}e_*^{z(\alpha{+}\frac{1}{i\h}u{\ctt}v)}
d{\sqrt{z{-}\sigma}}= 
\int_{C^2}e_*^{z(\alpha{+}\frac{1}{i\h}u{\ctt}v)}
\frac{dz}{2\sqrt{z{-}\sigma}}
\end{equation}
where $C$ is a closed path in $\mathbb C$, which may include 
other singular points  inside, and $C^2$ is the same path 
rounding twice on the same path so that 
the integrand to be closed.

Next Theorem together with Cauchy integral theorem shows 
that the integral \eqref{resintres} depends only to  
the prescribed singular point $\sigma$: 
\begin{thm}\label{lemmaresidue}
If a singular point $\sigma$ is fixed, then the integral 
\begin{equation}\label{intequ}
\frac{1}{2\pi i}
\int_{{\tilde C}^2}e_*^{z(\alpha{+}\frac{1}{i\h}u{\ctt}v)}
d{\sqrt{z{-}\sigma}}
\end{equation}  
over a small circle ${\tilde C}$ centered at an arbitrary point is given by 
$$
\frac{1}{2\pi i}
\int_{{\tilde C}^2}e_*^{z(\alpha{+}\frac{1}{i\h}u{\ctt}v)}
d{\sqrt{z{-}\sigma}}{=}
\left\{
\begin{matrix}
{\rm{Res}}(\sigma,\alpha,K)& \text{if $\sigma$ is the inside of $\tilde C$}\\
\frac{1}{2}{\rm{Res}}(\sigma,\alpha,K)& \text{if $\sigma\in\tilde C$}\\
0 & \text{if $\sigma$ is the outside of  $\tilde C$}
\end{matrix}
\right.
$$  

On the other hand, if $\sigma$ is not a singular point, then the
integral \eqref{intequ} vanishes everywhere.
\end{thm} 

\noindent
{\bf Proof}\,\,The integral vanishes when there is no singular point inside.
Let $\sigma'$ be another singular point inside $\tilde C$. 
In a neighborhood of $\sigma'$, the Laurent series of 
${:}e_*^{z(\alpha{+}\frac{1}{i\h}u{\ctt}v)}{:}_{_K}$ 
is given in the form 
$$
\cdots{+}\frac{a_{-3}}{\sqrt{z-\sigma'}^3}{+}
\frac{a_{-1}}{\sqrt{z-\sigma'}}{+}a_1\sqrt{z-\sigma'}{+}\cdots
$$
without terms of even degree w.r.t. $\sqrt{z-\sigma'}$. 
By setting $s^2=z-\sigma'$, we have
$\sqrt{z{-}\sigma}=\sqrt{s^2{+}\sigma'{-}\sigma}$ and the integral  
we have to consider is  
$$
\cdots{+}\int_{\tilde C}\frac{a_{-3}}{s^3}
\frac{sds}{\sqrt{s^2{+}\sigma'{-}\sigma}}
{+}\int_{\tilde C}\frac{a_{-1}}{s}
\frac{sds}{\sqrt{s^2{+}\sigma'{-}\sigma}}
{+}\int_{\tilde C}a_1s
\frac{sds}{\sqrt{s^2{+}\sigma'{-}\sigma}}{+}\cdots=0
$$
for the Taylor series of 
$\frac{1}{\sqrt{s^2{+}\sigma'{-}\sigma}}$ has
no term of odd degree. 
The second statement is proved by applying the argument above to the 
case that $\sigma'$ is a regular point. \hfill $\Box$ 

It is remarkable that this theorem makes us possible to concern  
with individual/subjective singular point depending 
on the individual expression parameters.

\subsubsection{Residues and Laurent coefficients}
Recall the formula \eqref{genericparam00}
\begin{equation*}
{:}e_*^{t\frac{1}{i\h}2u{\ctt}v}{:}_{_{K}}{=}
\frac{2}{\sqrt{\Delta^2{-}(e^t{-}e^{-t})^2\delta\delta'}}
\,\,e^{\frac{1}{i\h}
\frac{e^t-e^{-t}}{\Delta^2{-}(e^t{-}e^{-t})^2\delta\delta'}
\big((e^t-e^{-t})(\delta u^2{+}\delta' v^2){+}2\Delta uv\big)},
\end{equation*}
where $\Delta{=}e^t{+}e^{-t}{-}c(e^t{-}e^{-t})$. Using this, we
compute the Laurent coefficients $a_{2k{-}1}(\sigma,K)$, where 
$K{=}
\footnotesize
{\begin{bmatrix}
\delta&c\\
c&\delta'
\end{bmatrix}}
$ 
and $\sigma$ is a singular point. 

\bigskip
Now, we set 
$$
s^2=e^t{+}e^{-t}{-}(c{+}\sqrt{\delta\delta'})(e^t{-}e^{-t}),\quad 
{\hat s}^2=e^t{+}e^{-t}{-}(c{-}\sqrt{\delta\delta'})(e^t{-}e^{-t})
$$
so that the singular points are obtained by $s^2=0$ or ${\hat s}^2=0$.
 Since $\delta\delta'\not=0$ 
is assumed in general, singular points satisfy 
$s^2=0$ or ${\hat s}^2=0$ exclusively.  
One may set 
${\hat s}^2={\hat s}^2(s)$, ${\hat s}^2(0)\not=0$, and similarly  
$s^2=s^2({\hat s}), s^2(0)\not=0$.

Hence, 
$$
{\hat s}^2{-}s^2=2\sqrt{\delta\delta'}(e^t{-}e^{-t}),\quad 
{\hat s}^2{+}s^2=2(e^{t}{+}e^{-t}){-}c\frac{{\hat s}^2{-}s^2}{\sqrt{\delta\delta'}}.
$$
It follows that $2\Delta={\hat s}^2{+}s^2$. Plugging these we have 
\begin{equation}\label{formula0101}
{:}e_*^{t\frac{1}{i\h}2u{\ctt}v}{:}_{_{K}}{=}
\frac{2}{s{\hat s}}{\exp}\Big(\frac{1}{4i\h}(\frac{{\hat s}^2}{s^2}{-}\frac{s^2}{{\hat s}^2})
(\frac{u}{\sqrt{\delta'}}{+}\frac{v}{\sqrt{\delta}})^2\Big)
{\exp}(-\frac{1}{2i\h}((\frac{u}{\sqrt{\delta'}})^2{+}(\frac{v}{\sqrt{\delta}})^2)).
\end{equation}
Note that the variable $s$ (resp. ${\hat s}$) may be viewed as the local coordinate
around the singular point given by $s=0$ (resp. ${\hat s}=0$). 
It follows that 
$$
{\rm{Res}}_{s=0}{:}e_*^{(\sigma{+}s^2)\frac{1}{i\h}2u{\ctt}v}{:}_{_{K}}{=}
\frac{1}{2\pi i}
e^{-\frac{1}{2i\h}((\frac{u}{\sqrt{\delta'}})^2{+}(\frac{v}{\sqrt{\delta}})^2)}
\frac{2}{{\hat s}(0)}
\sum_{k=0}^{\infty}
\frac{(-1)^k}{k!k!}\big(\frac{1}{4i\h}(\frac{u}{\sqrt{\delta'}}{+}\frac{v}{\sqrt{\delta}})\big)^{2k}.
$$
${\rm{Res}}_{{\hat s}=0}{:}e_*^{(\sigma{+}{\hat s}^2)\frac{1}{i\h}2u{\ctt}v}{:}_{_{K}}$ is 
obtained by replacing ${\hat s}(0)$ by ${s}(0)$. 
Explicit formula of Laurent coefficients is not easy to write down, but it
is clear that there is no term of even degree.

\bigskip
\noindent
{\bf Secondary residues do not appear}\,\,\,
If there is a nontrivial term $\frac{1}{s^2}$ in the 
Laurent series of $f_*(\sigma{+}s^2, u,v)$ at a singular point 
$\sigma$, then the residue-like integral gives $2\pi ia_{-2}$. 
\begin{equation}\label{secndary}
a_{-2}(\sigma,K)=\frac{1}{4\pi i}\int_{C^2}{:}f_*(\sigma{+}s^2,u,v){:}_{_K}ds^2 
\end{equation}
will be called the {\bf secondary residue} at $\sigma$, 
where $C^2$ is a small circle 
with center at $\sigma$ round twice. We  
denote this by ${\rm{Res}_2}(\sigma,\alpha, K)$.
The secondary residues are much easier to 
calculate as we have only to use double loops, 
but the secondary residue does not appear in our  
case  $e_*^{z(\alpha{+}\frac{1}{i\h}u{*}v)}$. 
These observation may be summarized as follows:
\begin{thm}\label{2ndresidue}
If a singular point $\sigma$ is fixed, then the integral 
$\frac{1}{2\pi i}
\int_{C^2}e_*^{z(\alpha{+}\frac{1}{i\h}u{\ctt}v)}dz$
on a small circle centered at $\sigma$ vanishes always.   

Another word, the secondary residue vanishes identically at any 
isolated $\sqrt{{\,\, \,}}$ branching singular point.
\end{thm}

To define the such an integral at a singular point 
$\sigma$, one has to use $d(s^2)$ instead of $ds$ 
in the definition \eqref{residuform} of $a_k(K)$, that is, 
$$
\int_{\tilde{C}}\!s^{k{-}1}
{:}e_*^{(\sigma{+}s^2)\frac{1}{i\h}u{\ctt}v}{:}_{_K}d(s^2)
=
2\int_{\tilde{C}}\!s^{k}
{:}e_*^{(\sigma{+}s^2)\frac{1}{i\h}u{\ctt}v}{:}_{_K}ds
=2a_{-(k+1)},
$$
but the integral may be simply written as 
$\int_{C^2}{:}
\sqrt{z{-}\sigma}^{k-1}e_*^{z\frac{1}{i\h}u{\ctt}v}{:}_{_K}dz{:}$. 
For $k=1$, this integral is much easier to treat. But at 
the singular point $\sigma$, this must give $4\pi ia_{-2}(K)$  
and \eqref{onlyeven} shows there is no term of even degree. 

\medskip
\begin{prop}
The integral 
$\int_C{:}e_*^{t(\frac{1}{i\h}u{\ctt}v{+}\alpha)}{:}_{_K}dt$ 
is obtained as the difference of two different inverses of 
$\frac{1}{i\h}u{\ctt}v{+}\alpha$.
%
%
%
In contrast, the residue ${\text{\rm{Res}}}(\sigma,\alpha)$ can 
not be given by such a way.
\end{prop}
\bigskip
\noindent
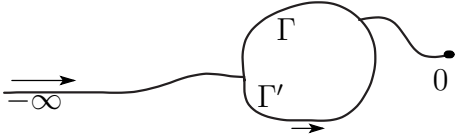
\unitlength 0.1in
\begin{picture}( 23.1900,  6.6400)(  1.2000, -12.00)
%
\special{pn 13}%
\special{pa 120 704}%
\special{pa 152 704}%
\special{pa 184 702}%
\special{pa 216 702}%
\special{pa 248 702}%
\special{pa 280 702}%
\special{pa 312 702}%
\special{pa 344 702}%
\special{pa 376 702}%
\special{pa 408 702}%
\special{pa 440 702}%
\special{pa 472 702}%
\special{pa 504 702}%
\special{pa 536 702}%
\special{pa 568 702}%
\special{pa 600 702}%
\special{pa 632 702}%
\special{pa 664 704}%
\special{pa 696 704}%
\special{pa 728 704}%
\special{pa 760 702}%
\special{pa 792 700}%
\special{pa 824 694}%
\special{pa 856 688}%
\special{pa 886 680}%
\special{pa 918 672}%
\special{pa 948 662}%
\special{pa 980 652}%
\special{pa 1010 640}%
\special{pa 1040 630}%
\special{pa 1072 620}%
\special{pa 1102 614}%
\special{pa 1134 608}%
\special{pa 1166 604}%
\special{pa 1196 606}%
\special{pa 1228 608}%
\special{pa 1260 612}%
\special{pa 1292 618}%
\special{pa 1324 620}%
\special{pa 1356 622}%
\special{pa 1364 622}%
\special{sp}%
%
\special{pn 13}%
\special{pa 1364 632}%
\special{pa 1364 666}%
\special{pa 1364 700}%
\special{pa 1370 730}%
\special{pa 1380 758}%
\special{pa 1400 782}%
\special{pa 1424 802}%
\special{pa 1454 818}%
\special{pa 1486 830}%
\special{pa 1520 838}%
\special{pa 1552 844}%
\special{pa 1584 848}%
\special{pa 1616 848}%
\special{pa 1648 848}%
\special{pa 1680 848}%
\special{pa 1712 848}%
\special{pa 1744 850}%
\special{pa 1776 850}%
\special{pa 1808 848}%
\special{pa 1840 842}%
\special{pa 1870 832}%
\special{pa 1898 816}%
\special{pa 1924 796}%
\special{pa 1948 774}%
\special{pa 1970 746}%
\special{pa 1988 716}%
\special{pa 2006 684}%
\special{pa 2020 652}%
\special{pa 2030 616}%
\special{pa 2038 580}%
\special{pa 2042 544}%
\special{pa 2042 508}%
\special{pa 2040 472}%
\special{pa 2032 438}%
\special{pa 2020 406}%
\special{pa 2006 376}%
\special{pa 1986 348}%
\special{pa 1964 322}%
\special{pa 1940 300}%
\special{pa 1914 280}%
\special{pa 1886 264}%
\special{pa 1856 250}%
\special{pa 1826 240}%
\special{pa 1794 234}%
\special{pa 1764 232}%
\special{pa 1732 232}%
\special{pa 1700 234}%
\special{pa 1666 240}%
\special{pa 1634 248}%
\special{pa 1600 258}%
\special{pa 1568 270}%
\special{pa 1536 282}%
\special{pa 1504 298}%
\special{pa 1474 314}%
\special{pa 1448 334}%
\special{pa 1424 354}%
\special{pa 1402 376}%
\special{pa 1384 402}%
\special{pa 1370 428}%
\special{pa 1362 456}%
\special{pa 1356 488}%
\special{pa 1354 518}%
\special{pa 1356 552}%
\special{pa 1358 586}%
\special{pa 1362 620}%
\special{pa 1364 640}%
\special{sp}%
%
\special{pn 13}%
\special{pa 1364 640}%
\special{pa 1364 640}%
\special{pa 1364 640}%
\special{pa 1364 640}%
\special{pa 1364 640}%
\special{pa 1364 640}%
\special{pa 1364 640}%
\special{pa 1364 640}%
\special{pa 1364 640}%
\special{sp}%
%
\special{pn 13}%
\special{pa 1958 324}%
\special{pa 1994 316}%
\special{pa 2028 312}%
\special{pa 2060 310}%
\special{pa 2090 314}%
\special{pa 2118 324}%
\special{pa 2142 340}%
\special{pa 2164 364}%
\special{pa 2184 392}%
\special{pa 2204 420}%
\special{pa 2222 448}%
\special{pa 2242 474}%
\special{pa 2266 496}%
\special{pa 2290 510}%
\special{pa 2320 516}%
\special{pa 2352 518}%
\special{pa 2384 514}%
\special{pa 2420 508}%
\special{pa 2440 504}%
\special{sp}%
%
\special{pn 20}%
\special{sh 1}%
\special{ar 2432 504 10 10 0  6.28318530717959E+0000}%
\special{sh 1}%
\special{ar 2424 504 10 10 0  6.28318530717959E+0000}%
\special{sh 1}%
\special{ar 2416 504 10 10 0  6.28318530717959E+0000}%
\put(15.3000,-4.3000){\makebox(0,0)[lb]{$\Gamma$}}%
%
\special{pn 8}%
\special{pa 160 640}%
\special{pa 482 640}%
\special{fp}%
\special{sh 1}%
\special{pa 482 640}%
\special{pa 414 620}%
\special{pa 428 640}%
\special{pa 414 660}%
\special{pa 482 640}%
\special{fp}%
\put(1.2800,-8.0300){\makebox(0,0)[lb]{$-\infty$}}%
\put(23.4000,-6.9000){\makebox(0,0)[lb]{$0$}}%
%
\special{pn 8}%
\special{pa 1610 890}%
\special{pa 1770 890}%
\special{fp}%
\special{sh 1}%
\special{pa 1770 890}%
\special{pa 1704 870}%
\special{pa 1718 890}%
\special{pa 1704 910}%
\special{pa 1770 890}%
\special{fp}%
\put(14.3000,-7.8000){\makebox(0,0)[lb]{$\Gamma'$}}%
\end{picture}%
\hfill
\parbox[b]{.6\linewidth}
{{\bf Proof}\,\, Suppose ${\rm{Re}}\alpha>0$. Then the 
both integrals 
$\int_{-\infty}^0
{:}e_*^{t(\frac{1}{i\h}u{\ctt}v{+}\alpha)}{:}_{_K}dt$ along the 
path $\Gamma$ and $\Gamma'$ converge to give inverses of 
$\frac{1}{i\h}u{\ctt}v{+}\alpha$. Thee difference is 
the integral over $C$. 
If $\alpha<1$, then the integral $-\int_{0}^{\infty}
{:}e_*^{t(\frac{1}{i\h}u{\ctt}v{+}\alpha)}{:}_{_K}dt$
also gives an inverse.
Suppose a residue $a_{-1}(\alpha)={\rm{Res}}(\sigma,\alpha)$ 
is written as a difference} 
$$
a_{-1}(\alpha)=
(\frac{1}{i\h}u{\ctt}v{+}\alpha)_{*\mu}^{-1}{-}
(\frac{1}{i\h}u{\ctt}v{+}\alpha)_{*\nu}^{-1}.
$$ 
Then, we must have
$a_1(\alpha)=
(\frac{1}{i\h}u{\ctt}v{+}\alpha){*}a_{-1}(\alpha)=0$, but 
$a_1(\alpha)\not=0$ for generic $\alpha$.
${}$\hfill$\Box$

By vanishing of $a_{-2}(K)$, we have no need to care about residue-like
quantity in the computation of principal values:
\begin{prop}
Let $I_{\ctt}(K)=(a,b)$ be the exchanging interval. 
In generic ordered expressions,  
$$
{\text{\rm{vp-}}}\int_0^{2\pi}\!\!
{:}e_*^{(a{+}i\tau)\frac{1}{i\h}u{*}v}{:}_{_K}d\tau=
2\pi\varpi_{00}
$$
In particular, 
${\text{\rm{vp-}}}\int_0^{2\pi}\!\!
{:}e_*^{(a{+}i\tau)\frac{1}{i\h}u{*}v}{:}_{_K}d\tau$
satisfies the differential equation 
$(\frac{1}{i\h}u{*}v){*}f=0$.

Moreover, if $\ell$ is an integer, then the exponential law of 
the integrand shows 
$$
{\text{\rm{vp-}}}\int_0^{2\pi}\!\!
{:}e_*^{(a{+}i\tau)(\frac{1}{i\h}u{*}v{+}\ell)}{:}_{_K}d\tau=
\left\{
\begin{matrix}
0& \ell>0\\
2\pi{\varpi}_{00}& \ell=0\\
\infty& \ell<0.
\end{matrix}
\right.
$$
Similarly, taking the growth order into account, we have 
$$
{\text{\rm{vp-}}}\int_0^{2\pi}\!\!
{:}e_*^{(b{+}i\tau)(\frac{1}{i\h}u{*}v{+}\ell)}{:}_{_K}d\tau=
\left\{
\begin{matrix}
0& \ell<1\\
2\pi{\overline{\varpi}}_{00}&\ell=1\\
\infty& \ell>1.
\end{matrix}
\right.
$$
\end{prop}

Let $\sigma$ be a moving singular point which we want
to concern. Theorem\,\ref{remarkableresd} below shows that the residues of 
${:}e_*^{z(\alpha{+}\frac{1}{i\h}u{\ctt}v)}{:}_{_K}$ at $\sigma$ can be computed  
by the integral along a fixed big closed path.

\bigskip
\begin{thm}\label{remarkableresd}
Even though other singular points may lie inside $C$, the 
integral \\
$\int_{C^2}{:}e_*^{z(\alpha{+}\frac{1}{i\h}u{\ctt}v)}{:}_{_K}d{\sqrt{z{-}\sigma}}$ 
on the closed path $C^2$ gives 
$\ell{\times}{\rm{Res}}(\sigma,\alpha,K)$, where $\ell$ is the winding number of 
$C$ around $\sigma$.

On the other hand, if $\sigma$ is not a singular point, then the
integral 
$\int_{C^2}{:}e_*^{z(\alpha{+}\frac{1}{i\h}u{\ctt}v)}{:}_{_K}d{\sqrt{z{-}\sigma}}$ 
on the closed path $C^2$ vanishes. 
\end{thm}
  
\noindent
{\bf Proof}\,\,\,It is obvious by the Cauchy  
integral theorem if $C$ does not involve other singular
point inside. Suppose that $C$ is a simple closed curve. 
Let $\sigma_i$, $i=0,1,\dots,n$ be singular points 
inside $C$, and $\sigma=\sigma_0$. The proof is given by 
showing that $C^2$ is homologically equivalent to 
$C_0^2\cup C_1^2\cup\cdots\cup C_n^2$. This is trivial if there 
is no slit. But, Theorem\,\ref{lemmaresidue} and the next figure for $\sigma_i$, $i\not=0$, 
show that the same proof is valid. This proves also the second statement.  
 
\unitlength 0.1in
\begin{picture}( 52.5000, 14.8000)(  1.4000,-16.1000)
%
\special{pn 8}%
\special{pa 636 896}%
\special{pa 1612 896}%
\special{fp}%
%
\special{pn 20}%
\special{sh 1}%
\special{ar 628 896 10 10 0  6.28318530717959E+0000}%
\special{sh 1}%
\special{ar 628 896 10 10 0  6.28318530717959E+0000}%
%
\special{pn 20}%
\special{pa 1124 176}%
\special{pa 1124 760}%
\special{fp}%
\special{pa 1124 760}%
\special{pa 970 888}%
\special{fp}%
\special{pa 970 888}%
\special{pa 1140 996}%
\special{fp}%
\special{pa 1140 1006}%
\special{pa 1140 1552}%
\special{fp}%
%
\special{pn 13}%
\special{pa 806 176}%
\special{pa 806 760}%
\special{da 0.070}%
\special{pa 816 760}%
\special{pa 970 888}%
\special{da 0.070}%
\special{pa 970 888}%
\special{pa 806 1032}%
\special{da 0.070}%
\special{pa 806 1032}%
\special{pa 806 1562}%
\special{da 0.070}%
%
\special{pn 8}%
\special{pa 798 650}%
\special{pa 474 650}%
\special{dt 0.045}%
%
\special{pn 8}%
\special{pa 474 1178}%
\special{pa 798 1178}%
\special{dt 0.045}%
\special{pa 474 540}%
\special{pa 1124 540}%
\special{dt 0.045}%
\special{pa 1132 1298}%
\special{pa 474 1298}%
\special{dt 0.045}%
%
\special{pn 8}%
\special{ar 474 914 236 264  1.5707963 1.6187963}%
\special{ar 474 914 236 264  1.7627963 1.8107963}%
\special{ar 474 914 236 264  1.9547963 2.0027963}%
\special{ar 474 914 236 264  2.1467963 2.1947963}%
\special{ar 474 914 236 264  2.3387963 2.3867963}%
\special{ar 474 914 236 264  2.5307963 2.5787963}%
\special{ar 474 914 236 264  2.7227963 2.7707963}%
\special{ar 474 914 236 264  2.9147963 2.9627963}%
\special{ar 474 914 236 264  3.1067963 3.1547963}%
\special{ar 474 914 236 264  3.2987963 3.3467963}%
\special{ar 474 914 236 264  3.4907963 3.5387963}%
\special{ar 474 914 236 264  3.6827963 3.7307963}%
\special{ar 474 914 236 264  3.8747963 3.9227963}%
\special{ar 474 914 236 264  4.0667963 4.1147963}%
\special{ar 474 914 236 264  4.2587963 4.3067963}%
\special{ar 474 914 236 264  4.4507963 4.4987963}%
\special{ar 474 914 236 264  4.6427963 4.6907963}%
%
\special{pn 8}%
\special{ar 474 914 334 374  1.5658428 1.5997891}%
\special{ar 474 914 334 374  1.7016278 1.7355741}%
\special{ar 474 914 334 374  1.8374128 1.8713591}%
\special{ar 474 914 334 374  1.9731978 2.0071441}%
\special{ar 474 914 334 374  2.1089828 2.1429291}%
\special{ar 474 914 334 374  2.2447678 2.2787141}%
\special{ar 474 914 334 374  2.3805528 2.4144991}%
\special{ar 474 914 334 374  2.5163379 2.5502841}%
\special{ar 474 914 334 374  2.6521229 2.6860691}%
\special{ar 474 914 334 374  2.7879079 2.8218541}%
\special{ar 474 914 334 374  2.9236929 2.9576391}%
\special{ar 474 914 334 374  3.0594779 3.0934241}%
\special{ar 474 914 334 374  3.1952629 3.2292091}%
\special{ar 474 914 334 374  3.3310479 3.3649942}%
\special{ar 474 914 334 374  3.4668329 3.5007792}%
\special{ar 474 914 334 374  3.6026179 3.6365642}%
\special{ar 474 914 334 374  3.7384029 3.7723492}%
\special{ar 474 914 334 374  3.8741879 3.9081342}%
\special{ar 474 914 334 374  4.0099729 4.0439192}%
\special{ar 474 914 334 374  4.1457579 4.1797042}%
\special{ar 474 914 334 374  4.2815429 4.3154892}%
\special{ar 474 914 334 374  4.4173280 4.4512742}%
\special{ar 474 914 334 374  4.5531130 4.5870592}%
\special{ar 474 914 334 374  4.6888980 4.7123890}%
%
\special{pn 8}%
\special{pa 2416 906}%
\special{pa 3392 906}%
\special{fp}%
%
\special{pn 20}%
\special{sh 1}%
\special{ar 2408 906 10 10 0  6.28318530717959E+0000}%
\special{sh 1}%
\special{ar 2408 906 10 10 0  6.28318530717959E+0000}%
%
\special{pn 8}%
\special{pa 2578 660}%
\special{pa 2254 660}%
\special{dt 0.045}%
%
\special{pn 13}%
\special{pa 2254 1188}%
\special{pa 2578 1188}%
\special{da 0.070}%
\special{pa 2254 550}%
\special{pa 2904 550}%
\special{da 0.070}%
\special{pa 2912 1306}%
\special{pa 2254 1306}%
\special{da 0.070}%
%
\special{pn 13}%
\special{ar 2254 924 236 264  1.5707963 1.8107963}%
\special{ar 2254 924 236 264  1.9547963 2.1947963}%
\special{ar 2254 924 236 264  2.3387963 2.5787963}%
\special{ar 2254 924 236 264  2.7227963 2.9627963}%
\special{ar 2254 924 236 264  3.1067963 3.3467963}%
\special{ar 2254 924 236 264  3.4907963 3.7307963}%
\special{ar 2254 924 236 264  3.8747963 4.1147963}%
\special{ar 2254 924 236 264  4.2587963 4.4987963}%
\special{ar 2254 924 236 264  4.6427963 4.7462743}%
%
\special{pn 20}%
\special{ar 2254 924 334 374  1.5658428 4.7123890}%
%
\special{pn 13}%
\special{pa 2262 660}%
\special{pa 2586 660}%
\special{da 0.070}%
\special{pa 2262 1306}%
\special{pa 2286 1306}%
\special{da 0.070}%
\special{pa 2286 1306}%
\special{pa 2912 1306}%
\special{da 0.070}%
\special{pa 2912 1306}%
\special{pa 2262 1306}%
\special{da 0.070}%
\special{pa 2262 550}%
\special{pa 2912 550}%
\special{da 0.070}%
\special{pa 2912 550}%
\special{pa 2262 550}%
\special{da 0.070}%
\special{pa 2262 1188}%
\special{pa 2262 1188}%
\special{da 0.070}%
%
\special{pn 20}%
\special{pa 2912 1306}%
\special{pa 2912 996}%
\special{fp}%
\special{pa 2912 996}%
\special{pa 2742 906}%
\special{fp}%
\special{pa 2912 540}%
\special{pa 2912 796}%
\special{fp}%
\special{pa 2912 796}%
\special{pa 2750 896}%
\special{fp}%
%
\special{pn 13}%
\special{pa 2594 668}%
\special{pa 2594 824}%
\special{da 0.070}%
\special{pa 2594 824}%
\special{pa 2732 906}%
\special{da 0.070}%
\special{pa 2732 906}%
\special{pa 2586 988}%
\special{da 0.070}%
\special{pa 2586 988}%
\special{pa 2586 1198}%
\special{da 0.070}%
\special{pa 140 148}%
\special{pa 140 130}%
\special{da 0.070}%
%
\special{pn 8}%
\special{pa 4416 896}%
\special{pa 5390 896}%
\special{fp}%
%
\special{pn 20}%
\special{sh 1}%
\special{ar 4408 896 10 10 0  6.28318530717959E+0000}%
\special{sh 1}%
\special{ar 4408 896 10 10 0  6.28318530717959E+0000}%
%
\special{pn 8}%
\special{pa 4578 650}%
\special{pa 4252 650}%
\special{dt 0.045}%
%
\special{pn 13}%
\special{pa 4252 1178}%
\special{pa 4578 1178}%
\special{da 0.070}%
\special{pa 4252 540}%
\special{pa 4902 540}%
\special{da 0.070}%
\special{pa 4912 1298}%
\special{pa 4252 1298}%
\special{da 0.070}%
%
\special{pn 13}%
\special{ar 4252 914 236 264  1.5707963 1.8107963}%
\special{ar 4252 914 236 264  1.9547963 2.1947963}%
\special{ar 4252 914 236 264  2.3387963 2.5787963}%
\special{ar 4252 914 236 264  2.7227963 2.9627963}%
\special{ar 4252 914 236 264  3.1067963 3.3467963}%
\special{ar 4252 914 236 264  3.4907963 3.7307963}%
\special{ar 4252 914 236 264  3.8747963 4.1147963}%
\special{ar 4252 914 236 264  4.2587963 4.4987963}%
\special{ar 4252 914 236 264  4.6427963 4.7462743}%
%
\special{pn 20}%
\special{ar 4252 914 334 374  1.5658428 4.7123890}%
%
\special{pn 13}%
\special{pa 4268 660}%
\special{pa 4594 660}%
\special{da 0.070}%
\special{pa 4594 660}%
\special{pa 4594 176}%
\special{da 0.070}%
\special{pa 4586 1178}%
\special{pa 4586 1570}%
\special{da 0.070}%
%
\special{pn 20}%
\special{pa 4260 550}%
\special{pa 4902 550}%
\special{fp}%
\special{pa 4902 550}%
\special{pa 4902 176}%
\special{fp}%
\special{pa 4252 1288}%
\special{pa 4902 1288}%
\special{fp}%
%
\special{pn 4}%
\special{pa 1164 476}%
\special{pa 1164 294}%
\special{fp}%
\special{sh 1}%
\special{pa 1164 294}%
\special{pa 1144 362}%
\special{pa 1164 348}%
\special{pa 1184 362}%
\special{pa 1164 294}%
\special{fp}%
\special{pa 1180 1480}%
\special{pa 1180 1298}%
\special{fp}%
\special{sh 1}%
\special{pa 1180 1298}%
\special{pa 1160 1364}%
\special{pa 1180 1350}%
\special{pa 1200 1364}%
\special{pa 1180 1298}%
\special{fp}%
\special{pa 840 1298}%
\special{pa 840 1288}%
\special{fp}%
\special{sh 1}%
\special{pa 840 1288}%
\special{pa 820 1356}%
\special{pa 840 1342}%
\special{pa 860 1356}%
\special{pa 840 1288}%
\special{fp}%
\special{pa 840 468}%
\special{pa 840 286}%
\special{fp}%
\special{sh 1}%
\special{pa 840 286}%
\special{pa 820 352}%
\special{pa 840 338}%
\special{pa 860 352}%
\special{pa 840 286}%
\special{fp}%
%
\special{pn 4}%
\special{pa 2360 496}%
\special{pa 2522 496}%
\special{fp}%
\special{sh 1}%
\special{pa 2522 496}%
\special{pa 2454 476}%
\special{pa 2468 496}%
\special{pa 2454 516}%
\special{pa 2522 496}%
\special{fp}%
\special{pa 2530 1224}%
\special{pa 2368 1224}%
\special{fp}%
\special{sh 1}%
\special{pa 2368 1224}%
\special{pa 2434 1244}%
\special{pa 2420 1224}%
\special{pa 2434 1204}%
\special{pa 2368 1224}%
\special{fp}%
\special{pa 2530 1362}%
\special{pa 2368 1362}%
\special{fp}%
\special{sh 1}%
\special{pa 2368 1362}%
\special{pa 2434 1382}%
\special{pa 2420 1362}%
\special{pa 2434 1342}%
\special{pa 2368 1362}%
\special{fp}%
%
\special{pn 4}%
\special{pa 4618 440}%
\special{pa 4618 258}%
\special{fp}%
\special{sh 1}%
\special{pa 4618 258}%
\special{pa 4598 326}%
\special{pa 4618 312}%
\special{pa 4638 326}%
\special{pa 4618 258}%
\special{fp}%
%
\special{pn 4}%
\special{pa 4944 458}%
\special{pa 4944 276}%
\special{fp}%
\special{sh 1}%
\special{pa 4944 276}%
\special{pa 4924 344}%
\special{pa 4944 330}%
\special{pa 4964 344}%
\special{pa 4944 276}%
\special{fp}%
\special{pa 4944 1552}%
\special{pa 4944 1370}%
\special{fp}%
\special{sh 1}%
\special{pa 4944 1370}%
\special{pa 4924 1438}%
\special{pa 4944 1424}%
\special{pa 4964 1438}%
\special{pa 4944 1370}%
\special{fp}%
%
\special{pn 8}%
\special{pa 4618 1552}%
\special{pa 4618 1380}%
\special{fp}%
\special{sh 1}%
\special{pa 4618 1380}%
\special{pa 4598 1446}%
\special{pa 4618 1432}%
\special{pa 4638 1446}%
\special{pa 4618 1380}%
\special{fp}%
\put(5.0000,-8.5){$\sigma_i$}
\put(11.6400,-8.5900){\makebox(0,0)[lb]{$slit$}}%
\put(17.6500,-9.4100){\makebox(0,0)[lb]{$+$}}%
\put(34.9600,-9.0500){\makebox(0,0)[lb]{$=$}}%
\put(23.0000,-8.5){$\sigma_i$}
\put(28.6300,-8.7800){\makebox(0,0)[lb]{$slit$}}%
\put(43.0000,-8.5){$\sigma_i$}
\put(45.2000,-9.0500){\makebox(0,0)[lb]{$slit$}}%
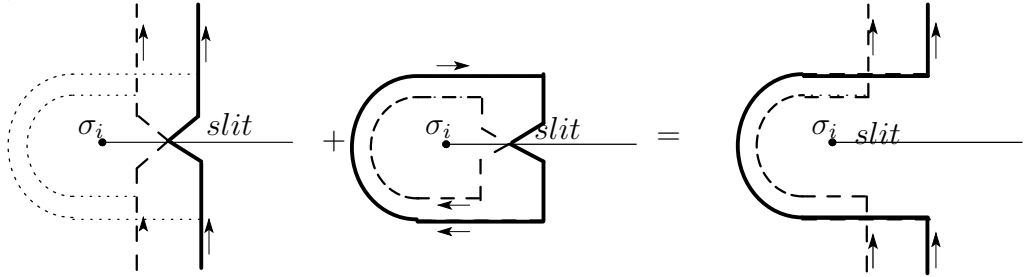
\put(15.7800,-17.8000){\makebox(0,0)[lb]{Homological nature in the double cover}}%
%
\special{pn 20}%
\special{pa 2260 550}%
\special{pa 2900 550}%
\special{fp}%
%
\special{pn 20}%
\special{pa 2260 1310}%
\special{pa 2910 1310}%
\special{fp}%
%
\special{pn 20}%
\special{pa 4260 1290}%
\special{pa 4900 1290}%
\special{fp}%
\special{pa 4900 1290}%
\special{pa 4900 1580}%
\special{fp}%
\end{picture}%

${}$\hfill $\Box$
\bigskip 

\section{The differential equation
$\frac{\nabla}{d\sigma}f_{\sigma}(u,v)
={:}(\frac{1}{i\h}u{\ctt}v{+}\alpha){:}_{_K}{*_{_K}}f_{\sigma}(u,v)$} 
\label{DiffEq}

In ordinary complex calculus, residues mainly relate 
to the global nature of Riemann surfaces. In contrast, 
$\int_{C^2}{:}e_*^{z(\alpha{+}\frac{1}{i\h}u{\ctt}v)}{:}_{_K}
d{\sqrt{z{-}\sigma}}$ appears to depend on the position 
$\sigma$ which moves by the expression parameter $K$. 
We see indeed the following: 
\begin{thm}\label{resmovement}
Suppose $C^2$ is fixed and the singular point $\sigma$ moves 
without crossing $C$, or $C$ is an infinitesimally small circle 
with center at $\sigma$, and moving together with $\sigma$. 
Then 
$$
\frac{d}{dt}\big|_{t=0}\int_{C^2}
{:}e_*^{z(\alpha{+}\frac{1}{i\h}u{*}v)}{:}_{_K}
d{\sqrt{z{-}\sigma{-}t}}=
{:}(\alpha{+}\frac{1}{i\h}u{*}v){:}_{_K}{*_{_K}}
\int_{C^2}{:}e_*^{z(\alpha{+}\frac{1}{i\h}u{*}v)}{:}_{_K}
d{\sqrt{z{-}\sigma}}. 
$$
\end{thm}

Suppose $C$ is any smooth closed curve 
in the complex plane ${\mathbb C}$ avoiding singular points. 
To consider the movement of a singular point $\sigma$, we 
use the integral 
$$
R_{-2k{-}1}(z,u,v;K)=
\int_{C^2}{:}s^{2k{+2}}e_*^{(z{+}s^2)(\alpha{+}\frac{1}{i\h}u{\ctt}v)}{:}_{_K}ds,
\quad k\in{\mathbb Z}. 
$$ 
If $z$ is an independent variable, then 
integration by parts gives that this must satisfy 
\begin{equation}\label{diffeqeq}
\partial_zR_{-2k{-}1}(z,u,v;K)=
{:}(\alpha{+}\frac{1}{i\h}u{\ctt}v){:}_{_K}{*_{_K}}
R_{-2k{-}1}(z,u,v;K).
\end{equation}
However, Theorems \,\,\ref{lemmaresidue},\,\,\ref{remarkableresd} show that the integral 
$$
\int_{C^2}{:}s^{2k{+}2}e_*^{(z{+}s^2)(\alpha{+}\frac{1}{i\h}u{\ctt}v)}{:}_{_K}ds
$$
vanishes for all regular points $z\not\in\Sigma_K$. 
$R_{-2k{-}1}(z,u,v;K)$ gives a nontrivial value only when $z\in
\Sigma_K$. That is, $z$ cannot be an independent variable in 
$R_{-2k{-}1}(z,u,v;K)$. 
Summarizing these observation we have the following by
using \eqref{residuform}:
\begin{thm}\label{formaldist}
For every Laurent polynomial $\psi(s^2,s^{-2})$, the integral 
$$
\Phi[\psi](z,K,u,v)=
\int_{C^2}{:}\psi(s^2,s^{-2})e_*^{(z{+}s^2)(\alpha{+}\frac{1}{i\h}u{\ctt}v)}{:}_{_K}ds
$$
is supported only on the set $\{(z,K); z{\in}\Sigma_K\}$, where $C$ is any smooth closed curve 
in the complex plane ${\mathbb C}$ avoiding singular points.
\end{thm}

Let ${\mathfrak G}$ be the totality of $K$ such that $\Sigma_K$ is
a set of simple singularities. Then, 
${\mathcal S}= \{(z,K); z{\in}\Sigma_K, K\in {\mathfrak G}\}$
is a holomorphic submanifold of codimension $1$. 
Note that $\Phi[\psi](z,K,*,*)$ is a holomorphic on the submanifold   
${\mathcal S}$. 

\bigskip
\noindent
{\bf Co-moving differentials}\,\,\,  
Let $\sigma$ be a variable $\sigma{\in}{\mathbb C}$ together with 
an expression parameter $K(\sigma)$ such that 
$\sigma\in \Sigma_{K(\sigma)}$. We define   
\begin{equation}\label{covderivative}
\frac{\nabla}{d\sigma}f(\sigma, K(\sigma); u,v)=
\partial_\sigma f(\sigma, K; u,v)\big|_{K=K(\sigma)}=
\partial_z(f(\sigma, K(\sigma); u,v)){-}\frac{i\h}{4}\dot{K}(\sigma)(f)
\end{equation}
where $\frac{i\h}{4}\dot{K}(\sigma)(f)$ is the infinitesimal
intertwiner (cf.\eqref{parallel}) given by  
\begin{equation}\label{infintner}
\frac{i\h}{4}\dot{K}(\sigma)(f)=
\frac{i\h}{4}\sum_{ij}\frac{dK^{ij}(\sigma)}{d\sigma}\partial_{u^i}\partial_{u^j}f,
\quad (u,v)=(u^1,u^2)
\end{equation} 
for every $H\!ol({\mathbb C}^2)$-valued function$f(\sigma,K; *,*)$. 
Intuitively, this may be written  as 
\begin{equation}\label{Intuitive}
f(\sigma{+}\delta,K(\sigma{+}\delta);u,v)=
I_{K(\sigma)}^{K(\sigma{+}\delta)}\Big({:}e_*^{\delta(z{+}\frac{1}{i\h}u{\ctt}v)}
{:}_{_K(\sigma)}{*_{_{K(\sigma)}}}f(\sigma,K(\sigma);u,v)\Big),
\end{equation}
 where $\delta$ is an infinitesimal. 
$\Phi[\psi](\sigma,K,u,v)$ given in Theorem\,\ref{formaldist} satisfies 
$$
\frac{\nabla}{d\sigma}\Phi[\psi](\sigma,K,u,v)=
{:}(\alpha{+}\frac{1}{i\h}u{\ctt}v){:}_{_{K(\sigma)}}
{*_{_{K(\sigma)}}}\Phi[\psi](\sigma,K,u,v).
$$
This is proved as follows:
$$
\begin{aligned}
\lim_{t\to 0}\frac{1}{t}&\Big(
I_{K(\sigma{+}t)}^{K(\sigma)}\int_{C^2}{:}\psi(s,s^{-1})
e_*^{(\sigma{+}t{+}s^2)(\alpha{+}\frac{1}{i\h}u{\ctt}v)}{:}_{_{K(\sigma{+}t)}}ds
-\int_{C^2}{:}\psi(s,s^{-1})
e_*^{(\sigma{+}s^2)(\alpha{+}\frac{1}{i\h}u{\ctt}v)}{:}_{_{K(\sigma)}}ds\Big)\\
&
=\lim_{t\to 0}\frac{1}{t}\Big(\int_{C^2}{:}\psi(s,s^{-1})
e_*^{(\sigma{+}t{+}s^2)(\alpha{+}\frac{1}{i\h}u{\ctt}v)}{:}_{_{K(\sigma)}}ds
-\int_{C^2}\psi(s,s^{-1})
{:}e_*^{(\sigma{+}s^2)(\alpha{+}\frac{1}{i\h}u{\ctt}v)}{:}_{_{K(\sigma)}}ds\Big)\\
&
=\int_{C^2}\psi(s,s^{-1})\lim_{t\to 0}\frac{1}{t}
\Big({:}e_*^{(\sigma{+}t{+}s^2)(\alpha{+}\frac{1}{i\h}u{\ctt}v)}{-}
e_*^{(\sigma{+}s^2)(\alpha{+}\frac{1}{i\h}u{\ctt}v)}{:}_{_{K(\sigma)}}\Big)ds\\
&={:}(\alpha{+}\frac{1}{i\h}u{\ctt}v){*}
\int_{C^2}\psi(s,s^{-1})
e_*^{(\sigma{+}s^2)(\alpha{+}\frac{1}{i\h}u{\ctt}v)}ds{:}_{_{K(\sigma)}}. 
\end{aligned}
$$ 
Since every Laurent coefficients of 
${:}e_*^{\delta(z{+}\frac{1}{i\h}u{\ctt}v)}{:}_{_K(\sigma)}$
is obtained by this integral, we see 
in particular, for every $\sigma$, the coefficient 
$a_{2k{-}1}(\sigma,K(\sigma))$ of Laurent series of  
${:}e_*^{(\sigma{+}s^2)(\alpha{+}\frac{1}{i\h}u{\ctt}v)}{:}_{_{K(\sigma)}}$ 
at a singular point $\sigma$ satisfies the equation 
\begin{equation}\label{EqEqres} 
\frac{\nabla}{d\sigma}f(\sigma,K(\sigma); u,v)=
{:}(\alpha{+}\frac{1}{i\h}u{\ctt}v){:}_{_K}{*_{_K}}f(z,K(\sigma);u,v).
\end{equation}

\bigskip
\noindent
{\bf Star-product integrals} \,\,
Viewing \eqref{EqEqres} as a differential equation, we can make 
the solution by the product integrals, if one can expect the convergence.  

Let $\sigma(t)$ be a piecewise smooth curve in ${\mathbb C}$ 
for $0\leq t\leq 1$. We assume 
that the expression parameter $K$ moves together with $\sigma$ and we
suppose $\sigma(t)\in \Sigma_{K(\sigma(t))}$. For a division  
$$
\Delta;\quad 0=t_0{<}t_1{<}t_2{<}{\cdots}{<}t_n=1
$$ 
of $[0,1]$, we first define the product integral 
${\mathcal P}_{\Delta}(\sigma(t))$ inductively by setting 
$$
{\mathcal P}_{\Delta}(\sigma(0))
=f(\sigma(0),K_{\sigma(0)},u,v) 
$$ 
and 
$$
{\mathcal P}_{\Delta}(\sigma(t))=
I_{K(\sigma_{i})}^{K(\sigma(t))}
\Big(\big({:}e_*^{(\sigma(t){-}\sigma(t_{i}))(\alpha{+}\frac{1}{i\h}u{\ctt}v)}
{:}_{K(\sigma(t_i))}\big){*_{K(\sigma(t_i))}}
{\mathcal P}_{\Delta}(\sigma(t_{i}))\Big),\quad 
\sigma_{i}<t{\leq}\sigma(t_{i{+}1}) 
$$
where  $I_{K(\sigma)}^{K(\sigma')}$ is the intertwiner defined by 
\eqref{intertwiner} (cf. also \eqref{parallel}). 
Note that ${\mathcal P}_{\Delta}(\sigma(t))$ is computed under 
a $K(\sigma(t)$-expression.

We say that the product integral converges, if 
by setting $|\Delta|=\max\{|t_i{-}t_{i{-}1}|\}$ the limit 
${\mathcal P}_{d}(\sigma(t))=\lim_{|\Delta|\to 0}{\mathcal P}_{\Delta}(\sigma(t))$ 
exists. It is not hard that if the product integral converges then it
satisfies 
$$
\frac{d}{dt}{\mathcal P}_{d}(\sigma(t))=
\frac{d\sigma}{dt}(t){:}(\alpha{+}\frac{1}{i\h}u{\ctt}v){:}_{K(\sigma(t))}{*_{K(\sigma(t))}}
{\mathcal P}_{d}(\sigma(t)),\quad 
{\mathcal P}_{d}(0)=f(\sigma(0),K_{\sigma(0)},u,v).
$$
Note that $\frac{d}{dt}f(z(t))=\frac{df}{dz}(z(t))\frac{dz(t)}{dt}$
for every holomorphic function $f(z)$.   
The above identity shows that $\frac{d{\mathcal P}_{d}(\sigma)}{d\sigma}$ is
well-defined and 
\begin{equation}\label{tentadiff}
\frac{d{\mathcal P}_{d}}{d\sigma}(\sigma(t))=
{:}(\alpha{+}\frac{1}{i\h}u{\ctt}v){:}_{K(\sigma(t))}{*_{K(\sigma(t))}}{\mathcal P}_{d}(\sigma(t)),\quad 
{\mathcal P}_{d}(\sigma(0))=f(\sigma(0),K_{\sigma(0)},u,v).
\end{equation}

It is remarkable that even if the starting point $\sigma(0)$ and the
ending point $\sigma(1)$ are fixed, the product integral may depend on
the path $\sigma(t)$ from $\sigma(0)$ to $\sigma(1)$.
There is no general rule to select a specific path from 
$\sigma(0)$ to $\sigma(1)$. The natural variational problem  
degenerates.  

\bigskip
Note that $\frac{d}{dt}{\mathcal P}_{d}(\sigma(t))$ is different from
an ordinary differentiation. This is given as 
\begin{equation*}\label{covariant00}
\lim_{\delta\to 0}\frac{1}{\delta}
\Big(I^{K(\sigma)}_{K(\sigma{+}\delta)}
{\mathcal P}_{d}(\sigma(t{+}\delta))
{-}{\mathcal P}_{d}(\sigma(t))\Big).
\end{equation*} 
Hence it is better to write \eqref{tentadiff} in the form   
\begin{equation}\label{covariant00}
\frac{\nabla}{d\sigma}{\mathcal P}_{d}(\sigma(t))
={:}(\alpha{+}\frac{1}{i\h}u{\ctt}v){:}_{K(\sigma(t))}
{*_{K(\sigma(t))}}{\mathcal P}_{d}(\sigma(t)),
\quad 
{\mathcal P}_{d}(\sigma(0))=
f(\sigma(0),K_{\sigma(0)},u,v).
\end{equation} 
The next one is fundamental
\begin{prop}
For every $s\not=0$, then 
${:}e_*^{(\sigma{+}s^2)(\alpha{+}\frac{1}{i\h}u{\ctt}v)}{:}_{_{K(\sigma)}}$
satisfies 
$$
\frac{\nabla}{d\sigma}{:}e_*^{(\sigma{+}s^2)(\alpha{+}\frac{1}{i\h}u{\ctt}v)}{:}_{_{K(\sigma)}}
={:}(\alpha{+}\frac{1}{i\h}u{\ctt}v){:}_{_{K(\sigma)}}{*_{_{K(\sigma)}}}
{:}e_*^{(\sigma{+}s^2)(\alpha{+}\frac{1}{i\h}u{\ctt}v)}{:}_{_{K(\sigma)}}.
$$
\end{prop}

\noindent
{\bf Proof}\,\,This is computed as follows:
$$
\begin{aligned}
\lim_{t\to 0}&\frac{1}{t}\Big(I_{K(\sigma{+}t)}^{K(\sigma)}
{:}e_*^{(\sigma{+}t{+}s^2)(\alpha{+}\frac{1}{i\h}u{\ctt}v)}{:}_{_{K(\sigma{+}t)}}
-{:}e_*^{(\sigma{+}s^2)(\alpha{+}\frac{1}{i\h}u{\ctt}v)}{:}_{_{K(\sigma)}}\Big)\\
&=
\lim_{t\to 0}\frac{1}{t}\Big(
{:}e_*^{(\sigma{+}t{+}s^2)(\alpha{+}\frac{1}{i\h}u{\ctt}v)}{:}_{_{K(\sigma)}}
-{:}e_*^{(\sigma{+}s^2)(\alpha{+}\frac{1}{i\h}u{\ctt}v)}{:}_{_{K(\sigma)}}\Big)\\
&={:}(\alpha{+}\frac{1}{i\h}u{\ctt}v){:}_{_{K(\sigma)}}{*_{_{K(\sigma)}}}
{:}e_*^{(\sigma{+}s^2)(\alpha{+}\frac{1}{i\h}u{\ctt}v)}{:}_{_{K(\sigma)}}.
\end{aligned}
$$

\medskip
\noindent
{\bf Parallel functions}\,\,
Viewing $\nabla$ the notion of co-moving derivative, 
we extend \eqref{covderivative} as the covariant/comoving differentiation 
not only for $f(\sigma, K(\sigma); u,v)$, but also for 
functions $f(\sigma,K(\sigma))$ without $u,v$ by 
$$
\nabla_{\sigma}f(\sigma,K(\sigma))=\partial_{\sigma}f(\sigma,K)\big|_{K=K(\sigma)}.
$$
Let $(a_{ij}(\sigma))$ be a matrix such that $\sum_{i,j}a_{ij}(\sigma)K^{ij}(\sigma)=0$.
Then $f(\sigma, K)=\int^{\sigma}\sum_{i,j}a_{ij}(\tau)d\tau K^{ij}$ is
a parallel function.

Given $K(\sigma)$, parallel functions forms a commutative algebra. We call these 
{\it parallel functions on} $K=K(\sigma)$ and denote this by 
${\mathcal P}[K(\sigma)].$

\subsection{Co-moving expression parameters}

Now consider a general expression parameter 
${K}=\footnotesize{
\begin{bmatrix}
\delta&c\\
c&\delta'\\
\end{bmatrix}
}$, and set $u{\ctt}v=\frac{1}{2}(u{*}v{+}v{*}u)$. 

In this section we use the variable $t$ instead of $\sigma$, and 
we think of the expression parameter $K$ as moving together with
the parameter $t$, indicating a specified singular point.  
Given $t$, we choose $K(t)$ so that  $t\in \Sigma_{K(t)}$, but
different from the case of one variable, we have 
a lot of choices of $K(t)$. Recall that $\Sigma_{K(t)}$ is the 
singular set of ${:}e_*^{z\frac{1}{i\h}u{\ctt}v}{:}_{K(t)}$.  
In what follows, we think of $\delta$, $\delta'$ and $c$ as functions of 
$t$. Under this notation, the infinitesimal intertwiner is given by  
\begin{equation}\label{niceinter}
\frac{i\h}{4}{\dot K}=\frac{i\h}{4}
({\dot\delta}(t)\partial_u^2{+}2{\dot c}(t)\partial_u\partial_v{+}{\dot{\delta}'}(t)\partial^2_v),
\end{equation}
where ${\dot a}=\frac{d}{dt}a(t)$.

As  $t\in \Sigma_{K(t)}$,  $t$ must satisfy 
\begin{equation}\label{sigequation}
(e^{t/2}{+}e^{-t/2}{-}c(t)(e^{t/2}{-}e^{-t/2}))^2-(e^{t/2}{-}e^{-t/2})^2\delta(t)\delta'(t)=0
\end{equation}
by \eqref{genericparam00}.
By this $c(t), \delta(t)\delta'(t)$ must be singular at $t=0$ and we have 

\begin{equation}\label{sigequation00}
c(t)=\pm\sqrt{\delta(t)\delta'(t)}{+}\frac{e^{t/2}{+}e^{-t/2}}{e^{t/2}{-}e^{-t/2}},
\quad t\not=0.
\end{equation}
On the other hand, noting that 
${:}\frac{1}{i\h}u{\ctt}v{:}_{_K}{=}\frac{1}{i\h}uv{+}\frac{1}{2}c$, 
 the $K$-ordered product of  
${:}(\frac{1}{i\h}u{\ctt}v{+}\alpha){:}_{_K}{*_{_K}}f(u,v)$
is written precisely as   
\begin{equation}\label{bigeqeq}
\begin{aligned}
(\frac{c}{2}{+}\alpha{+}&\frac{1}{i\h}uv)f(u,v){+}
((c{+}1)u{+}\delta v)\partial_uf(u,v){+}
(\delta'u{+}(c{-}1)v)\partial_vf(u,v)\\
{+}&
\frac{i\h}{4}\big(\delta(c{+}1)\partial_u^2f(u,v){+}
(\delta\delta'{+}c^2{-}1)\partial_u\partial_vf(u,v)
{+}\delta'(c{-}1)\partial_v^2f(u,v)\big).
\end{aligned}
\end{equation} 

By\eqref{bigeqeq} the r.h.s. of the equation \eqref{EqEqres} is rewritten as   
\begin{equation}\label{bigeqeq1}
\begin{aligned}
\frac{\nabla}{\partial\sigma}f(\sigma, u,v, K(\sigma))&=
\Big((\alpha{+}c/2{+}\frac{1}{i\h}uv){+}
((c{+}1)u{+}\delta v)\partial_uf{+}
(\delta'u{+}(c{-}1)v)\partial_v\\
&{+}
\frac{i\h}{4}\big(\delta(c{+}1)\partial_u^2{+}
(\delta\delta'{+}c^2{-}1)\partial_u\partial_v
{+}\delta'(c{-}1)\partial_v^2\big)\Big)f(\sigma,u,v,K(\sigma)).
\end{aligned}
\end{equation}

\noindent
{\bf Note} also that $a_{2k{+}1}(\sigma,K)$ satisfies also 
$$
\frac{\nabla}{d\sigma}a_{2k{+}1}(\sigma,K(\sigma))=
a_{2k{+}1}(\sigma,K(\sigma)){*_{_K(\sigma)}}
{:}(\alpha{+}\frac{1}{i\h}u{\ctt}v){:}_{_K(\sigma)}.
$$
The equation for this is slightly changed as follows:
\begin{equation}\label{bigeqeq2}
\begin{aligned}
\frac{\nabla}{\partial\sigma}f(\sigma, u,v, K(\sigma))&=
\Big((\alpha{+}c/2{+}\frac{1}{i\h}uv){+}
((c{-}1)u{+}\delta v)\partial_uf{+}
(\delta'u{+}(c{+}1)v)\partial_v\\
&{+}
\frac{i\h}{4}\big(\delta(c{-}1)\partial_u^2{+}
(\delta\delta'{+}c^2{-}1)\partial_u\partial_v
{+}\delta'(c{+}1)\partial_v^2\big)\Big)f(\sigma,u,v,K(\sigma)).
\end{aligned}
\end{equation}
The difference of solutions of 
these two equations must satisfies in particular the equation 
$$
[\frac{1}{i\h}u{\ctt}v,f(\sigma, u,v, K(\sigma))]_{_{*_{K(\sigma)}}}=0.
$$ 
This is a differential equation of order 2 in general:
\begin{equation}\label{preellip} 
{ }[\frac{1}{i\h}u{\ctt}v,f(\sigma, u,v,
K(\sigma))]_{_{*_{K(\sigma)}}}=
\Big(\big(u\partial_u{-}v\partial_v\big)
{+}\frac{i\h}{2}\big(\delta\partial_u^2{-}\delta'\partial_v^2\big)\Big)f(\sigma,u,v,K(\sigma))=0.
\end{equation} 
\bigskip
To eliminate the quadratic terms in the second line of 
\eqref{bigeqeq1}  by the infinitesimal intertwiner, we set 
$$
\left\{
\begin{matrix}
\medskip
\frac{d}{dt}\delta&=&\!\!\!\!-\delta(c{+}1)\\
\medskip
\frac{d}{dt}\delta'&=&\!\!\!\!-\delta'(c{-}1)\\
\medskip
\frac{d}{dt}c&=&-\frac{1}{2}(\delta\delta'{+}c^2{-}1)
\end{matrix}
\right.
$$
Setting $\xi(t)=\int_0^tc(s)ds$, we have 
\begin{equation}\label{112233}
\delta(t)=ae^{-{\xi}(t){-}t}, \quad \delta'(t)=a'e^{-{\xi}(t){+}t}, \quad 
\frac{d^2}{dt^2}{\xi}(t)=-\frac{1}{2}aa'e^{-2{\xi}(t)}
{-}\frac{1}{2}(\frac{d\xi}{dt})^2{+}\frac{1}{2}
\end{equation}
where $a, a'$ are arbitrary constants. 
Before solving this, we confirm that these are consistent with \eqref{sigequation00}.
Plugging the first two equality to \eqref{sigequation00}, we have 
\begin{equation}\label{korekore}
\frac{d\xi}{dt}=\pm\sqrt{aa'}e^{-\xi(t)}{+}\frac{e^{t/2}{+}e^{-t/2}}{e^{t/2}{-}e^{-t/2}}
\end{equation}
Differentiating this to obtain 
$$
\frac{d^2{\xi}}{dt^2}=
-aa'e^{-2{\xi}(t)}\pm\sqrt{aa'}e^{-{\xi}(t)}\frac{e^{t/2}{+}e^{-t/2}}{e^{t/2}{-}e^{-t/2}}
\,+\,\frac{d}{dt}\Big(\frac{e^{t/2}{+}e^{-t/2}}{e^{t/2}{-}e^{-t/2}}\Big).
$$
This is the third equation of \eqref{112233}. 
Note that setting $v=\frac{d\xi}{dt}$, the third equation of
\eqref{112233} is equivalent with 
$$
\frac{d}{dt}(v{+}\sqrt{aa'}e^{-\xi})=\frac{1}{2}{-}\frac{1}{2}(v{+}\sqrt{aa'}e^{-\xi})^2.
$$
It follows $v{+}\sqrt{aa'}e^{-\xi}=\frac{e^{t/2}{-}e^{-t/2}}{e^{t/2}{+}e^{-t/2}}$.

We have only to solve \eqref{korekore}. By setting 
$\eta(t)=e^{\xi(t)}$, \eqref{korekore} gives 
$$
\frac{d\eta(t)}{dt}=\eta(t)\frac{e^{t/2}{+}e^{-t/2}}{e^{t/2}{-}e^{-t/2}}\pm\sqrt{aa'}.
$$ 
For simplicity, choose the plus sign $\sqrt{aa'}$.  
Solving this, we have   
\begin{equation}\label{yyyt}
\eta(t)=(\gamma{-}\frac{\sqrt{aa'}}{4})e^t{+}(\gamma{+}\frac{\sqrt{aa'}}{4})e^{-t}{-}2\gamma,
\quad \gamma\in {\mathbb C}.
\end{equation}

Hence using $\eta(t)=e^{\xi(t)}$ we have 
\begin{equation}
\left\{
\begin{matrix}
\medskip
\delta(t)\,=ae^{{-}t}\eta(t)^{-1}=ae^{-\xi(t){-}t}\\
\medskip
\delta'(t)=a'e^{t}\eta(t)^{-1}=a'e^{-\xi(t){+}t}\\
\medskip
 c(t)=\frac{d}{dt}\log\eta(t)=\frac{d}{dt}\xi(t)
\end{matrix}
\right.
\end{equation}
where $a, a'$ and $\gamma$ are arbitrarily chosen.

To simplify the solution, we set $\gamma=0$, $a=a'=1$ and set in what follows  
$\eta(t)={-}\frac{1}{4}(e^t{-}e^{-t})$ by restricting the domain 
$t$ in ${\rm{Re}}\,t>0$.

\begin{thm}
By choosing a suitable path $K(t)$, 
equation \eqref{EqEqres} turns out to be a differential equation of
order one  
$$
\partial_{t}f(t,K(t);u,v)
=
\Big((\frac{c}{2}{+}\alpha{+}\frac{1}{i\h}uv){+}
((c{+}1)u{+}\delta v)\partial_u{+}
(\delta'u{+}(c{-}1)v)\partial_v\Big)f(t, K(t);u,v).
$$
\end{thm}

Changing variables $x=\frac{1}{\sqrt{i\h}}u,\,
y=\frac{1}{\sqrt{i\h}}v$ and setting 
$$
C(t)=(\alpha{+}\frac{c(t)}{2}),\quad 
L(t)=
\begin{bmatrix}
c(t){+}1&\delta'(t)\\
\delta(t)&c(t){-}1
\end{bmatrix}=
\begin{bmatrix}
0&1\\
1&0
\end{bmatrix}(K(t){+}J),
$$  
gives the equivalent equation 
\begin{equation}\label{eqtosolve}
\Big(\partial_{t}-
(x,y)L(t) 
\begin{bmatrix}
\partial_x\\ \partial_y\\
\end{bmatrix}\Big)f(t,K(t);x,y)
=(xy{+}C(t))f(t,K(t);x,y).
\end{equation}
As we set $\eta(t)={-}\frac{1}{4}(e^t{-}e^{-t})$, 
$$
C(t)=\alpha{+}\frac{1}{2}\frac{e^t{+}e^{-t}}{e^t{-}e^{-t}}, \quad  
L(t)=
\begin{bmatrix}
\frac{2}{1{-}e^{-2t}}&4\frac{1}{1{-}e^{2t}}\\
4\frac{-1}{1{-}e^{-2t}}&\frac{2}{1{-}e^{2t}}
\end{bmatrix}
$$  

To solve \eqref{eqtosolve}, we set   
$\psi_{\lambda}(t){=}\int_{\lambda}^{t}\frac{1}{e^{2s}{-}1}ds$
for every positive $\lambda$. Easily,  
$\int_\lambda^{t}\frac{1}{1{-}e^{-2s}}ds{=}t{-}\lambda{+}\psi_{\lambda}(t)$,
and if $t$ is real then
$$
(t{-}{\lambda}{+}\psi_{\lambda}(t))\psi_{\lambda}(t)\geq 0, \quad
{\text{and}}\quad >0 , (t\not=\lambda).
$$

Choosing different positive real numbers $\lambda$ and $\mu$, we set  
$$
{\tilde L}(t)=
\begin{bmatrix}
t{-}\mu{+}\psi_{\mu}(t)& \psi_{\lambda}(t) \\
{-}t{+}\lambda{-}\psi_{\lambda}(t) &\psi_{\mu}(t)
\end{bmatrix}.
$$
Then, we see $\frac{d}{dt}{\tilde L}(t)=L(t)$ and $\det{\tilde L}(t)$
is positive definite on $t>0$. 

Letting 
$g(t,x,y){=}f\big(t,(x,y){\tilde L}(t)\big)$, we see 
$$
\frac{d}{dt}g(t,x,y)=(xy{+}C(t))g(t,x,y), \quad
{\text{hence}}\quad g(t,x,y)=e^{txy{+}{\tilde C}(t)}G(x,y)
$$ 
where ${\tilde C}(t){=}\log(e^{\alpha t}(e^t{-}e^{-t})$ and 
$G(x,y)$ is an arbitrary holomorphic function.  
Viewing $\phi_t={\tilde L}^{-1}(t): {\mathbb C}^2\to {\mathbb C}^2$
as a linear diffeomorphism,  $f(t,x,y)$ is obtained by  
the pull-back of $g(t,x,y)$:
$$
f(t,x,y)=\phi_t^*\big(e^{txy{+}{\tilde C}(t)}G(x,y)\big).
$$

Since 
${:}e_*^{(t{+}s^2)(\alpha{+}\frac{1}{i\h}u{\ctt}v)}{:}_{_{K(t)}}$ 
satisfies this equation,  
$$
{:}e_*^{(t{+}s^2)(\alpha{+}\frac{1}{i\h}u{\ctt}v)}{:}_{_{K(t)}}=
\phi_t^*\big(e^{t\frac{1}{i\h}uv{+}{\tilde C}(t)}G(u,v)\big).
$$

For adjusting the initial data, setting $t=-s^2$, we see 
$$
1=\phi_{-s^2}^*\big(e^{-s^2\frac{1}{i\h}uv{+}{\tilde C}(-s^2)}G(u,v)\big)
$$
$$
1={\tilde L}(-s^2)(1)=e^{-s^2\frac{1}{i\h}uv{+}{\tilde C}(-s^2)}G(u,v)
$$
It follows that 
$G(u,v)=e^{s^2\frac{1}{i\h}uv{-}{\tilde C}(-s^2)}$.
Plugging this we have 
$$
{:}e_*^{(t{+}s^2)(\alpha{+}\frac{1}{i\h}u{\ctt}v)}{:}_{_{K(t)}}=
\phi_t^*\big(e^{(t{+}s^2)\frac{1}{i\h}uv{+}{\tilde C}(t){-}{\tilde C}(-s^2)}\big)
=\phi_t^*
\big(e^{(t{+}s^2)(\frac{1}{i\h}uv{+}\alpha){+}\frac{1}{2}\xi(t)-\frac{1}{2}\xi(-s^2)}\big).
$$
This may be written as 
$$
{:}e_*^{(t{+}s^2)(\alpha{+}\frac{1}{i\h}u{\ctt}v)}{:}_{_{K(t)}}=
\sqrt{\frac{\eta(t)}{\eta(-s^2)}}\phi_t^*\big(e^{(t{+}s^2)(\frac{1}{i\h}uv{+}\alpha)}\big)
=\sqrt{\frac{e^{-t}{-}e^{t}}{e^{s^2}{-}e^{-s^2}}}
\phi_t^*\big(e^{(t{+}s^2)(\frac{1}{i\h}uv{+}\alpha)}\big).
$$

\begin{center}
\fbox{\parbox[t]{.8\linewidth}{
If we give an attention to a movement of specific singular point and
use a suitable expression parameter moving together with the 
singular point, then the $*$-exponential function looks as if it were
a classical function $e^{(t{+}s^2)(\frac{1}{i\h}uv{+}\alpha)}$.}}
\end{center}

\section{Matrix elements in generic ordered expressions}
\label{matrix elem}

So far, we are concerned with individual/ subjective singular point. This depends
on the individual expression parameters. However, there are a lot of 
mathematical facts which can be stated without specifying the
expression parameters. Such one may be viewed as ``objective'' object  
that is commonly accepted by many observers. That is, we are thinking
that the ``objective'' object is nothing but objects which almost all 
expression parameters accept.    
In this section, we discuss such common objects. 

\medskip
First of all, we note several identities that ensure the associativity:
\begin{lem}
\label{vacvac25} 
If $p\not=0$, then 
$\varpi_{00}{*}(u^p{*}\varpi_{00}){=}0, and\,\,\,
 (\varpi_{00}{*}v^p){*}\varpi_{00}{=}0$.
\end{lem}

\noindent
{\bf Proof}\,\,
By taking the formal power series
expansion with respect to $i\h$ for $e_*^{su{*}v}$, the formal 
associativity theorem (cf.\cite{ommy5}) together with the bumping identity gives 
the following:  
$$
e_*^{su{*}v}{*}(u^{p}{*}e_*^{tu{*}v})
{=}(e_*^{su{*}v}{*}u^{p}){*}e_*^{tu{*}v}
{=}
u^p{*}e_*^{(s{+}t)u{*}v{+}i\h ps}.
$$
The right hand side of the above equality is continuous 
in $s, t$. In particular, 
$$
\lim_{t{\to}a}e_*^{su{*}v}{*}(u^{p}{*}e_*^{tu{*}v}){=}
e_*^{su{*}v}{*}\lim_{t{\to}a}(u^{p}{*}e_*^{tu{*}v}).
$$ 
Using the bumping identity, we have 
$$
\begin{aligned}
e_*^{su{*}v}{*}(u^{p}{*}\lim_{t{\to}{-}\infty}e_*^{tu{*}v})
{=}&e_*^{su{*}v}{*}\lim_{t{\to}{-}\infty}u^{p}{*}e_*^{tu{*}v}
{=}\lim_{t{\to}{-}\infty}u^{p}{*}e_*^{(s{+}t)u{*}v+i\h ps}\\
{=}&u^{p}{*}\lim_{t{\to}{-}\infty}e_*^{(s{+}t)u{*}v+i\h ps}
{=}u^{p}e^{i\h ps}{*}\varpi_{00}.
\end{aligned}
$$
It follows that
$$
\varpi_{00}{*}(u^p{*}\varpi_{00}){=}
\lim_{s{\to}{-}\infty}e_*^{s\frac{1}{i\h}u{*}v}
{*}(\lim_{t{\to}{-}\infty}{u^p}{*}e_*^{t\frac{1}{i\h}u{*}v})
{=}
\lim_{s{\to}{-}\infty}u^{p}e^{ps}{*}\varpi_{00}{=}0.
$$  
Similarly, we also have $(\varpi_{00}{*}v^p){*}\varpi_{00}{=}0$. 

\bigskip
These are proved also by using the integral expressions.
$$
\begin{aligned}
&\int_0^{2\pi}\!\!e_*^{(s{+}it)\frac{1}{i\h}u{*}v}dt {*}u^p{*} 
\!\int_0^{2\pi}\!\!e_*^{(s{+}i\tau)\frac{1}{i\h}u{*}v}d\tau 
=
u^p{*}\int_0^{2\pi}\!\!e_*^{(s{+}it)(\frac{1}{i\h}u{*}v{+}p)}dt*\! 
\int_0^{2\pi}\!\!e_*^{(s{+}i\tau)\frac{1}{i\h}u{*}v}d\tau \\
&=
u^p{*}\!\!\int_0^{2\pi}\!\!\!e^{(s{+}it)p}e_*^{(s{+}it)\frac{1}{i\h}u{*}v}dt*\! 
\int_0^{2\pi}\!\!e_*^{(s{+}i\tau)\frac{1}{i\h}u{*}v}d\tau{=}
u^p{*}\!\!\int_0^{2\pi}\!\!\!\int_0^{2\pi}\!\!e^{(s{+}it)p}
e_*^{(2s{+}it{+}i\tau)\frac{1}{i\h}u{*}v}dtd\tau.
\end{aligned}
$$
Thus, the changing variables gives that this vanishes by 
$\int_0^{2\pi}\!\!e^{(s{+}it)p}dt=0$.
${}$ \hfill$\Box$ 

\begin{lem}
  \label{vacpolyvac}
For every polynomial $f(u,v){=}\sum a_{ij}u^i{*}v^{j}$,   
$$
\varpi_{00}{*}(f(u,v){*}\varpi_{00}){=}f(0,0)\varpi_{00}
{=}(\varpi_{00}{*}f(u,v)){*}\varpi_{00}.
$$
Consequently, associativity holds for 
$\varpi_{00}{*}f(u,v){*}\varpi_{00}$ for a polynomial $f(u,v)$.  
\end{lem}

By the formal associativity theorem (cf.\cite{ommy5}), we have easily  
$$
(e_*^{su{*}v}{*}v^q){*}(u^{p}{*}e_*^{tu{*}v})
{=}e_*^{su{*}v}{*}(v^q{*}u^{p}{*}e_*^{tu{*}v})
{=}e^{(q-p)ti\h}e_*^{(s{+}t)u{*}v}{*}v^q{*}u^{p}
\quad\mbox{ for } q\geq p,
t$$
$$
(e_*^{su{*}v}{*}v^q){*}(u^{p}{*}e_*^{tu{*}v})
{=}e_*^{su{*}v}{*}(v^q{*}u^{p}{*}e_*^{tu{*}v})
{=}v^q{*}u^{p}{*}e_*^{(s{+}t)u{*}v}{*}e^{(p-q)si\h} 
\quad\mbox{ for } q\leq p.
$$
Replacing $s, t$ by $\frac{1}{i\h}s, \frac{1}{i\h}t$ and 
taking $\lim_{t\to {-}\infty}, \lim_{s\to {-}\infty}$ 
for the case $p\geq q, q\geq p$ respectively,  we have  
\begin{equation}
(\varpi_{00}{*}v^q){*}(u^p{*}\varpi_{00}){=}
\delta_{p,q}p!(i\h)^p{=}
\varpi_{00}{*}(v^q{*}u^p{*}\varpi_{00}){=}
(\varpi_{00}{*}v^q{*}u^p){*}\varpi_{00}.
\end{equation}
Since
$\varpi_{00}{*}v^q{*}u^{p}{*}\varpi_{00}{=}
\delta_{p,q}p!(i\h)^p\varpi_{00}$, we have the following:
\begin{prop}
  \label{vacvac4}
In generic ordered expressions,
$E_{p,q}=
\frac{1}{\sqrt{p!q!(i\h)^{p{+}q}}}u^p{*}\varpi_{00}{*}v^q$ 
is the 
$(p,q)$-matrix element, that is 
$E_{p,q}{*}E_{r,s}=\delta_{q,r}E_{p,s}$. The $K$-expression 
${:}E_{p,q}{:}_{_K}$ of $E_{p,q}$ will be denoted by 
$E_{p,q}(K)$. Note that $E_{0,0}(K){=}{:}\varpi_{00}{:}_{_K}$.
\end{prop}

Similar calculation caring the $\pm$ sign shows also 
\begin{prop}
\label{barvacvac4}
${\overline E}_{p,q}=
\frac{\sqrt{-1}^{p+q}}{\sqrt{p!q!(i\h)^{p{+}q}}}
v^p{*}{\overline{\varpi}}_{00}{*}u^q$ is the 
$(p,q)$-matrix element in generic ordered expressions.
 The $K$-expression of 
$\overline{E}_{p,q}$ will be denoted by $\overline{E}_{p,q}(K)$.
Note that $\overline{E}_{0,0}(K){=}{:}{\overline{\varpi}}_{00}{:}_{_K}$.
\end{prop}

Proposition\,\ref{control} gives  
\begin{equation}\label{Ebar E}
E_{p,q}{*}\overline{E}_{r,s}=0=\overline{E}_{r,s}{*}E_{p,q}.
\end{equation}

\bigskip
By a similar computation, we can consider the idempotent element 
$\varpi_{*}(0){=}\frac{1}{2\pi}\!\!\int_0^{2\pi}\!\!\!
e_*^{it(\frac{1}{i\h}u{\ctt}v)}dt$ in Theorem\,\ref{thepseudovacuum},
called pseudo-vacuum. 
We first recall some formulas which will be used below.  The bumping identity gives  
$$
v{*}(v{*}u){*}u=v{*}(u{\ctt}v{+}\frac{1}{2}i\h){*}u
=(u{\ctt}v{+}\frac{1}{2}i\h){*}(u{\ctt}v{+}\frac{3}{2}i\h).
$$
Repeating this we see that 
\begin{equation}\label{cttctt}
\begin{aligned}
\frac{1}{(i\h)^n}v^n{*}u^n=&(\frac{1}{i\h}u{\ctt}v{+}\frac{1}{2}){*}
(\frac{1}{i\h}u{\ctt}v{+}\frac{3}{2}){*}\cdots{*}
(\frac{1}{i\h}u{\ctt}v{+}\frac{2n{-}1}{2})\\
=&(\frac{1}{i\h}v{*}u){*}
(\frac{1}{i\h}v{*}u{+}1){*}\cdots{*}
(\frac{1}{i\h}v{*}u{+}n{-}1)= \{\frac{1}{i\h}v{*}u\}_{*n},
\end{aligned}
\end{equation}
where $\{A\}_{*n}=A{*}(A{+}1){*}\cdots{*}(A{+}n{-}1)$,
$\{A\}_{*0}=1$. Next formulas are very useful in our computations
$$
\{\frac{1}{i\h}u{*}v\}_{*n}{*}u=u{*}\{\frac{1}{i\h}u{*}v{+}1\}_{*n},\quad 
\{\frac{1}{i\h}u{*}v\}_{*n}{*}v=v{*}\{\frac{1}{i\h}u{*}v{-}1\}_{*n}.
$$

Note that the identity $(u{\ctt}v){*}\varpi_{*}(0)=0$ gives  
\begin{equation}\label{DDD}
(\frac{1}{i\h}u{*}v{+}\ell){*}\varpi_{*}(0)=
(\frac{1}{i\h}u{\ctt}v{+}\ell{-}\frac{1}{2}){*}\varpi_{*}(0)=
(\ell{-}\frac{1}{2})\varpi_{*}(0).
\end{equation}
It follows that 
$$
\begin{aligned}
\varpi_{*}(0){*}(\frac{1}{i\h})^nu^n{*}v^n{*}\varpi_{*}(0)
=&({1}/{2})_{n}\varpi_{*}(0)\\
\varpi_{*}(0){*}(\frac{1}{i\h})^nv^n{*}u^n{*}\varpi_{*}(0)
=&({1}/{2})_{-n}\varpi_{*}(0),
\end{aligned}
$$
where $(a)_n=a(a{+}1)\cdots(a{+}n{-}1)$, $(a)_0=1$ and 
$(a)_{-n}=(a{-}1)(a{-}2)\cdots(a{-}n)$.
If we use the convention \eqref{pmconvention}, then  this is written
by 
\begin{equation}\label{unifiednot}
\varpi_{*}(0){*}(\frac{1}{i\h})^n\hat\zeta^n{*}\zeta^n{*}\varpi_{*}(0)
=(\frac{1}{2})_n\varpi_{*}(0),\quad n\in{\mathbb Z}, 
\end{equation}
where $\zeta^n,~\hat{\zeta}^n$ are given by \eqref{pmconvention}.

\begin{lem}\label{quasivac}
If $K{\in}{\mathfrak K}_0$, then    
${:}e_*^{it(\frac{1}{i\h}u{\ctt}v)}{:}_{_K}$ is
$2\pi$-periodic and 
$$
D_{k,\ell}(K)= 
\frac{1}
{\sqrt{(\frac{1}{2})_{k}(\frac{1}{2})_{\ell}(i\h)^{|k|+|\ell|}}}
{\zeta}^k{*}{:}\varpi_*(0){:}_{_K}{*}{\hat\zeta}^{\ell}, \quad 
{:}\varpi_*(0){:}_{_K}{=}
\frac{1}{2\pi}\!\!\int_0^{2\pi}\!\!\!
{:}e_*^{it(\frac{1}{i\h}u{\ctt}v)}dt{:}_{_K}dt
$$
are matrix elements for every $k,\ell{\in}{\mathbb Z}$. Note that 
$D_{n,n}(K){=}\frac{1}{2\pi}\int_0^{2\pi}{:}e_*^{it(\frac{1}{i\h}u{\ctt}v{-}n)}{:}_{_K}dt$.
\end{lem}

\noindent
{\bf Proof}\,\, Different from the ordinary 
vacuum or bar-vacuum, 
we see $u{*}\varpi_*(0){\not=}0$, 
$v{*}\varpi_*(0){\not=}0$. But note 
that the bumping identity gives 
\begin{equation}\label{slide}
u^n{*}e_*^{it(\frac{1}{i\h}u{\ctt}v)}{=}
e_*^{it(\frac{1}{i\h}u{\ctt}v{-}n)}{*}u^n,
 \quad 
v^n{*}e_*^{it(\frac{1}{i\h}u{\ctt}v)}{=}
e_*^{it(\frac{1}{i\h}u{\ctt}v{+}n)}{*}v^n.
\end{equation}
Moreover, if $k{\not=}\ell$, then the exponential law 
and the change of variables gives 
$$
\int_{0}^{2\pi}
e_*^{is(\frac{1}{i\h}u{\ctt}v{+}k)}ds
*\!\int_{0}^{2\pi}
e_*^{it(\frac{1}{i\h}u{\ctt}v{+}\ell)}dt
{=}
\int_{0}^{2\pi}e^{it(k{-}\ell)}dt
\!\int_{0}^{2\pi}
e_*^{is(\frac{1}{i\h}u{\ctt}v{+}\ell)}ds{=}0, 
$$ 
and 
$$
\frac{1}{2\pi}
\int_{0}^{2\pi}
e_*^{it(\frac{1}{i\h}u{\ctt}v{+}k)}dt
{*}
\frac{1}{2\pi}
\int_{0}^{2\pi}
e_*^{is(\frac{1}{i\h}u{\ctt}v{+}k)}ds
{=}\frac{1}{2\pi}\int_{0}^{2\pi}
e_*^{it(\frac{1}{i\h}u{\ctt}v{+}k)}dt.
$$
Theorem\,\ref{thepseudovacuum}, \eqref{circlezero}  and \eqref{shiftvacuum} 
show that the $*$-product 
$P(u,v){*}\varpi_*(0){*}Q(u,v)$ by any polynomials 
$P(u,v)$, $Q(u,v)$ is reduced to the shape 
$\phi{*}\varpi_*(0){*}\psi$ where $\phi$, $\psi$ are 
polynomials of single variable $u$ or $v$.

Using the above formula, \eqref{unifiednot}, \eqref{slide},  
we have the desired result. \hfill ${\Box}$

\medskip
Since $\frac{1}{2\pi}\int_0^{2\pi} 
{:}e_*^{is(\frac{1}{i\h}u{\ctt}v-n)}{:}_{_K}ds =D_{n,n}(K)$, 
the Fourier expansion of 
$e_*^{it\frac{1}{i\h}u{\ctt}v}$ is written as  
\begin{equation}\label{FFexpand}
{:}e_*^{it\frac{1}{i\h}u{\ctt}v}{:}_{_K}
=\frac{1}{2\pi}\sum_n\int_0^{2\pi}{:}e_*^{is(\frac{1}{i\h}u{\ctt}v-n)}{:}_{_K}ds
e^{int}
{=}\sum_{n{\in}\mathbb Z}D_{n,n}(K)e^{int},\quad
K{\in}{\mathfrak K}_0.
\end{equation}
Hence we see by the exponential law for every $\alpha$, the Fourier
series  
\begin{equation}\label{FFexpansion}
{:}e_*^{it(\frac{1}{i\h}u{\ctt}v{+}\alpha)}{:}_{_K}
=\sum_{n{\in}\mathbb Z}D_{n,n}(K)e^{it(n{+}\alpha)},\quad K{\in}{\mathfrak K}_0
\end{equation}
converges uniformly1y in $C^{\infty}(S^1,H\!ol({\mathbb C}^2))$,  
the $C^{\infty}$-topology of the space of $H\!ol({\mathbb C}^2)$-valued
smooth functions on $S^1$. 

\subsection{Matrix representations}
Now back to the case $\varpi_{00}$, we note that one can set $w{=}e^t$ in the l.h.s. of 
\eqref{genericparam00}.
As $u{*}v=u{\ctt}v{-}\frac{1}{2}i\h$, Proposition\,\,\ref{existvacuum} gives 
the convergence 
$$
\lim_{w\to 0}{:}e_*^{\log w\frac{1}{i\h}2u{*}v}{:}_{_{K}}
={:}e_*^{\log 0\frac{1}{i\h}2u{*}v}{:}_{_{K}}
{=}{:}\varpi_{00}{:}_{_K}.
$$
Hence, we get a holomorphic function of $w$ defined on an open 
neighborhood of $w=0$. 
$$
f_{_{K}}(w)={:}e_*^{\log w\frac{1}{i\h}u{*}v}{:}_{_{K}}.
$$
Using the bumping identity, we have
$$
\partial_{w}\big|_0e_*^{(\log w)\frac{1}{i\h}u{*}v}{=}
\lim_{t{\to}-\infty}e^{-t}\partial_t
e_*^{t\frac{1}{i\h}u{*}v}
{=}\lim_{t{\to}-\infty}
\frac{1}{i\h}u{*}v{*}
e_*^{t(\frac{1}{i\h}u{*}v{-}1)}
{=}\lim_{t{\to}{-}\infty}
\frac{1}{i\h}u{*}
e_*^{t(\frac{1}{i\h}u{*}v)}{*}v.
$$
Repeating this procedure, we have the following remarkable 
formulas:
\begin{equation}
\frac{1}{n!}f_{_{K}}^{(n)}(0){=}
\frac{1}{n!(i\h)^n}{:}u^n{*}\varpi_{00}{*}v^n{:}_{_{K}}.
\end{equation}
The convergence of Taylor series gives 
$f_{_K}(w){=}\frac{w^n}{n!}f_{_K}^{(n)}(0)$. 
These are in $(n,n)$-matrix elements, denoted by 
${:}E_{n,n}{:}_{_{K}}$. 
It follows in generic ordered expressions,  
$$
e_*^{\log w\frac{1}{i\h}u{*}v}=
\sum_{n=0}^{\infty}\frac{1}{n!}
(\frac{1}{i\h}u{*}v)^n{*}e^{-\frac{1}{i\h}u{*}v}w^n
=\sum_{n=0}^{\infty}
\frac{1}{n!(i\h)^n}u^n{*}\varpi_{00}{*}v^n w^n
=\sum_{n\geq0}w^nE_{n,n}
$$
on a neighborhood of $w=0$ depending on the expression 
parameter $K$. 
In what follows we use notations 
\begin{equation}\label{EE}
{:}\frac{1}{n!(i\h)^n}u^n{*}\varpi_{00}{*}v^n{:}_{_K}
={:}E_{n,n}{:}_{_K}=E_{n,n}(K).
\end{equation}
In particular, 
$$
{:}E_{0,0}{:}_{_K}=E_{0,0}(K)={:}\varpi_{00}{:}_{_K}
$$

Here we apply the following general fact:
\begin{lem}
If $f(z)$ is a holomorphic mapping from a complex open 
disk $D(R)$ of radius $R$ centered at $0$  
into a Fr{\'e}chet space over 
${\mathbb C}$, then the Taylor series of $f(z)$ at $z{=}0$ 
converges uniformly on any closed disk of radius $r$, $0<r<R$.   
\end{lem}

\begin{prop}\label{nice}
If the singular points of 
${:}e_*^{t\frac{1}{i\h}u{*}v}{:}_{_K}$ lies on the open 
right half plane, then we have in the space 
$H\!ol({\mathbb C}^2)$ 
$$
1{=}\sum_{n=0}^{\infty}{:}E_{n,n}{:}_{_K},\quad 
{:}e_*^{\tau\frac{1}{i\h}u{*}v}{:}_{_K}=
\sum_{n=0}^{\infty}e^{n\tau}{:}E_{n,n}{:}_{_K},\quad 
({\rm{Re}}\,\tau\leq 0).
$$
Since $\frac{1}{i\h}u{*}v=u{\ctt}v{-}\frac{1}{2}$, the second equality
may be written as 
$$
{:}e_*^{\tau\frac{1}{i\h}u{\ctt}v}{:}_{_K}=
\sum_{n=0}^{\infty}e^{\tau(n{+}\frac{1}{2})}E_{n,n}(K),\quad 
({\rm{Re}}\,\tau\leq 0).
$$
\end{prop}

\bigskip
Similarly, set $w=e^{t}$ in the l.h.s. of 
\eqref{genericparam00} 
by noting 
$v{*}u=u{\ctt}v{+}\frac{1}{2}i\h$.
Proposition\,\,\ref{existvacuum} gives 
the convergence  
$$
\lim_{w\to\infty}{:}e_*^{\log w\frac{1}{i\h}2v{*}u}{:}_{_{K}}
$$
which we called a {\bf bar-vacuum} and denoted by 
${\overline{\varpi}}_{00}$.
Hence we get a holomorphic function of $\hat{w}=w^{-1}$ 
defined on a neighborhood of $\hat{w}=0$ depending on 
$K$. Set as follows:
$$
g_{_{K}}(\hat w)=
{:}e_*^{-\log\hat{w}\frac{1}{i\h}v{*}u}{:}_{_{K}}
$$
Using the bumping identity, we have
$$
\partial_{\hat{w}}\big|_0
e_*^{-(\log\hat{w})\frac{1}{i\h}v{*}u}{=}
-\lim_{t{\to}\infty}e^{t}\partial_t
    e_*^{t\frac{1}{i\h}v{*}u}
{=}-\lim_{t{\to}\infty}
\frac{1}{i\h}v{*}u{*}
e_*^{t(\frac{1}{i\h}v{*}u{+}1)}
{=}-\lim_{t{\to}\infty}\frac{1}{i\h}v{*}
e_*^{t(\frac{1}{i\h}v{*}u)}{*}u.
$$
Repeating this procedure, we have the following 
remarkable formulas:
\begin{equation}\label{EE2}
\frac{1}{n!}g_{_{K}}^{(n)}(0){=}
\frac{(-1)^n}{n!(i\h)^n}{:}v^n{*}
{\overline{\varpi}}_{00}{*}u^n{:}_{_{K}}=
{:}\overline{E}_{n,n}{:}_{_K} (=\overline{E}_{n,n}(K)).
\end{equation}

It follows in generic ordered expression 
$$
e_*^{-\log\hat w\frac{1}{i\h}v{*}u}= 
\sum_{n=0}^{\infty}
\Big(\frac{(-1)^n}{n!(i\h)^n}
{:}v^n{*}{\overline{\varpi}}_{00}{*}u^n{:}_{_{K}}\Big)\hat{w}^{n},\quad
\hat w=w^{-1}. 
$$

\begin{thm}\label{maeda}
Let $I_{\ctt}(K)=[a,b]$ be the exchanging interval. 
Then the radius of convergence of the Taylor series in the Fr{\'e}chet space 
$H\!ol({\mathbb C}^2)$ of 
$$
f_{_K}(w){=}
\sum \frac{1}{n!}f_{_K}^{(n)}(0)w^n,\quad
({\rm{resp.}}\,\, g_{_K}(\hat w){=}
\sum \frac{1}{n!}g_{_K}^{(n)}(0){\hat w}^n, 
\quad \hat{w}{=}w^{-1})
$$
is $e^a$ $($resp. $e^{-b}$$)$. 
If $K{\in}{\mathfrak K_+}$ $({\rm{resp.}}\,\,{\mathfrak K}_-)$, then 
the radius of convergence is bigger than $1$.
\end{thm}

\begin{prop}\label{surprise}
In the $K$-ordered expression for $K{\in}{\mathfrak K_+}$, we have 
$$
1{=}
\sum_{n{=}0}^{\infty}E_{n,n}(K),\quad
{:}e_*^{\frac{\tau}{i\h}(u{\ctt}v{+}\lambda)}{:}_{_K}
{=}
\sum_{n{=}0}^{\infty} 
e^{\tau(n{+}\frac{1}{2})+\tau\lambda}E_{n,n}(K),\quad 
({\rm{Re}}\,\,\tau\leq 0).
$$
The former converges in the space $H\!ol({\mathbb C}^2)$.
If ${\rm{Re}}\,\tau\leq 0$, the latter converges uniformly 
on every compact subset w.r.t. $\tau$.
Similarly, if $K{\in}{\mathfrak K_-}$, then 
$$
1{=}\sum_{n=0}^{\infty}\overline{E}_{n,n}(K),\quad 
{:}e_*^{\frac{\tau}{i\h}(u{\ctt}v{+}\lambda)}{:}_{_K}
{=}
\sum_{n{=}0}^{\infty} 
e^{\tau(-n{-}\frac{1}{2})+\tau\lambda}\overline{E}_{n,n}(K),\quad 
({\rm{Re}}\,\,\tau\geq 0). 
$$
The former converges in the space $H\!ol({\mathbb C}^2)$.
If ${\rm{Re}}\,\tau\geq 0$, the latter converges uniformly 
on every compact subset w.r.t. $\tau$.
\end{prop}

Note that the conditions $K{\in}{\mathfrak K}_{+}$  and ${\rm{Re}}\,\,\tau\leq 0$
are used only to ensure that $\sum_{n{=}0}^{\infty}E_{n,n}(K)$ and  
$\sum e^{\tau(n{+}\frac{1}{2})+\tau\lambda}E_{n,n}(K)$ 
in the r.h.s.\! converge in the topology of 
$H\!ol({\mathbb  C}^2)$. However, each term of 
$\sum_{n{=}0}^{\infty} 
e^{\tau(n{+}\frac{1}{2})+\tau\lambda}{E}_{n,n}(K)$ is 
an entire function of $\tau$, although 
${:}e_*^{\tau\frac{1}{i\h}u{\ctt}v{+}\tau\lambda}{:}_{_K}$ may be singular at
some $\tau{=}\tau_0$. 
Singular point $\tau{=}\tau_0$ means simply that 
$\sum_{n{=}0}^{\infty} 
e^{\tau_0(n{+}\frac{1}{2})+\tau_0\lambda}{E}_{n,n}(K)$ diverges in 
the topology of $H\!ol({\mathbb C}^2)$.

\medskip
Note that without conditions such as 
$K\in {\mathfrak K}_{+}$  and ${\rm{Re}}\,\,t\leq 0$, 
${:}e_*^{\log w\frac{1}{i\h}u{*}v}{:}_{_{K}}$ is holomorphic on a
neighborhood of $w=0$. Hence setting  
$e_*^{c\frac{1}{i\h}u{*}v}{*}e_*^{(z{-}c)\frac{1}{i\h}u{*}v}$ by 
choosing $c$ sufficiently large, the Taylor series 
$e_*^{(z{-}c)\frac{1}{i\h}u{*}v}=\sum_{n\geq 0}e^{(z-c)(n{+}\frac{1}{2})}E_{n,n}$
of $e_*^{(z{-}c)\frac{1}{i\h}u{*}v}$ at $w=0$ converges in $H\!ol({\mathbb C}^2)$.
Hence, applying $e_*^{c\frac{1}{i\h}u{*}v}$, we have 
$$
e_*^{z\frac{1}{i\h}u{*}v}=
e_*^{c\frac{1}{i\h}u{*}v}{*}\sum_{n\geq 0}e^{(z-c)(n{+}\frac{1}{2})}E_{n,n}.
$$
Although $e_*^{c\frac{1}{i\h}u{*}v}{*}$ of r.h.s. is not continuous
in $H\!ol({\mathbb C}^2)$, the componentwise computation of matrices 
 may be applied to yield    
\begin{equation*}
e_*^{z\frac{1}{i\h}u{*}v}=e_*^{c\frac{1}{i\h}u{*}v}{*}
\sum_{n\geq 0}e^{(z-c)(n{+}\frac{1}{2})}E_{n,n}=
\sum_{n\geq 0}e^{z(n{+}\frac{1}{2})}E_{n,n}
\end{equation*}
as $\frac{1}{i\h}u{*}v*E_{n,n}=nE_{n,n}$ gives 
$e_*^{c\frac{1}{i\h}u{*}v}{*}E_{n,n}=e^{cn}E_{n,n}$. We denote this by 
\begin{equation}\label{asmatrix}
{:}e_*^{z\frac{1}{i\h}u{*}v}{:}_{_{E(K)mat}}=
\sum_{n\geq 0}e^{z(n{+}\frac{1}{2})}E_{n,n}(K).
\end{equation}

Similarly, we have 
\begin{equation*}
e_*^{z\frac{1}{i\h}v{*}u}=e_*^{-c\frac{1}{i\h}v{*}u}{*}
\sum_{n\geq 0}e^{-(z+c)(n{+}\frac{1}{2})}\overline{E}_{n,n}=
\sum_{n\geq 0}e^{-z(n{+}\frac{1}{2})}\overline{E}_{n,n}
\end{equation*}
and this is denoted by 
\begin{equation}\label{asmatrix00}
{:}e_*^{z\frac{1}{i\h}v{*}u}{:}_{_{\overline{E}(K)mat}}=
\sum_{n\geq 0}e^{-z(n{+}\frac{1}{2})}\overline{E}_{n,n}(K).
\end{equation}

\subsubsection{Representations by Laurent expansions}

If $a<{\rm{Re}}\,z<b$, then setting $w=e^z$ in \eqref{genericparam00},  
Laurent expansion 
of ${:}e_*^{\log w\frac{1}{i\h}u{\ctt}v}{:}_{_K}$ by $w$ gives that 
\begin{equation}\label{Fexp}
{:}e_*^{z(\frac{1}{i\h}u{\ctt}v)}{:}_{_K}
=\sum_{\ell=-\infty}^{\infty}\tilde{D}_{\ell,\ell}(K)\,\,e^{z\ell},\quad 
a<{\rm{Re}}\,z<b 
\end{equation}
and the r.h.s. converges in $H\!ol({\mathbb C}^2)$. Note that 
$$
\tilde{D}_{n,n}(K)=
\frac{1}{2\pi}
\int_0^{2\pi}{:}e_*^{(s{+}it)(\frac{1}{i\h}u{\ctt}v)}{:}_{_K}e^{-(s{+}it)n}dt,\quad 
a<s<b.
$$
Cauchy's integral theorem shows that $\tilde{D}_{n,n}(K)$ is
independent of $s$ whenever $s\in I_{\ctt}(K)$.

\bigskip
In particular, if $a<0<b$, we see that
$\tilde{D}_{n,n}(K){=}{D}_{n,n}(K)$ the $(n,n)$
diagonal matrix element given in Lemma\,\ref{quasivac}. Hence   
\begin{prop}\label{that'sall} 
If $K\in {\mathfrak K}_0$, then $\tilde{D}_{n,n}(K)$ is an $(n,n)$
diagonal matrix element and 
for $\tau$ such that $a<{\rm{Re}}\,\tau<b$,
$$
{:}e_*^{\tau\frac{1}{i\h}(u{\ctt}v{+}\lambda)}{:}_{_K}{=}
\sum_{n=-\infty}^{\infty}e^{\tau n{+}\tau\lambda}\tilde{D}_{n,n}(K),\quad
\lambda{\in}{\mathbb C},
$$ 
and the r.h.s. converges in $H\!ol({\mathbb C}^2)$.
In particular, 
$1{=}\sum_{n=-\infty}^{\infty}\tilde{D}_{n,n}(K)$.
\end{prop} 
Although $\tilde{D}_{n,n}(K){*_{_K}}\tilde{D}_{n,n}(K)$ is
not defined in general, we show in what follows that  
$\tilde{D}_{n,n}(K)$ may be viewed as a diagonal matrix element.

Note now 
$$
\frac{1}{i\h}(u{\ctt}v){*}\int_{0}^{2\pi}
e_*^{(s{+}i\tau)\frac{1}{i\h}u{\ctt}v}e^{-(s{+}i\tau)n}d\tau
=n\int_{0}^{2\pi}e_*^{(s{+}i\tau)\frac{1}{i\h}u{\ctt}v}e^{-(s{+}i\tau)n}d\tau.
$$
It follows that 
${:}e_*^{-s\frac{1}{i\h}u{\ctt}v}{:}_{_K}{*_{_K}}\tilde{D}_{n,n}(K)
=e^{-ns}\tilde{D}_{n,n}(K)$. Hence one may write 
$$
{:}e_*^{it\frac{1}{i\h}u{\ctt}v}{:}_{_K}{=}
\sum_{n=-\infty}^{\infty}\tilde{D}_{n,n}(K)e^{int}
$$
though r.h.s. does not converge in $H\!ol({\mathbb C}^2)$. 
To avoid the confusion, we denote this by 
\begin{equation}\label{DKmat}
{:}e_*^{it\frac{1}{i\h}u{\ctt}v}{:}_{_{D(K)mat}}{=}
\sum_{n=-\infty}^{\infty}\tilde{D}_{n,n}(K)e^{int}
\end{equation}
and regard this as a formal element.
We see that 
$\tilde{D}_{n,n}(K){=}\frac{1}{2\pi}\int_{0}^{2\pi}
{:}e_*^{it\frac{1}{i\h}u{\ctt}v}{:}_{_{D(K)mat}}e^{-itn}dt$. 
Suppose the exponential law holds formally on the l.h.s., then one
must set 
$$
\tilde{D}_{m,m}(K){*_{mat}}\tilde{D}_{n,n}(K){=}\delta_{m,n}\tilde{D}_{n,n}(K)
$$
where ${*_{mat}}$ indicates the formal product which allows the formal
exponential low.  By this observation, we define  
${:}e_*^{it\frac{1}{i\h}u{\ctt}v}{:}_{_{D(K)mat}}$ as 
a diagonal matrix, and $\tilde{D}_{n,n}(K)$ as a diagonal matrix element.

\bigskip
By \eqref{asmatrix}, \eqref{asmatrix00} and \eqref{DKmat},
$e_*^{it\frac{1}{i\h}u{\ctt}v}$ is expressed as diagonal matrices,
which will be called {\bf diagonal matrix expressions}.

\subsubsection{Fourier series of alternating $2\pi$-periodic functions}
Beside these, we note that the matrix representation via Taylor
expansion in the previous section can be obtained by using the Fourier
expansion along a closed curve parallel to the pure imaginary axis
(cf.\eqref{defvacuum}). 

Every  $4\pi$-periodic function $f(\theta)$ is written as the sum of
a $2\pi$-periodic and a alternating $2\pi$-periodic functions: 
$$
f(\theta)=f_0(\theta){+}f_{-}(\theta),\quad 
f_0(\theta){=}\frac{1}{2}(f(\theta){+}f(\theta{+}2\pi)),\quad 
f_{-}(\theta){=}\frac{1}{2}(f(\theta){-}f(\theta{+}2\pi)).
$$ 
The Fourier series of $f(\theta)$ is given as 
$$
\begin{aligned}
f(\theta)=&\sum_{n}\frac{1}{4\pi}\int_0^{4\pi}f(t)e^{-\frac{1}{2}in t}dt
e^{i\frac{n}{2}\theta}\\
=&\frac{1}{4\pi}\sum_{n}\frac{1}{4\pi}\int_0^{4\pi}f_0(t)e^{-\frac{1}{2}i2nt}dte^{in\theta}+
\sum_{n}\frac{1}{4\pi}\int_0^{4\pi}f_{-}(t)
e^{-\frac{1}{2}i(2n{+}1)t}dte^{i(n{+}\frac{1}{2})\theta}.
\end{aligned}
$$

As the periodicity of ${:}e_*^{(s{+}it)\frac{1}{i\h}u{\ctt}v}$
w.r.t. $t$ depends on $s$, we have to use Fourier basis 
depending on $s$. Hence we have 
\begin{thm}\label{unifiedrep} 
$e_*^{(s{+}it)(\frac{1}{i\h}u{\ctt}v{+}z)}$ is expressed by diagonal
matrices 
$$
\begin{aligned}
\begin{matrix}
\medskip
{:}e_*^{(s{+}it)(\frac{1}{i\h}u{\ctt}v{+}z)}{:}_{_{E(K)mat}}=
\sum_{k=0}^{\infty}E_{k,k}(K)e^{(s{+}it)(k{+}\frac{1}{2}{+}z)},& 
\quad z{\in}{\mathbb C}\\
\medskip
{:}e_*^{(s{+}it)(\frac{1}{i\h}u{\ctt}v{+}z)}{:}_{_{D(K)mat}}=
\sum_{n=-\infty}^{\infty}\tilde{D}_{n,n}(K)e^{(s{+}it)(n{+}z)},&\quad z{\in}{\mathbb C}\\
\medskip
{:}e_*^{(s{+}it)(\frac{1}{i\h}u{\ctt}v{+}z)}{:}_{_{\overline{E}(K)mat}}=
\sum_{k=0}^{\infty}\overline{E}_{k,k}(K)e^{(s{+}it)({-}k{-}\frac{1}{2}{+}z)},&
\quad z{\in}{\mathbb C}\\
\end{matrix}
\end{aligned}
$$
where ${E}_{k,k}(K)$, $\overline{E}_{k,k}(K)$ equal to the ones given by \eqref{EE}, \eqref{EE2}.
Let $I_{\ctt}(K)=[a,b]$ be the exchanging interval in generic ordered
expression $K$. Then, these converge 
for $s<a$, $a<s<b$ and $b<s$  respectively in
$C^{\infty}(S^1,H\!ol({\mathbb C}^2))$ 
$($$H\!ol({\mathbb C}^2)$-valued smooth 
functions on $S^1$ with $C^\infty$-topology$)$.   
\end{thm}

For every Schwartz distribution $f(t)$ on $S^1$, the integral  
$$
{:}\tilde{f}_s(z{+}\frac{1}{i\h}u{\ctt}v){:}_{_K}
=\int_{S^1}f(t){:}e_*^{(s{+}it)(z{+}\frac{1}{i\h}u{\ctt}v)}{:}_{_K}dt,\quad
\text{for generic}\quad K 
$$
gives an element of $H\!ol({\mathbb C}^2)$. $\tilde{f}_s(z{+}\frac{1}{i\h}u{\ctt}v)$ has a
diagonal matrix expression as follow:
$$
\begin{matrix}
\medskip
{:}\tilde{f}_s(z{+}\frac{1}{i\h}u{\ctt}v){:}_{_{E(K)mat}}=
\sum_{n=0}^{\infty}\int_{S^1}f(t)e^{(s{+}it)(z{+}\frac{1}{2}{+}n)}dtE_{n,n}(K),\\
\medskip
{:}\tilde{f}_s(z{+}\frac{1}{i\h}u{\ctt}v){:}_{_{D(K)mat}}=
\sum_{n=-\infty}^{\infty}\int_{S^1}f(t)e^{(s{+}it)(z{+}n)}dt\tilde{D}_{n,n}(K),\\
\medskip
{:}\tilde{f}_s(z{+}\frac{1}{i\h}u{\ctt}v){:}_{_{\overline{E}(K)mat}}=
\sum_{n=0}^{\infty}\int_{S^1}f(t)e^{(s{+}it)(z{-}\frac{1}{2}{-}n)}dt\overline{E}_{n,n}(K)
\end{matrix}
$$
The series of r.h.s. converge respectively $s<a$, $a<s<b$, $b<s$ 
 in the space $H\!ol({\mathbb C}^2)$, 
where $(a,b){=}I_{\ctt}(K)$ is  the exchanging interval of 
${:}e_*^{\tau\frac{1}{i\h}u{\ctt}v}{:}_{_K}$. 

More systematic treatment of the diagonal matrix
  expressions will be given in the next paper 

\subsection{Diagonal matrix expressions and applications}

Note that in Theorem\,\ref{unifiedrep} one may set $s=0$, when 
$0<a$, $a<0<b$ and $b<0$ respectively. 
Thus, by differentiating $k$-times and by setting $s{+}it{=}0$ in each
case, an element such as  
$(z{+}\frac{1}{i\h}u{\ctt}v)_*^{k}$, $k\geq 0$, 
has three expressions as diagonal matrices 
in the space $H\!ol({\mathbb C}^2)$: 
\begin{equation}\label{polymatrixrep}
{:}(z{+}\frac{1}{i\h}u{\ctt}v)_*^{k}{:}_{_K}
{=}
\left\{
\begin{matrix}
\medskip
\sum_{n=0}^{\infty}(z{+}n{+}\frac{1}{2})^{k}E_{n,n}(K),\quad K\in {\mathfrak K}_+\\
\medskip
\sum_{n=-\infty}^{\infty}(z{+}n)^{k}D_{n,n}(K),\quad K\in {\mathfrak K}_0\\
\medskip
\sum_{n=0}^{\infty}(z{-}n{-}\frac{1}{2})^{k}\overline{E}_{n,n}(K),\quad K\in {\mathfrak K}_-
\end{matrix}
\right.
\end{equation}  
depending on expression parameters. As it is noted, each of them 
converges  in $H\!ol({\mathbb C}^2)$. 

\bigskip
Consider now the componentwise calculation for r.h.s.. 
We define matrices for every $k\in {\mathbb Z}$ by 
\begin{equation}\label{invinvinv}
{:}(z{+}\frac{1}{i\h}u{\ctt}v)_*^{k}{:}_{Kmat}{=}
\left\{
\begin{matrix}
\medskip
\sum_{n=0}^{\infty}(z{+}n{+}\frac{1}{2})^{k}E_{n,n}(K),\quad K\in {\mathfrak K}_+\\
\medskip
\sum_{n=-\infty}^{\infty}(z{+}n)^{k}D_{n,n}(K),\quad K\in {\mathfrak K}_0\\
\medskip
\sum_{n=0}^{\infty}(z{-}n{-}\frac{1}{2})^{k}\overline{E}_{n,n}(K),\quad K\in {\mathfrak K}_-
\end{matrix}
\right.
\end{equation}
but we often omit the suffix ${K}$ if the expression
parameter is not strictly specified and denote simply by 
by $(z{+}\frac{1}{i\h}u{\ctt}v)^k_{mat}$.
The goal of this section is the following theorem:
\begin{thm}\label{analconti}
If $k=-1$, the r.h.s.of \eqref{invinvinv} defines respectively a holomorphic
mapping as follows:

\medskip
\noindent
(1) If $K{\in}{\mathfrak K}_+$, then 
$\sum_{n=0}^{\infty}(z{+}n{+}\frac{1}{2})^{-1}E_{n,n}(K)$ is a
 holomorphic mapping of 
${\mathbb C}{\setminus}
\{{-}(\mathbb N{+}\frac{1}{2})\}$ into $H\!ol({\mathbb C}^2)$, 

\medskip
\noindent
(2) If $K{\in}{\mathfrak K}_0$, then 
 $\sum_{n=-\infty}^{\infty}(z{+}n)^{-1}D_{n,n}(K)$ is a
 holomorphic mapping of 
${\mathbb C}{\setminus}\mathbb Z$ into $H\!ol({\mathbb C}^2)$, 

\medskip
\noindent
(3) If $K{\in}{\mathfrak K}_-$, then 
 $\sum_{n=0}^{\infty}(z{-}n{-}\frac{1}{2})^{-1}\overline{E}_{n,n}(K)$ is a
 holomorphic mapping of 
${\mathbb C}{\setminus}\{\mathbb N{+}\frac{1}{2}\}$ into $H\!ol({\mathbb C}^2)$. 
\end{thm}

\bigskip
Now suppose $K\in{\mathfrak K}_{+}$. Theorem\,\ref{unifiedrep} gives 
${:}e_*^{s(z{+}\frac{1}{i\h}u{\ctt}v)}{:}_{_K}=
\sum_{k\geq 0}E_{k,k}(K)e^{s(z{+}k{+}\frac{1}{2})}$.  
If ${\rm{Re}}\,z{>}{-}\frac{1}{2}$, then the termwise integration
gives 
$$
\int_{-\infty}^0{:}e_*^{s(z{+}\frac{1}{i\h}u{\ctt}v)}{:}_{_K}ds 
{=}
\sum_{k=0}^{\infty}\int_{-\infty}^0E_{k,k}(K)
    e^{s(z{+}k{+}\frac{1}{2})}ds{=}\sum_{k=0}^{\infty}E_{k,k}(K)(z{+}k{+}\frac{1}{2})^{-1}
$$
As it is assumed in generic ordered expression,
$e_*^{s\frac{1}{i\h}u{\ctt}v}$ has no singular point on $s{\in}{\mathbb R}$, 
and $e^{-\frac{1}{2}|s|}$-growth in a generic ordered expression.

To extend this to the domain ${\rm{Re}}\,z{>}{-}n{-}\frac{1}{2}$,
we subtract first divergent terms 
$$
\sum_{k\geq n}E_{k,k}(K)e^{s(z{+}k{+}\frac{1}{2})}
={:}e_*^{s(z{+}\frac{1}{i\h}u{\ctt}v)}{:}_{_K}
{-}\sum_{k=0}^{n-1}E_{k,k}(K)e^{s(z{+}k{+}\frac{1}{2})},
$$
 and set on the domain ${\rm{Re}}\,z{>}{-}n{-}\frac{1}{2}$ 
$$
\sum_{k=0}^{\infty}(z{+}k{+}\frac{1}{2})^{-1}E_{k,k}(K)=
\sum_{k=0}^{n-1}(z{+}k{+}\frac{1}{2})^{-1}E_{k,k}(K){+}
\int_{-\infty}^0(\sum_{k\geq n}E_{k,k}(K)e^{s(z{+}k{+}\frac{1}{2})})ds
$$
The r.h.s. converges in $H\!ol({\mathbb C}^2)$ in the sense of partial
fractions for 
${\rm{Re}}\,z{>}{-}n{-}\frac{1}{2}$
$$
\int_{-\infty}^0(\sum_{k\geq n}E_{k,k}(K)e^{s(z{+}k{+}\frac{1}{2})})ds{=}
\sum_{k\geq n}E_{k,k}(K)(z{+}k{+}\frac{1}{2})^{-1}
$$  
under the assumption $K\in{\mathfrak K}_{+}$.
Hence we have $(1)$ of Theorem\,\ref{analconti}. 
A similar proof gives $(3)$ of Theorem\,\ref{analconti}.

\bigskip
For $(2)$ of Theorem\,\ref{analconti}, we show the next one:
\begin{prop}\label{existence}
If $K\in{\mathfrak K}_0$, then
$$
D_{K}^{-1}(z{+}\frac{1}{i\h}u{\ctt}v)=\sum_{n=-\infty}^{\infty}\frac{1}{z{+}n}D_{n,n}(K)
$$
defines a holomorphic mapping from  $z\in{\mathbb C}{\setminus}{\mathbb Z}$
into $H\!ol({\mathbb C}^2)$, and this gives an inverse of 
$z{+}\frac{1}{i\h}u{\ctt}v$ i.e.
$$
{:}(z{+}\frac{1}{i\h}u{\ctt}v){:}_{_K}{*_{_K}}D_{K}^{-1}(z{+}\frac{1}{i\h}u{\ctt}v)=1,
\quad z\in{\mathbb C}{\setminus}{\mathbb Z}.
$$
\end{prop}

\noindent
{\bf Proof}\,\,\, Note that  $D_{n,n}(K),\,\, (z{+}n)D_{n,n}(K),\,\, 
\frac{1}{z{+}n}D_{n,n}(K) \in H\!ol({\mathbb C}^2)$ for
every $n$. For every $z{\in}{\mathbb C}$, there is an integer $n(z)$
such that $n(z){-}1<{\rm{Re}}\,z <n(z){+}1$. (This may not be unique.) 

As $e_*^{t(z{+}\frac{1}{i\h}u{\ctt}v)}=\sum_n e^{t(z{+}n)}D_{n,n}(K)$
in the topology of $H\!ol({\mathbb C}^2)$, 
we set as follows
$$
e_*^{t(z{+}\frac{1}{i\h}u{\ctt}v)}{=}
\!\!\!\sum_{n\leq n(z){-}1}\!\!\!e^{t(z{+}n)}D_{n,n}(K){+}\!\!\sum_{n(z){-}1<n<n(z){+}1}
\!\!\!\!\!\!e^{t(z{+}n)}D_{n,n}(K)
{+}\!\! \sum_{n\geq n(z){+}1}\!\!\!e^{t(z{+}n)}D_{n,n}(K).
$$
These three terms are members of $H\!ol({\mathbb  C}^2)$. 

Consider inverses of each three term of the r.h.s.. The inverses of
the first and the third term can be replaced by using integrals. Hence   
$D_{K}^{-1}(z{+}\frac{1}{i\h}u{\ctt}v)$ is written as follows: 
$$
-\int_{0}^{\infty}\!\!\!\!\!\sum_{n\leq  n(z){-}1}\!\!\!\!\!\!e^{t(z{+}n)}dtD_{n,n}(K){+}
\!\!\sum_{n(z){-}1<n<n(z){+}1}
\!\!\!\!\!\!\!\!\!\!(z{+}n)^{-1}D_{n,n}(K)
{+}\int_{-\infty}^0\!\sum_{n\geq n(z){+}1}\!\!\!\!e^{t(z{+}n)}D_{n,n}(K).
$$
By Theorem\,\ref{unifiedrep}, this converges in $H\!ol({\mathbb  C}^2)$ to give 
$\sum_{n=-\infty}^{\infty}\frac{1}{z{+}n}D_{n,n}(K)$.\hfill $\Box$

By the definition, we see 
\begin{prop}
${\rm{Res}}((D_{_K}^{-1}(z{+}\frac{1}{i\h}u{\ctt}v); n)= D_{n,n}(K)$,
$n\in{\mathbb Z}$.
\end{prop}

\bigskip
\noindent
{\bf Note for $(1)$ and $(3)$}.\,\,Besides the concrete form of inverses,  
next two integrals 
$$
(z{+}\frac{1}{i\h}u{\ctt}v)_{+*}^{-1}{=}
\int_{-\infty}^0e_*^{s(z{+}\frac{1}{i\h}u{\ctt}v)}ds \quad 
({\rm{Re}}\,z{>}-\frac{1}{2}), 
$$
$$
(z{+}\frac{1}{i\h}u{\ctt}v)_{-*}^{-1}{=}
-\int_0^{\infty}e_*^{s(z{+}\frac{1}{i\h}u{\ctt}v)}ds \quad 
({\rm{Re}}\,z{<}\frac{1}{2}). 
$$
converges in generic ordered expression on each domain, and these give  
inverses of  $z{+}\frac{1}{i\h}u{\ctt}v$ respectively. 
In the next section we give analytic continuations of these.  

\subsubsection{Analytic continuation of inverses}
\label{analconi}

Using the half-inverse $v^{\ctt}$ given by \eqref{half-inv},  
we can give the analytic continuation of inverses. 
First, we see that 
$$
v^{\ctt}{*}(z{+}\frac{1}{i\h}u{\ctt}v)_{*+}^{-1}=
u{*}(\frac{1}{2}{+}\frac{1}{i\h}u{\ctt}v)_{*+}^{-1}{*}
(z{+}\frac{1}{i\h}u{\ctt}v)_{*+}^{-1}=
u{*}\frac{1}{z{-}\frac{1}{2}}
\Big((\frac{1}{2}{+}\frac{1}{i\h}u{\ctt}v)_{*+}^{-1}{-}(z{+}\frac{1}{i\h}u{\ctt}v)_{*+}^{-1}\Big).
$$
Hence $v^{\ctt}{*}(z{+}\frac{1}{i\h}u{\ctt}v)_{*+}^{-1}{*}v$ is
well-defined, if $z{\not=}\frac{1}{2}$. More directly we have  
\begin{equation*}
\begin{aligned}
v^{\ctt}{*}(z{+}\frac{1}{i\h}u{\ctt}v)_{*+}^{-1}{=}&
\Big(u*\!\int_{{-}\infty}^{0}
e_*^{t(\frac{1}{i\h}u{\ctt}v{+}\frac{1}{2})}dt\Big)*\!
\int_{{-}\infty}^{0}\!\!e_*^{s(z{+}\frac{1}{i\h}u{\ctt}v)}ds   
{=}
u*\!\int_{{-}\infty}^{0}\int_{{-}\infty}^{0}
\!\!e_*^{t(\frac{1}{i\h}u{\ctt}v{+}\frac{1}{2})}{*}
e_*^{s(z{+}\frac{1}{i\h}u{\ctt}v)}dtds \\
{=}&
\int_{{-}\infty}^{0}\int_{{-}\infty}^{0}\!\!e^{t\frac{1}{2}{+}sz}
u{*}e_*^{(t{+}s)\frac{1}{i\h}u{\ctt}v}dtds
\end{aligned}
\end{equation*}
Furthermore, if 
$(z{-}1{+}\frac{1}{i\h}u{\ctt}v)_{*+}^{-1}$ is already 
defined, then we continue to compute as follows:

\begin{equation*}
\begin{aligned}
{=}
\int_{{-}\infty}^{0}\int_{{-}\infty}^{0}
e^{t\frac{1}{2}{+}sz-(t{+}s)}
e_*^{(t{+}s)\frac{1}{i\h}u{\ctt}v}{*}u dtds.
\end{aligned}
\end{equation*}
Hence, we have the identity whenever both sides are defined: 
$$
\begin{aligned}
(v^{\ctt}{*}&(z{+}\frac{1}{i\h}u{\ctt}v)_{*+}^{-1}){*}v{=}
\int_{{-}\infty}^{0}\int_{{-}\infty}^{0}
e^{-t\frac{1}{2}{+}s(z{-}1)}
e_*^{(t{+}s)\frac{1}{i\h}u{\ctt}v}{*}(u{*}v)dtds\\
{=}&
\int_{{-}\infty}^{0}(u{*}v){*}e_*^{t\frac{1}{i\h}u{*}v}dt*\!
\int_{{-}\infty}^{0}e_*^{s(z{-}1{+}\frac{1}{i\h}u{\ctt}v)}ds
{=}(1{-}\varpi_{00}){*}
(z{-}1{+}\frac{1}{i\h}u{\ctt}v)_{*+}^{-1}.
\end{aligned}
$$
Noting that 
$$
\varpi_{00}{*}(z{-}1{+}\frac{1}{i\h}u{\ctt}v)_{*+}^{-1}{=}
(z{-}1{+}\frac{1}{i\h}u{\ctt}v)_{*+}^{-1}{*}\varpi_{00}{=}
(z{-}\frac{1}{2})^{-1}\varpi_{00}, 
$$  
whenever $(z{-}1{+}\frac{1}{i\h}u{\ctt}v)_{*+}^{-1}$ is defined, 
we have   
\begin{equation}
  \label{eq:inverse}
\big(v^{\ctt}{*}(z{+}\frac{1}{i\h}u{\ctt}v)_{*+}^{-1}\big){*}v
{+}(z{-}\frac{1}{2})^{-1}\varpi_{00}
=\big(z{-}1{+}\frac{1}{i\h}u{\ctt}v\big)_{*+}^{-1}.  
\end{equation}
Since $(z{-}\frac{1}{2})^{-1}\varpi_{00}$ is defined for $z{\not=}1/2$, 
we see that \eqref{eq:inverse} gives the formula for analytic 
continuation. Namely, replacing $z$ by $z{+}1$, we define 
the r.h.s. by the l.h.s. 
$$
\big(v^{\ctt}{*}(z{+}1{+}\frac{1}{i\h}u{\ctt}v)_{*+}^{-1}\big){*}v
{+}(z{+}\frac{1}{2})^{-1}\varpi_{00}
=\big(z{+}\frac{1}{i\h}u{\ctt}v\big)_{*+}^{-1},\quad {\rm{Re}}\,z>-\frac{3}{2}.
$$

Repeating this, we have the following formula: For 
${\rm{Re}}\,z>-(n{+}\frac{1}{2})$, 
\begin{equation}\label{weakmatrixrep}
\begin{aligned}
\big(z{+}\frac{1}{i\h}u{\ctt}v\big)_{*+}^{-1}
=&\sum_{k=0}^{n-1}(z{+}k{+}\frac{1}{2})^{-1}
(v^{\ctt})^k{*}\varpi_{00}{*}v^k{+}
(v^{\ctt})^n{*}
\big(z{+}n{+}\frac{1}{i\h}u{\ctt}v\big)_{*+}^{-1}{*}v^n,\\
\big(z{-}\frac{1}{i\h}u{\ctt}v\big)_{*-}^{-1}
=&-\sum_{k=0}^{n-1}(-z{+}k{+}\frac{1}{2})^{-1}
(u^{\btt})^k{*}\varpi_{00}{*}u^k{-}
(u^{\btt})^n{*}
\big(-z{+}n{-}\frac{1}{i\h}u{\ctt}v\big)_{*-}^{-1}{*}u^n.
\end{aligned}
\end{equation}
By using \eqref{eq:powerinv2} these may be written for 
$z$ such as ${\rm{Re}}\,z>-n{-}\frac{1}{2}$ 
\begin{equation*}
\begin{aligned}
\big(z{+}\frac{1}{i\h}u{\ctt}v\big)_{*+}^{-1}
&=\sum_{k=0}^{n-1}(z{+}k{+}\frac{1}{2})^{-1}E_{k,k}{+} 
(v^{\ctt})^n{*}
\big(z{+}n{+}\frac{1}{i\h}u{\ctt}v\big)_{*+}^{-1}{*}v^n,\\
\big(z{-}\frac{1}{i\h}u{\ctt}v\big)_{*-}^{-1}
&=\sum_{k=0}^{n-1}(z{-}k{-}\frac{1}{2})^{-1}\overline{E}_{k,k}{+}
(u^{\btt})^n{*}
\big(z{-}n{-}\frac{1}{i\h}u{\ctt}v\big)_{*-}^{-1}{*}u^n.
\end{aligned}
\end{equation*}
(See also a comment after Proposition\,\ref{twoinv}.)

Note that if $z{=}z_0$ is fixed then for a sufficiently large 
$n$, $(v^{\ctt})^n{*}
\big(z{+}n{+}\frac{1}{i\h}u{\ctt}v\big)_{*+}^{-1}{*}v^n$ 
is holomorphic on a neighborhood of $z_0$. 

\medskip
The residue at a singular point $z_0$ is given as usual by  
$\frac{1}{2\pi i}
\int_{C_{z_0}}(z{+}\frac{1}{i\h}u{\ctt}v)_{*\pm}^{-1}dz$, where 
$C_{z_0}$ is a small circle with the center at $z_0$. 
The analytic continuation formula \eqref{weakmatrixrep} gives the following:   

\begin{thm}\label{resres}
${\rm{Res}}((z{+}\frac{1}{i\h}u{\ctt}v)_{*+}^{-1}, 
{-}(n{+}\frac{1}{2}))$ 
is $\frac{1}{(i\h)^nn!}u^{n}{*}\varpi_{00}{*}v^n$ in generic 
ordered expressions. This is $E_{n,n}$ by \eqref{EE}.

Similarly, 
${\rm{Res}}((z{-}\frac{1}{i\h}u{\ctt}v)_{*-}^{-1}, 
{-}(n{+}\frac{1}{2}))$ 
is $-\frac{1}{(i\h)^nn!}v^{n}{*}
{\overline{\varpi}}_{00}{*}u^n$ in generic 
ordered expressions. This is $-\overline{E}_{n,n}$ $($cf.\eqref{EE2}$)$.
\end{thm}
It is remarkable that the singular points depend only on the growth
order of $e_*^{t\frac{1}{i\h}u{\ctt}v}$ which is independent 
of the expression parameters.  

Since 
$\big(z{+}\frac{1}{i\h}u{\ctt}v\big)_{*-}^{-1}=
{-}\big({-}z{-}\frac{1}{i\h}u{\ctt}v\big)_{*-}^{-1}$, 
Theorem\,\ref{resres} shows also 
\begin{thm}  \label{contnuation}
In generic ordered expressions, 
the inverses $(z{+}\frac{1}{i\h}u{\ctt}v)_{*+}^{-1}$,
$(z{-}\frac{1}{i\h}u{\ctt}v)_{*-}^{-1}$ extend to 
$H{\!o}l({\mathbb C}^2)$-valued  holomorphic functions of $z$ on 
${\mathbb C}{\setminus}\{-({\mathbb N}{+}\frac{1}{2})\}$
with simple poles. Namely, for every $n$
$$
{:}(z{+}\frac{1}{i\h}u{\ctt}v)_{*+}^{-1}{:}_{_K}
{-}\!\sum_{k=0}^{n}(z{+}k{+}\frac{1}{2})^{-1}E_{k,k}(K),\quad 
{:}(z{+}\frac{1}{i\h}u{\ctt}v)_{*-}^{-1}{:}_{_K}
{-}\!\sum_{k=0}^{n}(z{-}k{-}\frac{1}{2})^{-1}\overline{E}_{k,k}(K)
$$
are holomorphic on the domain ${\rm{Re}}\,z>-n{-}\frac{1}{2}$. 
However, 
$$
\sum_{k=0}^{\infty}(z{+}k{+}\frac{1}{2})^{-1}E_{k,k}(K), \quad 
\sum_{k=0}^{\infty}(z{-}k{-}\frac{1}{2})^{-1}\overline{E}_{k,k}(K)
$$ 
may not converge in $H\!ol({\mathbb C}^2)$ in the sense of partial
fractions. They converge only for $K\in{\mathfrak K}_+$, $K\in{\mathfrak K}_-$ 
respectively.
\end{thm}

The next result may sound strange   
\begin{thm}\label{continv}
If $K{\in}{\mathfrak K}_0$, then those three elements are inverse of 
$z{+}\frac{1}{i\h}u{\ctt}v$:
$$
{:}(z{+}\frac{1}{i\h}u{\ctt}v)_{*+}^{-1}{:}_{_K},\quad
\sum_{n=-\infty}^{\infty}(z{+}n)^{-1}D_{n,n}(K),\quad
{:}(z{+}\frac{1}{i\h}u{\ctt}v)_{*-}^{-1}{:}_{_K}
$$
They are holomorphic mappings respectively of  
${\mathbb C}{\setminus}\{-(\mathbb N{+}\frac{1}{2})\}$,
${\mathbb C}{\setminus}\{\mathbb Z\}$,  
${\mathbb C}{\setminus}\{\mathbb N{+}\frac{1}{2}\}$ into $H\!ol({\mathbb C}^2)$.
\end{thm}

Now, for a fixed $K{\in}{\mathfrak K}_0$, consider 
$z\in{\mathbb C}$ where 
${:}(z{+}\frac{1}{i\h}u{\ctt}v){:}_{_K}$ fails to be invertible,
which may be called the ``spectre'' of 
$\frac{1}{i\h}u{\ctt}v$ as in the operator theory. 
Theorem\,\ref{continv} shows that ${:}\frac{1}{i\h}u{\ctt}v{:}_{_K}$
cannot be viewed as a single element from a view point of operator 
representations, as this has three different kinds of specters

\end{document}